\documentclass[11pt,a4paper,parskip=half,english]{scrartcl}

\usepackage[utf8]{inputenc}
\usepackage[T1]{fontenc}
\usepackage{lmodern}
\usepackage{authblk}

\usepackage{geometry}
\usepackage{microtype}
\newcommand{\dd}{\mathop{}\!\mathrm{d}}
\usepackage{physics}            
\usepackage{amsmath, amssymb, amsthm}

\usepackage{graphicx}
\usepackage[format=plain, labelfont=bf, justification=justified, singlelinecheck=false, font=small]{caption}
\usepackage{subcaption}
\usepackage{setspace}
\usepackage{float}
\usepackage{booktabs}
\usepackage{array}
\usepackage{makecell}           

\usepackage[hidelinks]{hyperref}
\usepackage[backend=biber, style=nature, defernumbers=true]{biblatex}
\DeclareBibliographyCategory{methodsonly}
\addtocategory{methodsonly}{kim2023chiral,Exfol1,Exfol2}
\usepackage{siunitx}
\usepackage{xcolor}
\usepackage{enumitem}

\makeatletter
\newcommand{\equalcontrib}{\textsuperscript{\@fnsymbol{1}}}
\makeatother

\title{Topological Control of Polaritonic Flatbands in Anisotropic van der Waals Metasurfaces\\}
\renewcommand\Authfont{\normalsize\sffamily}
\renewcommand\Affilfont{\footnotesize\itshape\rmfamily}
\author[1,2]{Connor Heimig\thanks{These authors contributed equally to this work.}}
\author[2]{Thomas Weber\equalcontrib}
\author[3]{Cristina Cruciano}
\author[3,4]{Armando Genco\thanks{armando.genco@unipi.it}}
\author[2]{Thomas Possmayer}
\author[2]{Luca Sortino}
\author[3,5]{Gianluca Valentini}
\author[5]{Cristian Manzoni}
\author[6,7]{Maxim V. Gorkunov}
\author[3,5]{Giulio Cerullo}
\author[2]{Alexander A. Antonov\thanks{alexander.antonov@tuhh.de}}
\author[1,2]{Andreas~Tittl\thanks{andreas.tittl@tuhh.de}}

\affil[1]{Institute of Photonics, Hamburg University of Technology, Hamburg, Germany}
\affil[2]{Chair in Hybrid Nanosystems, Nanoinstitute Munich, Faculty of Physics, Ludwig-Maximilians-Universit{\"a}t M{\"u}nchen, Munich, Germany}	
\affil[3]{Dipartimento di Fisica, Politecnico di Milano, Milano, Italy}	
\affil[4]{Dipartimento di Fisica, Università di Pisa, 56127 Pisa, Italy}
\affil[5]{IFN-CNR, Istituto di Fotonica e Nanotecnologie, Milano, Italy}
\affil[6]{National Research Nuclear University 'MEPhI' (Moscow Engineering Physics Institute), Moscow, Russia}	
\affil[7]{Theoretical Physics and Quantum Technologies Department, National University of Science and Technology ‘MISIS’, Moscow, Russia}
\date{}

\makeatletter
\let\SI@maketitle\maketitle
\let\SI@author\@author   
\let\SI@thanks\thanks
\let\SI@and\and
\makeatother

\makeatletter
\newcommand{\startSI}{%
  \clearpage
  \setcounter{figure}{0}\setcounter{table}{0}%
  \setcounter{equation}{0}\setcounter{section}{0}%
  \setcounter{footnote}{0}%
  \setcounter{page}{1}\renewcommand{\thepage}{S\arabic{page}}
  \renewcommand{\thefigure}{S\arabic{figure}}%
  \renewcommand{\thetable}{S\arabic{table}}%
  \renewcommand{\theequation}{S\arabic{equation}}%
  \renewcommand{\thesection}{S\arabic{section}}%
  \renewcommand{\theHfigure}{S\arabic{figure}}%
  \renewcommand{\theHtable}{S\arabic{table}}%
  \renewcommand{\theHequation}{S\arabic{equation}}%
  \renewcommand{\theHsection}{S\arabic{section}}%
  \clubpenalty=150 \widowpenalty=150
  \everydisplay{\@displaytrue}\SetSinglespace{1.15}%
  \setcounter{secnumdepth}{3}
  \addtocontents{toc}{\protect\setcounter{tocdepth}{3}}
  \let\maketitle\SI@maketitle
  \let\thanks\SI@thanks\let\and\SI@and
  \let\@author\SI@author
  \title{Supporting Information: Topological Control of Polaritonic Flatbands in Anisotropic van der Waals Metasurfaces}%
  \renewcommand\Authfont{\footnotesize\sffamily}%
  \renewcommand\Affilfont{\scriptsize\itshape\rmfamily}%
  \date{}%
  \maketitle
}
\makeatother

\begin{document}

\addtocontents{toc}{\protect\setcounter{tocdepth}{-10}}
\begin{refsection}[ReS2BIC_references.bib]
\maketitle
\vspace*{-1.2 cm}
\begin{abstract}

Anisotropic van der Waals (vdW) materials exhibit direction-dependent optical and electronic properties, enabling tailored directional light-matter interactions. Rhenium disulfide (ReS\textsubscript{2}) stands out among them for its strong in-plane anisotropy and thickness-independent direct-bandgap excitons. These excitons can hybridize with light to form exciton-polaritons, rendering ReS\textsubscript{2} a compelling material for nanophotonic integration. Metasurfaces are arrays of subwavelength resonators and can support ultra-sharp quasi-bound state in the continuum (qBIC) resonances. Topological transformations of their bandstructure can give rise to flatbands, characteristic for dispersionless states with quenched kinetic energy and vanishing group velocity. 
\noindent Here we show that structuring ReS\textsubscript{2} into a resonant metasurface reshapes its photonic band topology, splitting the qBIC's integer topological charge into two half-integer polarization singularities and driving the formation of directional far-field flatbands. Tuned into resonance with ReS\textsubscript{2}'s linearly polarized excitonic transitions these flatband states hybridize into distinct directionally polarized exciton-polaritons. This points toward a scalable route to polarization-resolved quantum emitters; flatband-enhanced nonlinear optics; and low-threshold polariton devices.
\end{abstract}

\newpage

\section{Introduction}

Anisotropic materials exhibit direction-dependent optical, electronic, and mechanical properties, giving rise to phenomena from optical birefringence to anisotropic electron transport in topological semimetals and direction-dependent mechanical responses in DNA under tension.~\cite{nye_physical_properties_of_crystals, Principles_of_optics, Weyl-Fermions, DNA_Bending} In nanophotonics, this anisotropy offers a route to tailor light-matter interactions, enhance optical selectivity, and engineer quantum states of light and matter.~\cite{zhu2024strong}

Among 2D materials, transition metal dichalcogenides (TMDCs) are particularly well suited to exploiting anisotropy at the atomic scale. While commonly studied group-VI TMDCs like MoS\textsubscript{2} and WS\textsubscript{2} are optically isotropic in-plane due to their hexagonal symmetry~\cite{cihan2018silicon, blundo2020evidence}, the group-VII TMDC rhenium disulfide (ReS\textsubscript{2}) exhibits strong in-plane anisotropy from its distorted triclinic 1\textit{T'} structure~\cite{Jariwala2016}, giving rise to multiple linearly polarized excitonic resonances with axis-aligned transition dipoles~\cite{rydberg, aslan2015linearly}. Uniquely, these excitonic and direct-bandgap properties are nearly layer-independent~\cite{tongayReS2, stacking}, allowing anisotropy and direct bandgap characteristics to persist from monolayer to bulk and simplifying integration into thickness-tolerant photonic architectures~\cite{corbet2015field}.

Photonic metasurfaces -- subwavelength resonator arrays with engineered optical properties at the nanoscale~\cite{kildishev} -- can support symmetry-protected, non-radiative bound states in the continuum (BICs), which are fully decoupled from the far-field and can possess theoretically infinite Q-factors~\cite{hsuBIC}. Introducing controlled symmetry perturbations converts BICs into quasi-BICs (qBICs), which acquire finite linewidths and become radiatively accessible~\cite{tittl2018imaging,koshelev2018asymmetric}. In C\textsubscript{4}-symmetric metasurfaces, fourfold rotational symmetry enforces polarization-invariant qBICs~\cite{vaity2022polarization,yang2025polarization}. More broadly, qBIC metasurfaces are attractive for strong coupling to excitonic transitions due to their narrow linewidths~\cite{koshelev2018strong}.
Fabricating the metasurface directly from an excitonic material co-localizes photonic and excitonic components, enabling self-hybridized strong coupling within a single material system; routinely yielding Rabi splittings exceeding 100 meV at room temperature, simplified fabrication, and access to polaritons with various polarizations~\cite{weber2023intrinsic, heimig2026chiral}.


Topological concepts offer a robust framework for manipulating metasurface resonant properties. In qBIC metasurfaces, topology controls the momentum-space singularities of the far-field polarization. True BICs coincide with so-called V-points -- isolated points in momentum space where the far-field polarization vector winds around a singularity, causing the radiation to vanish entirely; the winding number of this vortex defines an integer topological charge~\cite{zhen2014topological}. A second type of polarization singularity occurs at mode degeneracies, characteristic of C\textsubscript{4}- or C\textsubscript{6}-symmetric metasurfaces, where two orthogonal modes coincide in frequency and the far-field polarization cannot be uniquely defined, since any linear combination of the degenerate modes is an equally valid eigenstate~\cite{liu2021topological}. Reducing the metasurface's rotational symmetry splits such singularities (V-points, or points of degeneracy) into multiple singularities, including circularly polarized C-points~\cite{liu2019circularly} and half-integer Dirac points~\cite{chen2019observing}, while conserving the total topological charge. This provides design principles for engineering qBICs with tailored emission properties.

Flatbands, whose energy does not disperse with light momentum, are closely tied to interaction-driven and topological effects, from graphene edge excitations~\cite{yao2009edge} to fractional quantum Hall states~\cite{wang2011fractional}. Topological flatbands have been realized in photonic and polaritonic systems, where their enhanced density of states gives rise to unconventional transport phenomena~\cite{leykam2018artificial} and polariton condensation in various geometries~\cite{klembt2017polariton,harder2021kagome,scafirimuto2021tunable,jacqmin2014direct}.

Here, we demonstrate that a C\textsubscript{4}-symmetric photonic metasurface fabricated directly from anisotropic bulk ReS\textsubscript{2} hosts topological singularities with fractional charge and enables robust, spectrally scalable photonic flatbands, without requiring the ultra-subwavelength Reststrahlenband regime~\cite{nan2025angular}, detuned exciton-polariton band reshaping~\cite{mischok2024breaking}, delicate BIC merging~\cite{le2024super}, or finely-tuned superlattice interference~\cite{nguyen2018symmetry, qin2024optical}. Whereas a C\textsubscript{4}-symmetric structure made of isotropic material enforces polarization-degenerate qBICs, ReS\textsubscript{2}'s intrinsic in-plane anisotropy breaks this degeneracy, splitting the qBIC into two polarization-dependent modes and fragmenting its topological charge into displaced half-integer vortices. This topological reshaping yields directional qBIC flatbands that remain accessible in the far-field, each nearly dispersionless along one momentum direction while retaining parabolic dispersion along the orthogonal one. We verify this mechanism through full-wave simulations and Fourier-space imaging measurements, and analytically capture the topological transitions via resonant-state expansion.
Tuning these polarization-dependent photonic flatbands into ReS\textsubscript{2}'s anisotropic excitonic resonances yields polarization-resolved exciton-polariton regimes with directional strong coupling. The resulting polaritonic flatbands offer enhanced local density of states, long lifetimes, and increased nonlinear interactions, opening avenues for coherent directional light sources, polariton-based quantum logic, and topological photonic devices~\cite{st2017lasing, whittaker2018exciton, solnyshkov2021microcavity, klembt2018exciton}.

\section{Results}
\subsection{Anisotropy lifts Mode Degeneracy}

We first study a dielectric metasurface composed of square nanoresonators with side length $w_0$, arranged  on a silicon dioxide (SiO\textsubscript{2}) substrate to form a C$_4$-rotational symmetry.
Given this symmetry, the structure supports a doubly degenerate BIC at the $\Gamma$-point, provided the constituent medium is isotropic.

\begin{figure}[h!]
    \centering
    \includegraphics[width=1\linewidth]{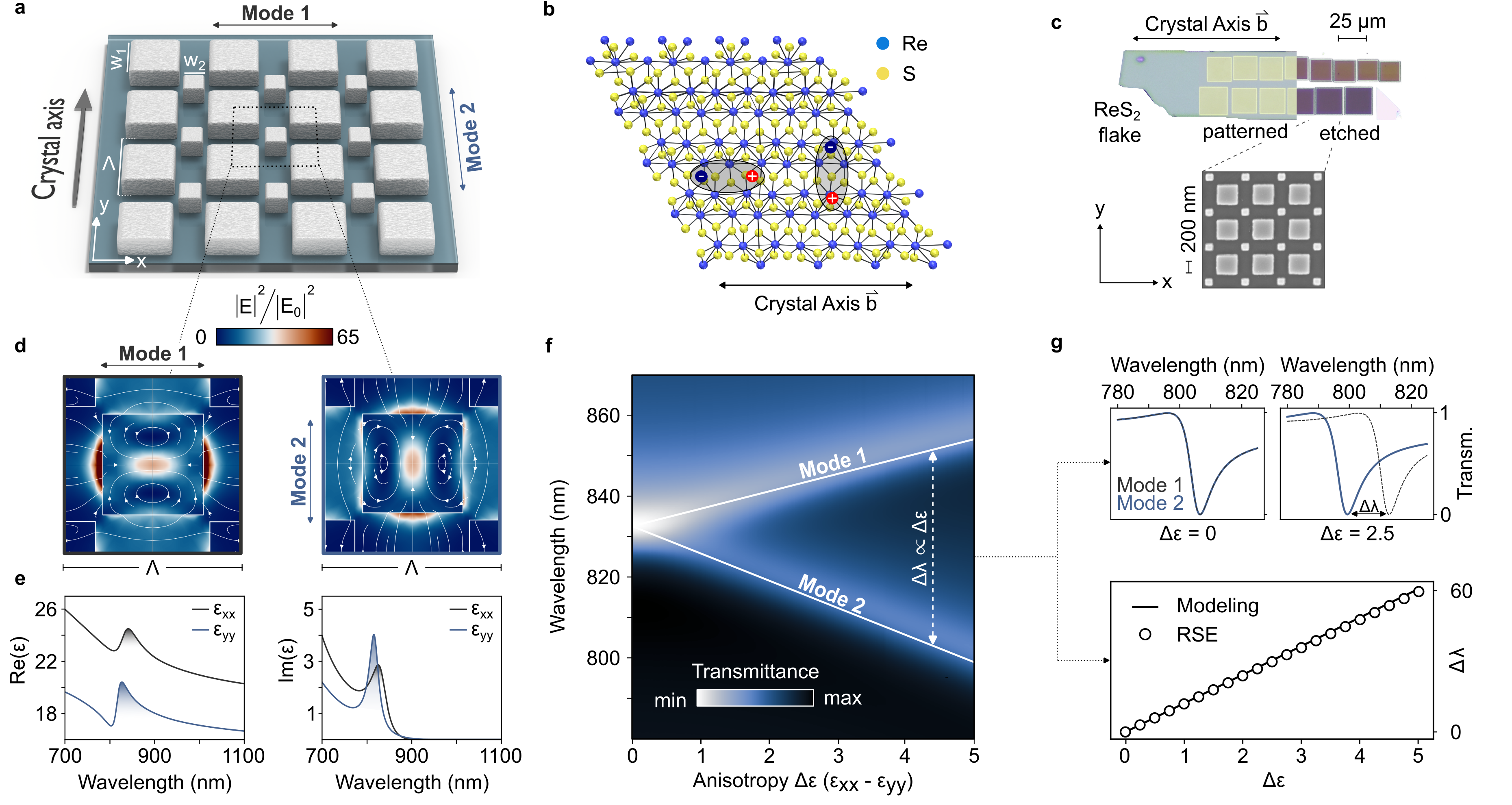}
    \caption{\setstretch{1}\textbf{Material Anisotropy-Induced qBIC Degeneracy Lifting}. 
    \textbf{a}~ReS\textsubscript{2} metasurface unit cell design with periodicity $\Lambda= 465$~nm, base side length $w_0$ = 220 nm, $\alpha$ = 0.6,  and resonator height $h = \SI{90}{\nano\metre}$. 
    \textbf{b}~Schematic of the crystal structure of ReS\textsubscript{2}, each monolayer consists of a central plane of rhenium atoms positioned between two layers of sulfur atoms. The rhenium atoms are coordinated in a distorted trigonal antiprismatic geometry, forming strong covalent bonds with the surrounding sulfur atoms.~\cite{rydberg}
    \textbf{c}~Optical microscope image of an exfoliated ReS\textsubscript{2} flake. The orientation of the crystal axes is identified from the flake geometry, providing guidance for the alignment of the metasurfaces (shown in SEM image) with respect to the in-plane anisotropy.
    \textbf{d}~Simulated electric field enhancement for Mode 1 (left) and Mode 2 (right), evaluated at the middle-cut of the resonators. $E_0$ is the excitation wave amplitude. 
    \textbf{e}~In-plane components of the real and imaginary parts of the permittivity $\varepsilon$ adapted to experiments (see Supplementary Note~1). 
    \textbf{f}~Simulated transmittance for both Mode 1 and Mode 2, showing the splitting of the qBICs as a function of dielectric anisotropy $\Delta \varepsilon = \varepsilon_{xx} - \varepsilon_{yy}$.  
    \textbf {g}~Simulated transmittance spectra for selected values of $\Delta \varepsilon$, illustrating the increasing spectral separation of the modes with anisotropy. Comparison between modeling and RSE of the spectral splitting $\Delta \lambda \propto\Delta\varepsilon$. 
    }
    \label{fig:fig1}
\end{figure}

To introduce asymmetry, thereby transforming the BIC into a qBIC, while preserving the overall C$_4$-symmetric {\color{black} shape}, we define a perturbation parameter $\alpha$, and express the new prisms side lengths (Fig.~\ref{fig:fig1}a):
\[
w_1 = w_0 \left( \frac{2}{1 + \alpha^2} \right)^{1/2}, \quad
w_2 = w_0 \left( \frac{2}{1 + \alpha^2} \right)^{1/2} \alpha,
\]
so that the in-plane footprint remains constant, while the unit cell increases in size by a factor of $\sqrt{2}$, leading to Brillouin zone folding (BZF). As a result, the degenerate BIC at the original $\Gamma$-point is transformed into two radiative qBICs (BZF-qBICs~\cite{wang2023brillouin}), which we denote as Mode 1 and Mode 2.
{\color{black}Note that C\textsubscript{4}-symmetry here refers to the shape of the resonator lattice: once the resonators are fabricated from ReS\textsubscript{2}, the anisotropic permittivity tensor reduces the overall symmetry of the full system to C\textsubscript{2}.}
Spectral tunability of the qBICs is enabled by applying an in-plane scaling factor $S$ to the structure.
The electric field profiles of the both modes are shown in Fig. \ref{fig:fig1}d. Both field distributions exhibit the same spatial shape, rotated by 90°, as expected from the orthogonal polarizations of the excitation.

The controlled alignment of the metasurface coordinate system ($x,y$) with the ReS\textsubscript{2} crystal axis \textit{b}, which defines the principal crystallographic direction (Fig.~\ref{fig:fig1}b, see Methods), establishes a direct correspondence between the metasurface geometry and the intrinsic optical axes of the material (Fig.~\ref{fig:fig1}c).
 As a result, polarization-selective excitation along these orthogonal directions probes the distinct spectral responses of the anisotropic permittivity tensor components $\varepsilon_{xx}$ and $\varepsilon_{yy}$ (Fig.~\ref{fig:fig1}e), thereby enabling anisotropy-induced degenerate mode splitting.
 We account for the in-plane optical anisotropy of ReS\textsubscript{2} by first employing a simplified non-dispersive material model that captures this splitting while neglecting (for now) excitonic contributions:
\begin{equation}
\begin{array}{l@{\quad\quad}l@{\quad\quad}l}
\varepsilon_{xx} = \varepsilon + \dfrac{\Delta \varepsilon}{2}, &
\varepsilon_{yy} = \varepsilon - \dfrac{\Delta \varepsilon}{2}.
\end{array}\label{eq:delta_epsilon}
\end{equation}
This simple model isolates the purely photonic behavior of the metasurface before exciton coupling is included through Lorentzian poles (Supplementary Note~1).
We assume $\varepsilon = 18$, $\Delta \varepsilon = \varepsilon_{xx} - \varepsilon_{yy}\approx1.7$ (equivalent to $\Delta n \approx 0.2$)  and constant out-of-plane component $\varepsilon_{zz}$ = 7.25. For $\Delta \varepsilon = 0$ the structure supports a doubly degenerate qBIC; as $\Delta \varepsilon$ becomes nonzero, the degeneracy of the modes lifts, with Mode~1 redshifting and Mode~2 blueshifting (Fig.~\ref{fig:fig1}f).

To further study this splitting mechanism, we apply resonant-state expansion (RSE), treating the in-plane anisotropy $\varepsilon$ as a perturbation,  which yields a pair of hybrid eigenfrequencies:
\begin{equation}
    \omega_{\pm}  \approx \omega_0 [1 \pm \varkappa \Delta \varepsilon]
    \label{eq:caption1}
\end{equation}

where $\varkappa$ is a constant factor calculated through overlap integrals between the modes and  $\omega_0$ is the eigenfrequency of the initial degenerate eigenstates (Supplementary Note~2). The perturbative result obtained via RSE closely matches the full numerical simulations, as shown in Fig. \ref{fig:fig1}e. Therefore, the anisotropy-induced spectral mode splitting 
$\Delta \lambda = 2\pi c(\omega_{-}^{-1} - \omega_{+}^{-1}) \approx 4\pi c\omega_0^{-1} \varkappa \Delta \varepsilon$  scales linearly with $\Delta \varepsilon$ (Fig. \ref{fig:fig1}g).

\subsection{Fractional Topological Charges in Momentum Space}

\begin{figure}[H]
    \centering
    \includegraphics[width=\linewidth]{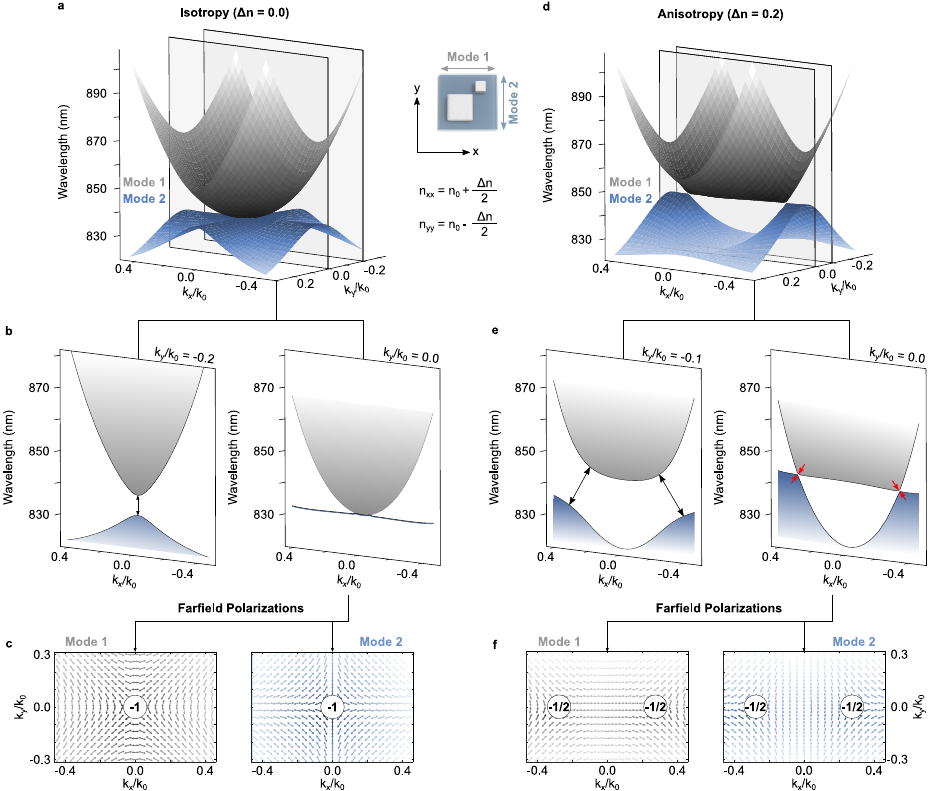}
    \caption{\textbf{Anisotropy-induced Modal Topological Transformation.} 
    \textbf{a} Simulated band structure of Mode~1 (gray) and Mode~2 (blue) in $k$-space for the isotropic case ($\Delta n = 0$). 
    \textbf{b} Band structure cuts at constant $k_y$.
    \textbf{c} Far-field polarization maps, showing topological charges of $-1$ at the $\Gamma$-point.
    \textbf{d} Simulated band structure for the anisotropic case ($\Delta n = 0.2$) with \textbf{e} the corresponding $k_y$-cuts, showing the Dirac points location by red arrows (points where the modes touch).
    \textbf{f} Far-field polarization maps show the splitting of the topological charge into two half-charges ($-1/2$), symmetrically displaced along the $k_x$-axis.  The metasurface with a period of $\Lambda = 535$~nm, base side length $w_0 = 250$~nm, $\alpha = 0.6$, and resonator height $h = 70$~nm was placed in vacuum to eliminate any parasitic diffraction associated with the substrate and to facilitate the qBIC tracking in momentum space. Its parameters were slightly adjusted so that the qBIC wavelength of $835$ nm at $\Delta n = 0$ lies between the two exciton resonances of ReS$_2$. All simulations were performed using the non-dispersive, non-excitonic material model with refractive indices $n_0=\sqrt{18}$, $n_{xx}=n_0+\Delta n/2$, and $n_{yy}=n_0-\Delta n/2$.}
    \label{fig:fig2}
\end{figure}

Starting in the isotropic case ($\Delta n = 0$) we trace the evolution of Mode 1 and Mode 2 in k-space with $k_0$ defined as a free-space wavevector (Fig.~\ref{fig:fig2}a). While Mode 1 exhibits an isotropic parabolic dispersion centered at the $\Gamma$-point (i.e. $k_x$ = $k_y$ = 0), Mode 2 remains nearly flat along the high-symmetry directions $k_x$ = 0 and $k_y$ = 0, bending slightly along the diagonals.  In the isotropic case, the metasurface supports a mode degeneracy at the $\Gamma$-point, which is lifted when moving to nonzero $k_x$ and $k_y$ values (Fig.~\ref{fig:fig2}b). 
The far-field polarization vector fields (Fig.~\ref{fig:fig2}c) reveal integer winding numbers of $-1$ centered at the $\Gamma$-point for both modes, indicating the presence of polarization vortices with an integer topological charge. Unlike true BICs, which exhibit vanishing far-field polarization at the $\Gamma$-point, the degenerate qBICs retain undetermined far field polarizations at the $\Gamma$-point.

A small in-plane anisotropy ($\Delta n = 0.2$, matching the value for ReS\textsubscript{2}) already substantially modifies the dispersions of both modes in $k$-space (Fig.~\ref{fig:fig2}d). 
In the $k_y=0$ ($k_x=0$) plane, the $x$-polarized Mode 1 ($y$-polarized Mode 2), exhibits a flatband, while the other mode develops a pronounced curvature.
Moreover, the $k_y = 0$ cut (Fig.~\ref{fig:fig2}e) reveals a band crossing between Mode~1 and Mode~2, forming a photonic Dirac cone at nonzero $k_x$ (Fig.~\ref{fig:fig2}d) that mirrors the graphene-like dispersion of two conical bands meeting at a single point~\cite{bravo2012enabling}.
A cut through the $k_y = -0.1k_0$ plane shows the evolution of the mode dispersions under anisotropy, with the flatband character gradually lost and the modes becoming clearly separated in energy (Fig.~\ref{fig:fig2}e).

These observations of anisotropy-induced directional flatbands are accompanied by a characteristic splitting of the topological charges of the underlying modes. The initial $\Gamma$-point integer topological charges (Fig.~\ref{fig:fig2}c) split into a pair of displaced fractional charges at nonzero $\Delta n$ (Fig.~\ref{fig:fig2}f). The fractional charges coincide with mode crossings, forming Dirac points (Fig.~\ref{fig:fig2}d). This topological rearrangement modifies the photonic band curvatures, giving rise to directional flatbands.

We capture the above modal topological transformations with remarkable accuracy by an analytical model that approximates the simulated dispersion at $\Delta n=0$ (Figs.~\ref{fig:fig2}a,b), with Mode~1 following a parabolic dispersion $\omega_1=\omega_0(1-\chi k^2/k_0^2)$, where $k=\sqrt{k_x^2+k_y^2}$, and Mode~2 being dispersionless, i.e. $\omega_2=\omega_0$. Their far-field polarizations are approximated by a pair of mutually orthogonal unit vectors, $\mathbf{e}_1^k=\hat{\mathbf{x}}\ k_y/k + \hat{\mathbf{y}}\ k_x/k$ and $\mathbf{e}_2^k = \hat{\mathbf{x}}\ k_x/k - \hat{\mathbf{y}}\ k_y/k$.

These assumptions enable us to apply the RSE formalism to track the mode evolution under the perturbation $\Delta\varepsilon$ from \eqref{eq:delta_epsilon} across the $k$-space (Supplementary Note~3), yielding the hybrid eigenfrequencies:
\begin{equation}
\omega_\pm(\mathbf{k}) \approx \omega_0\left[1 - \frac{\chi k^2}{2k_0^2} \pm \sqrt{\left( \frac{\chi k^2}{2k_0^2} - \beta\frac{k_x^2 - k_y^2}{2k^2} \right)^2 + \beta^2\frac{k_x^2 k_y^2}{k^4}}\ \right], \label{eq:3}
\end{equation}
where $\beta=\beta_0\Delta\varepsilon$ and $\beta_0$ is the modes' self-action (field self-overlap) integral.

Equation~\eqref{eq:3} directly explains the origin of the anisotropy-induced topological features. A Dirac point occurs when $\omega_+=\omega_-$, i.e., when both terms under the square root vanish simultaneously: the second term vanishes along $k_x=0$ or $k_y=0$, and for $\beta>0$ ($\Delta\varepsilon>0$) the first term also vanishes at $k_y=0$ and $k_x/k_0=\pm\sqrt{\beta/\chi}$, fixing the Dirac point's location in k-space. As for the flatbands, one of the modes stays dispersionless at $\omega_0[1+\beta/2]$ along $k_x=0$, or $\omega_0[1-\beta/2]$ along $k_y=0$.

\subsection{Experimental Photonic Flatbands}

\begin{figure}[]
    \centering
    \includegraphics[width=1\linewidth]{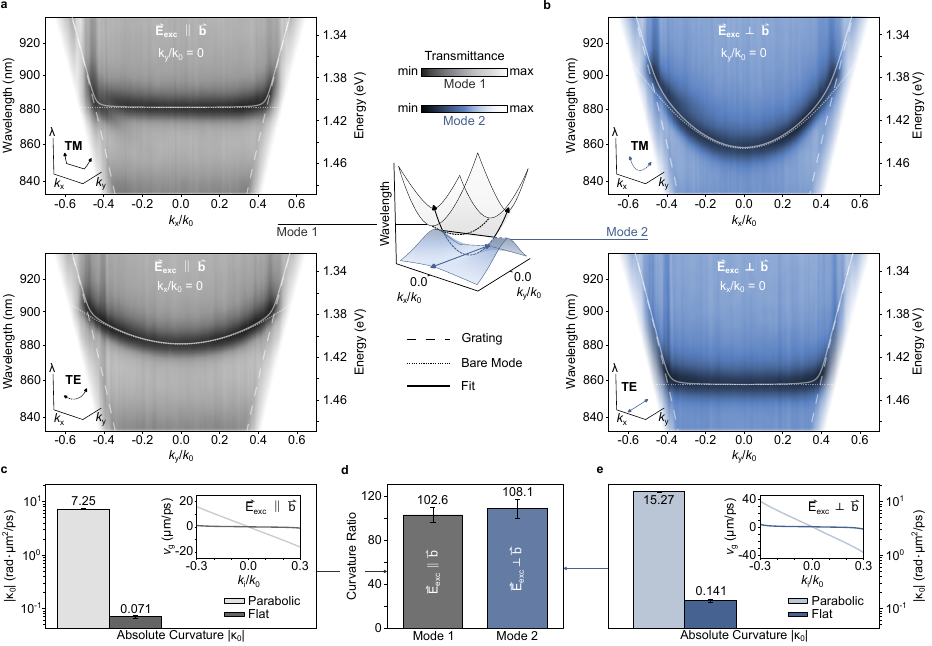}
    \caption{\color{black}\textbf{Experimental Photonic Flatbands.}  
    Angle-resolved Fourier-space transmittance for excitation polarization \textbf{a}
    $\vec{E}_\mathrm{exc}\parallel \vec{b}$ (Mode 1, grey) and \textbf{b}
    $\vec{E}_\mathrm{exc}\perp \vec{b}$ (Mode 2, blue), where $\vec{b}$ denotes the crystal axis (see Methods and Fig.~\ref{fig:fig1}c). 
    For each excitation configuration, TM (upper panels, $k_x/k_0$) and TE (lower panels, $k_y/k_0$) dispersions are shown. The metasurface parameters are $\Lambda=S\cdot465$ nm, $w_0=S\cdot220$ nm, $h=90$ nm, asymmetry $\alpha=0.6$ and in-plane scaling factor $S = 1$. The dashed lines indicate the grating mode associated with the substrate. 
    The dotted lines represent the intrinsic (bare) qBIC in the absence of substrate-induced radiative coupling. The solid lines correspond to a coupled-mode fit including hybridization between the intrinsic qBIC and the grating mode. 
    Depending on the propagation direction in $k$-space, the intrinsic qBIC exhibits either parabolic or strongly reduced curvature. 
    \textbf{c}~Absolute curvature $|\kappa_0|$ at the $\Gamma$-point for the two dispersions under $\vec{E}_\mathrm{exc}\parallel\vec{b}$, with the inset showing the corresponding calculated group velocity $v_g=\partial\omega/\partial k$ as a function of $k_x/k_0$ for flat and $k_y/k_0$ for parabolic branches (fitted values for parabolic $7.25 \pm 0.12$ and flat $0.071 \pm 0.005$).
    \textbf{d}~Curvature ratio between parabolic and flat dispersions (Mode 1: $102.6 \pm 7.0$ and Mode 2: $108.1 \pm8.9$). 
    \textbf{e}~Corresponding curvature and group-velocity analysis for $\vec{E}_\mathrm{exc}\perp\vec{b}$ (fitted values for parabolic $15.27 \pm 0.28$ and flat: $0.141 \pm 0.012$). Absolute curvature $|\kappa_0|$ and curvature ratio are presented as the mean fit values, $\pm$ the standard deviation of the fit.}
    \label{fig:fig3}
\end{figure}

To verify the theoretically predicted anisotropy-induced topological features, we fabricate the C\textsubscript{4}-symmetric ReS\textsubscript{2} metasurface and perform Fourier-space transmittance spectroscopy (see Methods and Supplementary Note~8). The measured dispersions (Figs.~\ref{fig:fig3}a,b) confirm the numerical predictions: for excitation polarization $\vec{E}_\mathrm{exc}\parallel\vec{b}$ (Mode~1, Fig.~\ref{fig:fig3}a), where $\vec{b}$ denotes the crystal axis (see Methods and Fig.~\ref{fig:fig1}c), the dispersion is pronouncedly anisotropic -- parabolic along one in-plane direction and strongly flattened along the orthogonal direction -- while for $\vec{E}_\mathrm{exc}\perp\vec{b}$ (Mode~2, Fig.~\ref{fig:fig3}b) the flat and parabolic directions are interchanged, confirming the predicted {\color{black}directional} photonic flatbands. 

Note that these directional flatbands are not two-dimensional flatbands, as the dispersion is quenched along a single momentum direction only.
To quantify the degree of band flatness, we analyze the curvature of the fitted dispersions extracted from Figs.~\ref{fig:fig3}a,b. The fits account for hybridization between the intrinsic qBICs and a substrate-induced grating mode via a minimal two-mode coupled-oscillator model (see Supplementary Note~6). From the fitted dispersions we compute the group velocity $v_g=\partial\omega/\partial k$ and curvature $\kappa=\partial^2\omega/\partial k^2$, which provide a direct quantitative measure of band flatness.
The resulting curvature analysis is summarized in Figs.~\ref{fig:fig3}c-e. Figure~\ref{fig:fig3}c shows the absolute curvature $|\kappa_0|$ at the $\Gamma$-point for the parabolic and flat Mode~1 dispersions observed for $\vec{E}_\mathrm{exc}\parallel\vec{b}$. The flat branch exhibits a curvature more than two orders of magnitude lower than the parabolic branch, highlighting the strongly reduced group velocity near the $\Gamma$-point (inset). The same analysis for $\vec{E}_\mathrm{exc}\perp\vec{b}$ is shown in Fig.~\ref{fig:fig3}e, confirming that the flattened dispersion persists under rotation of the excitation polarization.
Fig.~\ref{fig:fig3}d compares the curvature ratios between the parabolic and flat dispersions, revealing suppression factors exceeding two orders of magnitude for both modes. 
Note, that even for an intrinsically flat qBIC dispersion ($\chi=0$), hybridization with the grating channel introduces a residual curvature at the $\Gamma$-point, so the experimentally extracted flatband branches show a finite, strongly suppressed curvature rather than vanishing exactly. We also find that the flatbands extend across $\SI{93.7}{\percent}$ of the accessible $k$-space region for Mode 1 and $\SI{91.5}{\percent}$ for Mode 2, limited only by the onset of diffraction (see Supplementary Note~6).

\subsection{Experimental Dirac Points}

\begin{figure}[h!]
    \centering
    \includegraphics[width=1\linewidth]{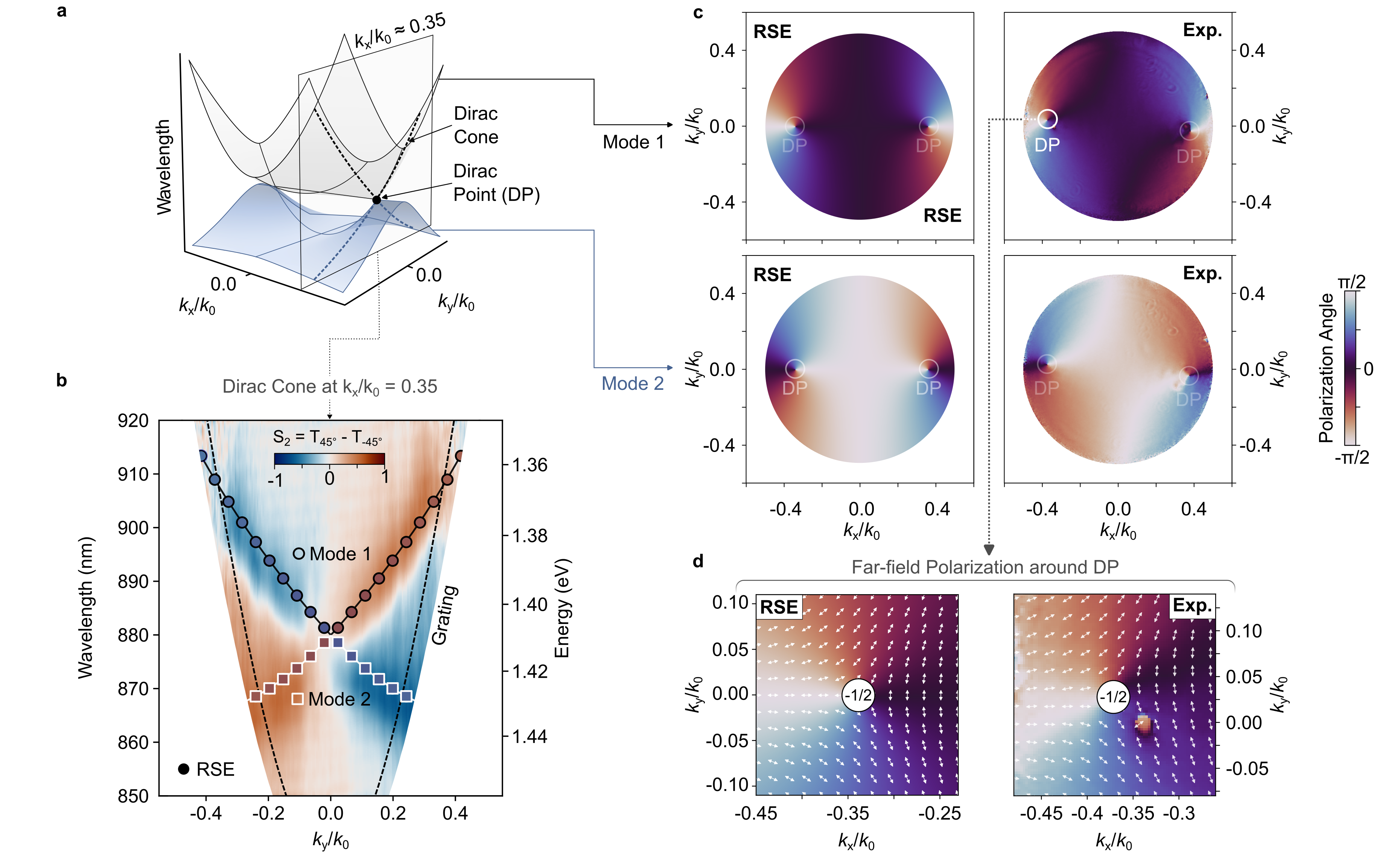}
    \caption{\textbf{Experimental Dirac Points and Far-Field Polarization Topology.}  
    \textbf{a} Simulated band structure in $k$-space, showing Mode~1 (gray) and Mode~2 (blue), with the dashed plane indicating the $k$-space cut through the Dirac point.
    \textbf{b} Measured Stokes parameter $S_2=T_{45^\circ}-T_{-45^\circ}$ spectra along the $k_x/k_0 \approx 0.35$ cut under TM excitation ($\vec{E}_{\mathrm{exc}} \perp \vec{b}$). 
    The wavelength-momentum dispersion reveals two modes forming a Dirac-like crossing. Markers indicate the RSE-fitted modal dispersion; their outline distinguishes Mode~1 and Mode~2, while their color encodes the $S_2$ parameter predicted by the RSE. Dashed lines denote the grating mode.
    \textbf{c} Polarization angle along the two dispersion branches obtained from the analytical RSE model (left) and experiment (right), shown for Mode~1 (top) and Mode~2 (bottom). Both theory and experiment reveal continuous polarization evolution and characteristic Dirac points (DPs), in quantitative agreement. The slight global rotation of the experimental maps can be attributed to a small misalignment of the excitation polarization with respect to the crystal $b$-axis (see Supplementary Note~4).
    \textbf{d} Far-field polarization field in the vicinity of a selected Dirac point. Vector plots show the local polarization orientation extracted from the analytical RSE model (left) and from experiment (right). The polarization vectors exhibit a half-integer winding around the singularity, corresponding to a topological charge of $-1/2$, in agreement with the the theoretical predictions in Fig.~\ref{fig:fig2}f. The measurements are performed on the same metasurface as in Figure~\ref{fig:fig3}.}
    \label{fig:fig4}
\end{figure}

We further experimentally probe the momentum-space region where anisotropy-induced transformation of the qBIC dispersion gives rise to a Dirac point, manifesting as a linear crossing of orthogonally polarized modes at $k_x/k_0\approx\pm0.35$ (Fig~\ref{fig:fig4}a).
To experimentally resolve the polarization structure of this crossing, we measure the hyperspectral Stokes parameters $\mathbf{S}=(S_0,S_1,S_2)$ in momentum space (see Methods and Supplementary Note~8), where $S_0$ is the total measured intensity, while $S_1=T_{0^\circ}-T_{90^\circ}$ and $S_2=T_{45^\circ}-T_{-45^\circ}$  are the transmittance differences between linearly polarized light at the respective angles. $S_3$ is neglected, since only linearly polarized components are measured.

Experimental $S_2$ spectra reveal two branches intersecting in a Dirac-like crossing at the $k_x/k_0=0.35$ cut, in excellent agreement with the theoretical prediction shown in Figure~\ref{fig:fig4}b. As demonstrated in Figure~\ref{fig:fig2}f, a Dirac point is a singularity in the far-field polarization, carrying a half-integer topological charge and marked by an orthogonal polarization flip along $k_x$. 
To experimentally demonstrate this, we extract the polarization angle $\phi=\arctan(S_2/S_1)/2$, which, because the two modes are orthogonal, relates directly to the mixing angle between Mode~1 and Mode~2 predicted by the RSE model (Supplementary Note~3). We evaluate this angle for each mode by fitting the analytical RSE band structure from Eq.~\eqref{eq:3} to the hyperspectral Stokes ($\textbf{S}$) data (Supplementary Note~4). The resulting momentum-resolved polarization-angle maps from both RSE and experiment, are shown in Fig.~\ref{fig:fig4}c. In both cases the polarization angle evolves continuously along the dispersion branches and exhibits singular points coinciding with the Dirac crossings.

Finally, we analyze the far-field polarization vectors around a selected Dirac point, comparing the local polarization orientation from the RSE and experiment (Fig.~\ref{fig:fig4}d; Supplementary Note~4). In both cases, the vectors exhibit a half-integer winding around the singularity, corresponding to a topological charge of $-1/2$, direct experimental evidence of the polarization vortex predicted by the RSE and full-wave simulations. Our platform also enables an additional degree of control without requiring further geometrical perturbations, since the orientation of the in-plane anisotropy axis provides a tuning parameter that revolves the Dirac points through momentum space while the underlying qBICs remain orthogonally polarized throughout (Supplementary Note~9). Note that at these Dirac points only the eigenfrequencies coalesce, unlike exceptional points~\cite{liu2026optical}, where both eigenvectors and eigenfrequencies coalesce.

\subsection{Formation of Anisotropic Polaritonic Flatbands}

Tuning the eigenfrequencies of the two orthogonally polarized qBICs into the range of the excitonic transition energies of ReS\textsubscript{2}, transfers topologically engineered dispersion landscape (including the directional flatbands) into the polaritonic regime.
In bulk ReS\textsubscript{2} crystals, the energetically separated excitonic transitions (Exciton~1 and Exciton~2) are linearly polarized along nearly orthogonal in-plane directions, based on ellipsometric measurements~\cite{munkhbat}, though other studies report slightly different polarization angles~\cite{aslan2015linearly}. This distinction is not critical here: we align the metasurface's two orthogonal polarization axes with the exciton axes and excite each mode accordingly, yielding good agreement between simulation, analytical theory, and experiment. Importantly, our C\textsubscript{4}-symmetric metasurface provides an ideal platform to isolate the optical consequences of this birefringence: varying the incident polarization selectively addresses each qBIC and the corresponding exciton (Mode~1/Exciton~1 and Mode~2/Exciton~2).

To treat exciton-polariton formation within the same RSE framework used above, we extend the formalism by introducing excitonic contributions to Eq.~\eqref{eq:delta_epsilon} as permittivity poles, so that close to the positive frequency half-axis the permittivity reads:
\begin{equation}
\varepsilon_{xx}(\omega) = \varepsilon + \frac{\Delta \varepsilon}{2} + \frac{f_x}{\omega_x - \omega}, \quad
\varepsilon_{yy}(\omega) = \varepsilon - \frac{\Delta \varepsilon}{2} + \frac{f_y}{\omega_y - \omega},
\label{eq:5}
\end{equation}
where, to comply with previous observations~\cite{munkhbat}, we assume that Exciton~1 ($\omega_x$) is redshifted relative to Exciton~2 ($\omega_y$), $(\mathrm{Re}\,\omega_x < \mathrm{Re}\,\omega_y)$, with similar oscillator strengths $f_x=f_y$. The anisotropic background term ($\Delta\varepsilon>0$) and the excitonic terms are all treated as a perturbation.

\begin{figure}[h!]
    \centering
    \includegraphics[width=1\linewidth]{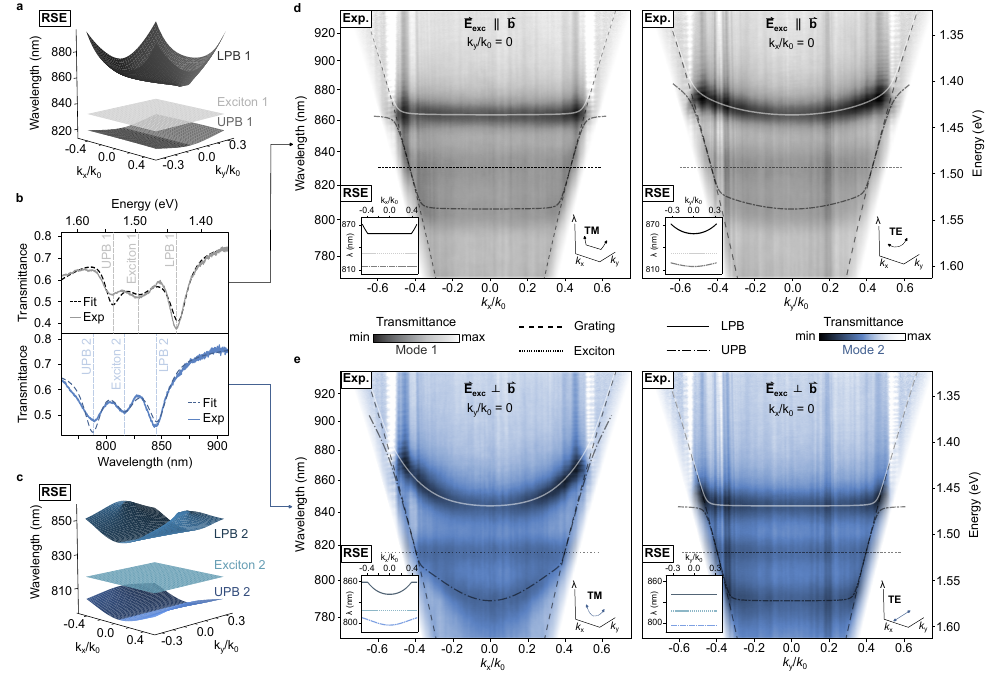}
    \caption{\setstretch{1}\textbf{Experimental Anisotropic Strong Coupling and Directional Polaritonic Flatbands.} 
    {\color{black} Upper (UPB) and lower (LPB) polariton band structures resulting from the coupling of \textbf{a} Mode~1 and Exciton~1 and \textbf{c} Mode~2 and Exciton~2, obtained from the nonlinear RSE (Supplementary Note~3).}
    \textbf{b} Experimental transmittance spectra at normal incidence of orthogonally polarized exciton-polaritons. Coupled-mode fit yields Rabi splittings of $\hbar\Omega_\mathrm{R1}=\SI{110}{\milli\electronvolt}$ and $\hbar\Omega_\mathrm{R2}=\SI{108}{\milli\electronvolt}$ (Supplementary Note~5). The UPBs comprise 43\%~Exciton~1/57\%~Mode~1 and 51\%~Exciton~2/49\%~Mode~2, respectively, as determined from the Hopfield coefficients (Supplementary Note~7); for the LPBs, these ratios are reversed.
    Experimental angle-resolved Fourier-space transmittance for excitation polarization \textbf{d}
    $\vec{E}_\mathrm{exc}\parallel \vec{b}$ (Mode 1, grey) and \textbf{e}
    $\vec{E}_\mathrm{exc}\perp \vec{b}$ (Mode 2, blue), where $\vec{b}$ denotes the crystal axis. Left and right panels correspond to TM ($k_x/k_0$) and TE ($k_y/k_0$) excitation configurations, respectively. The insets show the corresponding RSE solutions. 
    The dotted lines represent the excitonic resonance, and the dashed lines correspond to a grating mode. The solid and dot-dashed lines denote coupled-mode fit results for LPB and UPB, respectively (Supplementary Note~6).
    Depending on the propagation direction in $k$-space, the polaritonic branches exhibits either parabolic or flatband dispersion. Along the flatband direction, the Hopfield coefficients remain nearly constant over the accessible momentum range, consistent with their values at the $\Gamma$-point (Supplementary Note~7).  The metasurface parameters are $\Lambda=S\cdot465$ nm, $w_0=S\cdot220$ nm, $h=90$ nm, $\alpha=0.6$ and in-plane scaling factor $S=0.95$.  
    }
    \label{fig:fig5}
\end{figure}

Employing the nonlinear RSE formalism~\cite{muljarov2016resonant}, we obtain a sixth-order polynomial dispersion equation with a pair of dispersionless solutions ($\omega_x$, $\omega_y$), corresponding to uncoupled excitons, and four exciton-polariton branches (Figs.~\ref{fig:fig5}a,c). The resulting polaritons retain the momentum-space dispersion of the underlying photonic modes, as shown by the separate cross-sections, and the RSE further correctly predicts the far-field polarization of all solutions (Supplementary Note~3).

We begin the experimental verification of strong coupling between the photonic qBICs and the excitons at normal incidence. Figure~\ref{fig:fig5}b shows transmittance spectra for the two excitation polarizations, revealing clear splitting from hybridization between the photonic modes and excitons, with coupled-mode fits yielding Rabi splittings exceeding $100$~meV. A quantitative analysis of the coupling strengths and linewidths confirms that the system operates in the strong-coupling regime (Supplementary Note~5).

We next investigate the exciton-polariton dispersion in momentum space. Experimental Fourier-space transmittance measurements for $\vec{E}_\mathrm{exc}\parallel\vec{b}$ and $\vec{E}_\mathrm{exc}\perp\vec{b}$ are shown in Fig.~\ref{fig:fig5}d,e. As predicted by the analytical RSE model, the polaritonic branches inherit the anisotropic dispersion of the underlying photonic modes, exhibiting either a conventional parabolic dispersion or a strongly suppressed curvature corresponding to a polaritonic flatband, depending on the propagation direction in $k$-space.

To quantify these dispersions, we analyze the polaritonic branches' curvature (Supplementary Note~6), confirming that they retain the pronounced anisotropy of the underlying photonic dispersion. The corresponding Hopfield coefficients (Fig.~\ref{fig:fig5} caption; Supplementary Note~7) confirm a genuinely mixed, nearly momentum-independent light-matter composition along the flatband direction.

Taken together, these findings constitute an experimental demonstration of topologically engineered polaritonic flatbands arising from the interplay of C\textsubscript{4}-symmetry-imposed mode degeneracy and material anisotropy. They validate the theoretical framework developed above, confirming that in-plane anisotropy can be harnessed to control the dispersion of qBICs.

\section{Discussion}

We demonstrate that a photonic metasurface with a $\text{C}_4$-symmetric shape, realized directly in anisotropic bulk $\text{ReS}_2$, can be engineered to host topological singularities carrying fractional charges. Here, the dielectric tensor anisotropy of $\text{ReS}_2$ reshapes the topological structure of the qBICs by splitting the polarization vortex associated with the degenerate mode into two momentum-separated, half-integer topological charges -- analogous to topological charge splitting predicted in symmetry-broken photonic crystals [47], but induced purely by material anisotropy. This anisotropy-driven splitting produces a Dirac-like band structure and gives rise to photonic flatbands. Consequently, flat and parabolic dispersions coexist within the same metasurface and can be selectively accessed through excitation polarization.

Flatband photonics and polaritons have previously been explored in frustrated lattices such as Kagome and Lieb~\cite{klembt2017polariton, harder2021kagome, scafirimuto2021tunable}, as well as in strained honeycomb polariton lattices hosting Dirac cone and flatband physics through asymmetric hopping~\cite{milicevic2019type} \color{black} or superlattice and BIC-merging approaches~\cite{nguyen2018symmetry, qin2024optical, le2024super}. Such schemes rely on finely tuned structural interference or auxiliary ingredients such as detuned exciton-polaritons~\cite{mischok2024breaking}, making them sensitive to fabrication disorder and less scalable. By contrast, our anisotropy-based mechanism relies on a single resonance with no auxiliary quasiparticles, yielding flatbands that are directly accessible in the far-field, and spanning an exceptionally broad momentum range, ultimately limited only by the diffraction cut-off.
Importantly, this approach can be generalized to other anisotropic media, such as black phosphorus, 2D magnets~\cite{bae2022exciton} or emerging photocatalytic 2D materials~\cite{xu2020novel}, and potentially even mimicked in isotropic materials through engineered asymmetry, e.g. asymmetric unit cells or claddings that induce effective anisotropy~\cite{hu2024environmental}.

In hyperbolic media, anisotropic polaritons have been investigated in the shape of surface phonon polaritons, whose direction-dependent near-field propagation reflects structural and electronic anisotropy~\cite{ma2020anisotropic, li2018infrared, ma2018plane}. Distinct from these near-field approaches, we realize bulk exciton-polaritons with intrinsic anisotropy originating from the underlying excitonic transitions, extending into the far-field. By combining the topological texturing of the photonic modes with the naturally anisotropic excitons of ReS\textsubscript{2}, our strategy provides a platform for both polarization-controllable directional strong coupling and topologically-driven generation of polaritonic flatbands.

Such photonic and polaritonic flatbands offer suppressed group velocity, enhanced density of states, and intrinsic polarization selectivity. We highlight two target functionalities where we believe our platform can offer concrete application-level advantages. First, the combination of quenched kinetic energy, qBIC-enhanced photon lifetimes, and room-temperature excitons in a single monolithic material offers a route toward polarization-selective, low-threshold polariton lasing, without the disorder sensitivity of frustrated-lattice or BIC-merging schemes~\cite{nguyen2018symmetry, qin2024optical, le2024super}. 
Second, because the group-velocity suppression extends over more than 90\% of the far-field accessible momentum range rather than being confined to narrow band-edge regions as in photonic-crystal slabs~\cite{Baba2008}, the directional flatbands enable slow-light-enhanced nonlinear processes over a wide angular range. Combined with the exciton-driven nonlinearities of the self-hybridized polaritons, this opens a route toward polarization-selective, low-power nonlinear emission.

\subsection{Acknowledgements}
Funded by the European Union (EIC, OMICSENS, 101129734, NEHO, 101046329, ERC, METANEXT, 101078018, QUONDENSATE, 101130384). Views and opinions expressed are however those of the author(s) only and do not necessarily reflect those of the European Union or the European Research Council Executive Agency. Neither the European Union nor the granting authority can be held responsible for them. This project was funded by the Deutsche Forschungsgemeinschaft (DFG, German Research Foundation) under Germany’s Excellence Strategy EXC 3120/1–533771286 and EXC 2089/1–390776260 and the Emmy Noether Programme (TI 1063/1), the Bavarian program Solar Energies Go Hybrid (SolTech) and the Center for NanoScience (CeNS). The work of M.V.G. was supported by the Ministry of Science and Higher Education of the Russian Federation, project FSWU-2023-0075. C.C., A.G. and G.C. acknowledge the European Union’s NextGenerationEU Programme with the I-PHOQS Infrastructure [IR0000016, ID D2B8D520, CUP B53C22001750006] “Integrated Infrastructure Initiative in Photonic and Quantum Sciences”.

\subsection{Author contributions}
C.H., T.W., A.A.A. and A.T. conceived the idea and planned the research. C.H., T.W. fabricated the samples. C.H., T.W., C.C., A.G., T.P., L.S., G.V. and C.M. performed optical measurements. A.G. performed the polarization-resolved measurements and supervised the hyperspectral experiments. C.H., T.W. and A.A.A. conducted the numerical simulations and data processing. A.A.A.  and M.V.G. developed the theoretical background. M.V.G., G.C., A.A.A. and A.T. supervised the project. All authors contributed to the data analysis and to the writing of the paper.

\subsection{Competing interests}
The authors declare no competing interests.

\printbibliography[notcategory=methodsonly]

\section{Methods}
\subsection{Numerical simulations}
The refractive index of the SiO$_2$ substrate was set as 1.45, while that of the ReS\textsubscript{2} was adapted based on literature~\cite{munkhbat}. 
Simulations of transmittance spectra for the ReS\textsubscript{2} metasurfaces were conducted using CST Studio Suite 2021 with periodic Floquet boundary conditions. Transmittance behavior under oblique incidence was numerically investigated using the Electromagnetic Waves Frequency Domain module of COMSOL Multiphysics in 3D mode. The tetrahedral spatial mesh for FEM was automatically generated by COMSOL's physics-controlled preset. Simulations were performed within a rectangular spatial domain containing a single metasurface unit cell with periodic boundary conditions applied to its sides. Linearly polarized ports were set at the top and bottom to simulate excitation and registration. To evaluate the far-field polarizations of photonic eigenstate a previously developed approach based on calculating the overlap integrals between eigenstate's displacement current density and plane waves was used.~\cite{kim2023chiral}

\subsection{Sample fabrication}
ReS\textsubscript{2} flakes, sourced from commercial bulk crystals (HQ Graphene), were mechanically exfoliated onto glass substrates. Flake selection was guided by optical microscopy, prioritizing large-area flakes with uniform thickness and clearly oriented crystal axes. Flake height was subsequently characterized using a stylus profilometer (Bruker Dektak XT) to confirm suitable thickness.
During mechanical exfoliation, robust metal-metal (Re-Re) bonds cause the material to cleave along the \textit{b}-axis, resulting in flakes with elongated edges aligned to this crystallographic direction.~\cite{Exfol1, Exfol2} This fracture behavior allows for straightforward identification of the crystal axis using optical microscopy.~\cite{Exfol2} 
Deterministically orienting the metasurface with respect to the underlying crystal, we achieve a well-defined alignment between the metasurface resonator lattice and the \textit{b}-axis (Fig.~\ref{fig:fig1}c).
The thickness of the ReS\textsubscript{2} flake affects the overlap between the photonic modes and the excitonic medium, and therefore the effective coupling parameter $\varkappa$ (see Eq. \ref{eq:caption1}). A thickness of approximately 90 nm was chosen for the experiment to optimize the spectral alignment between the BIC resonances and the excitonic transitions.
To prepare the sample for lithographic patterning, a layer of PMMA 950k was spin-coated and thermally baked at 175°C for 3 minutes. A thin antistatic coating of Espacer 300Z was then applied to mitigate surface charging during electron-beam lithography (EBL). The metasurface layout was defined using a Raith eLine Plus system. After exposure, the resist was developed in an aqueous 80\% ethanol solution for 20 seconds and rinsed in isopropanol to halt the development process.
Subsequently, a gold hardmask was deposited by electron-beam evaporation, and a standard lift-off procedure was performed in acetone to reveal the patterned metal features. The underlying ReS\textsubscript{2} was etched using a reactive ion etching (RIE) process with SF\textsubscript{6}-based plasma chemistry, employing the gold layer as an etch mask. To complete fabrication, the hardmask was removed using a commercial gold etchant solution (Sigma-Aldrich), yielding the final patterned metasurface structures.

\subsection{Optical Measurements}
Optical characterization was carried out using a confocal transmittance microscope configured for bottom-side illumination. The samples were illuminated with collimated, linearly polarized white light. Transmitted light was collected through a 50× objective lens with a numerical aperture of 0.8, ensuring high spatial resolution and efficient light capture. The collected signal was spectrally resolved using a grating-based spectrometer (groove density: 150 mm$^{-1}$) and detected with a silicon-based CCD array.

\subsection{Hyperspectral k-space imaging in transmission}
Polarization-resolved hyperspectral $k$-space measurements are performed in a custom microscopy setup working in transmission (see Supplementary Note~8 for illustration). For the excitation branch, the tip of an optical fiber (core diameter: 200~\textmu m), illuminated by a tungsten lamp, is imaged on the metasurface plane from the back side of the sample, using a collimation lens and a $50\times$ objective lens (N.A.=0.5) with long working distance. For the detection path, the transmitted light is collected by a $100\times$ objective (N.A.=0.9). A Fourier lens conjugated to the back focal plane of the detection objective and a tube lens are used to form the $k$-space image on the entrance slit of a spectrometer. A grating inside the spectrometer diffracts a narrow vertical slice of the $k$-space image, spatially filtered by the entrance slit, forming a wavelength-$k_y$ intensity map on a CCD. The final transmission maps are calculated as the ratio between the transmitted intensity on the sample by the one on the substrate. The tube lens is mounted on a motorized translation stage moving orthogonally to the optical axis, which is used to do a tomographic scan of the $k$-space image along $k_x$. A rotating wire-grid polarizer placed before the spectrometer slit is used for the polarimetry measurements. All measurements were performed at room temperature under ambient conditions.
\color{black}

\subsection{Data availability}
The data that support the findings of this study are available via Zenodo at \url{ https://doi.org/10.5281/zenodo.22256282}\cite{heimig2026data}.

\printbibliography[category=methodsonly,title={Methods-only References}]
\end{refsection}

\startSI
\vspace*{-2.5cm}
\tableofcontents
\clearpage

\begin{refsection}[SI_references.bib]
\section{Material Model}
We use the values of diagonal in-plane components of the ReS$_2$ dielectric tensor $\varepsilon_{jj}=\varepsilon'_{jj}+i\varepsilon^{''}_{jj}$, with $j=x,y$, provided by the tabulated material model from Ref.~\cite{munkhbat}. In order to adapt the material parameters to our measurements, we fit a Tauc--Lorentz (TL) model to the imaginary part $\varepsilon^{''}_\mathrm{TL}$ of the permittivity, described by:
\begin{equation}
\varepsilon^{''}_\mathrm{TL}(\omega) =
\sum_{i = \mathrm{A,B,C}}
\begin{cases}
\dfrac{F_i \, \omega_{0,i} \Gamma_i \, (\omega - \omega_{\mathrm{g}})^2}
      {\left( \omega^2 - \omega_{0,i}^2 \right)^2 + \Gamma_i^2 \omega^2} \cdot \dfrac{1}{\omega},
& \omega > \omega_{\mathrm{g}} \\[10pt]
0, & \omega \leq \omega_{\mathrm{g}}
\end{cases}
\end{equation}
where $i = \mathrm{A, B, C}$ labels the three Tauc-Lorentz oscillators used (Table~\ref{table:TL_params}), and $F_i$ is the oscillator strength, $\omega_{0,i}$ the resonance frequency, $\omega_{\mathrm{g}}$ the band-gap energy, and $\Gamma_i$ the linewidth of oscillator $i$.

\begin{figure}[H]
    \centering
    \includegraphics[width=1\linewidth]{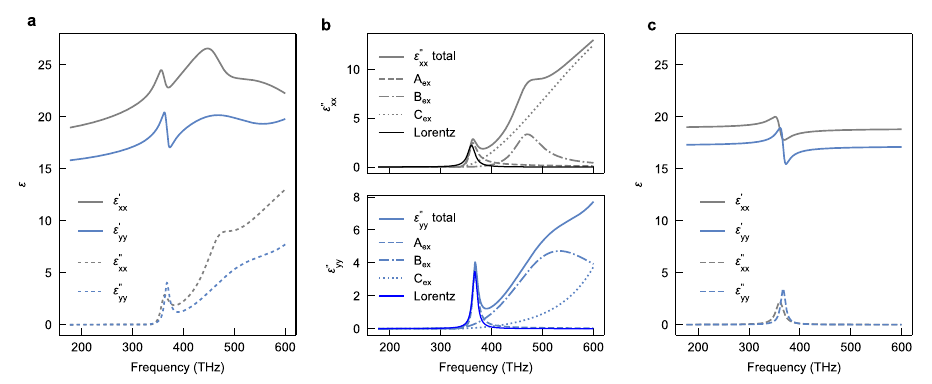}
    \caption{\textbf{Modelling of Dielectric Function of ReS$_2$.} \textbf{a} In-plane components of the real and imaginary parts of the permittivity $\varepsilon'$ adapted to experiments. \textbf{b} Decomposition of the imaginary parts of the permittivity $\varepsilon''$ into three TL oscillators. The main contribution governing the light-matter coupling is described by oscillator A. A simple Lorentzian peak described by Eq.~\eqref{eq:epsLorentz} captures the coupling physics of both excitons. \textbf{c} Complete dispersionless in-plane component of the simplified Lorentzian Model from panel \textbf{b}.}
    \label{fig:material_model}
\end{figure}

The real part $\varepsilon^{'}_\mathrm{TL}$ of the dielectric tensor can be retrieved by Kramers-Kronig relations. In our case, we use the analytical expression for a TL model defined in \cite{jellison1996parameterization}. We adapt the model by comparison with experimental data (Supplementary Note \ref{section:TCMT fit}), where we fix the spectral positions $\omega_0$ and damping rates $\Gamma$ with $\Gamma = 2\gamma_\mathrm{EX}$ of both excitons to our TCMT fits (Figure \ref{fig:material_model}a). In total, we use 3 Tauc--Lorentz oscillators in both x- and y-direction to describe the material.
The full set of parameters are listed in Table \ref{table:TL_params}. The oscillators corresponding to one spatial direction share the same band-gap energy $\omega_\mathrm{g}$, with $\omega_\mathrm{g,xx}/(2\pi)=\SI{331.74}{\tera\hertz}$ and $\omega_\mathrm{g,yy}/(2\pi)=\SI{319.38}{\tera\hertz}$. The constant dielectric offset of the real part of the permittivity are $\varepsilon_\mathrm{\infty,xx} = 6.5$ and $\varepsilon_\mathrm{\infty,yy}=9.62$. Note, that the values differ from the single Lorentzian model, as the additional oscillators give extra contributions to the real part of the permittivity. The out-of-plane component is assumed to be loss-free and constant with a value of $\varepsilon^{'}_\mathrm{zz} = 7.25$.

\begin{table}[H]
\centering
\renewcommand{\arraystretch}{1.2}
\setlength{\tabcolsep}{4pt} 
\begin{tabular}{l *{2}{S[table-format=5.2] S[table-format=3.2] S[table-format=3.2]}}
\toprule
 & \multicolumn{3}{c}{$\boldsymbol{\varepsilon_\mathrm{xx}}$} & \multicolumn{3}{c}{$\boldsymbol{\varepsilon_\mathrm{yy}}$} \\
\cmidrule(lr){2-4} \cmidrule(lr){5-7}
 & {\makecell{$F/(2\pi)^2$ \\ (THz$^2$)}} & {\makecell{$\omega_0/(2\pi)$ \\ (THz)}} & {\makecell{$\Gamma/(2\pi)$ \\ (THz)}} 
 & {\makecell{$F/(2\pi)^2$ \\ (THz$^2$)}} & {\makecell{$\omega_0/(2\pi)$ \\ (THz)}} & {\makecell{$\Gamma/(2\pi)$ \\ (THz)}} \\
\midrule
\textbf{A} & 6577.0 & 360.33 & 17.93  & 2500.1 & 366.95 & 11.54 \\
\textbf{B} & 2591.5 & 465.92 & 65.23  & 8269.0 & 499.75 & 246.73 \\
\textbf{C} & 52100  & 724.81 & 926.06 & 10350.0  & 740.93 & 174.77 \\
\bottomrule
\end{tabular}
\caption{Tauc--Lorentz model parameters for the adapted material model for x- and y-direction. $F$ is the oscillator strength of the resonance, $\omega_0$ the resonance frequency and $\Gamma$ the damping rate.}
\label{table:TL_params}
\end{table}

\begin{figure}[H]
    \centering
    \includegraphics[width=1\linewidth]{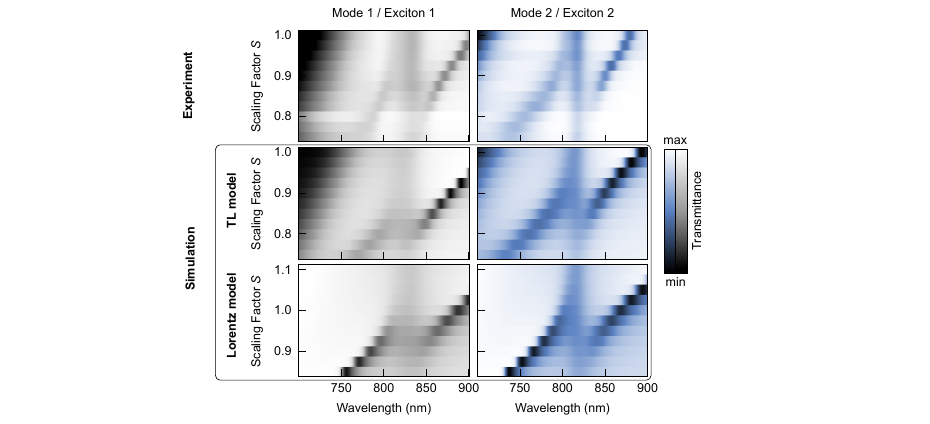}
    \caption{\textbf{Comparison of Simulated and Experimental Anticrossing.} Experimental transmittance spectra of metasurfaces with $h=\SI{90}{nm}$, $\Lambda=\SI{465}{\nano\metre}$, $w_0 = \SI{220}{\nano\metre}$, $\alpha=0.4$ and $S=1$. Simulations using the Tauc--Lorentz model show excellent agreement with the experiment. Simulations based on the simplified Lorentzian model were conducted without a substrate to mitigate the influence of grating modes. This leads to slightly different scaling factors but agrees qualitatively well with the experiment.}
    \label{fig:comp_exp_sim}
\end{figure}

In order to study the isolated light-matter coupling of the quasi-BICs with their respective excitons, we further simplify our model by isolating their major excitonic components (Fig.~\ref{fig:material_model}c). These can now be described by a simple Lorentzian material model:
\begin{equation}\label{eq:epsLorentz}
\varepsilon_{L}(\omega) 
= \varepsilon_{\infty} + \frac{F_L}{\omega_{0}^2 - \omega^2 - i \, \omega\Gamma } \, ,
\end{equation}
where $F_L/(2\pi)^2 = \SI{14414}{\tera\hertz\squared}$ for both excitons. The resonance positions and damping rates are the same as previously stated. As in the main text, the analytical model uses refractive index $n=\sqrt{18}$ with $\Delta n=0.2$, yielding $\varepsilon_\mathrm{\infty, xx} = 18.86$ and $\varepsilon_\mathrm{\infty, yy} = 17.16$ (Fig.~\ref{fig:material_model}c).
Overall, both models show good agreement with experimental data (Fig.~\ref{fig:comp_exp_sim}). The complete dispersive model can be used to accurately engineer the spectral response of the metasurfaces and the simplified Lorentzian model to investigate the intricate anisotropic coupling physics in both analytical and numerical approaches without the influence of band-gap absorption or grating modes.

\clearpage
\section{Resonant State Expansion for Anisotropy-Induced Degenerate Modes Splitting}

To gain deeper insight into the mechanism of eigenmode splitting due to material anisotropy, we implement Resonant State Expansion (RSE) theory. We consider a metasurface with C\textsubscript{4} rotational symmetry, consisting of square prisms with a base side length $w_0 = 220$ nm, height $h = 90$ nm, and a period of $\Lambda = 465$ nm. The prisms are placed on a transparent glass-like substrate with a refractive index $n = 1.45$, and are made of an intrinsically anisotropic material, whose diagonal dielectric permittivity tensor $\varepsilon$ has the following components: $\varepsilon_{xx} = \varepsilon_{yy} = \varepsilon = 18$, $\varepsilon_{zz} = 7.25$.

By introducing the asymmetry parameter $\alpha$, we reshape the prisms into two other square prisms with base side lengths:
\[ w_1 = w_0 \left[ \frac{2}{1 + \alpha^2} \right]^{1/2}, \quad w_2 = w_0 \alpha \left[ \frac{2}{1 + \alpha^2} \right]^{1/2}, \]
ensuring that the volume of the modified prisms remains equal to the original. The asymmetry increases the square unit cell period by a factor of $\sqrt{2}$, forming a pair of Brillouin Zone Folding (BZF) qBICs.

The C\textsubscript{4} rotational symmetry of the structure dictates that coupling to the farfield is only possible if the modes are at least doubly degenerate \cite{sakoda1995symmetry}. Introducing in-plane anisotropy $\Delta \varepsilon$ into the dielectric permittivity tensor as:
\begin{equation}
\hat{\varepsilon} = \begin{bmatrix}
\varepsilon + \Delta \varepsilon/2 & 0 & 0 \\
0 & \varepsilon - \Delta \varepsilon/2 & 0 \\
0 & 0 & \varepsilon_{zz}
\end{bmatrix},
\end{equation}
breaks the C\textsubscript{4} symmetry and results in spectral splitting of the BZF qBICs.

RSE is a powerful tool for describing the hybridization of a finite number of metasurface eigenmodes arising due to a permittivity perturbation $\Delta \varepsilon$. To analyze the hybridization of the BZF qBICs, the states with eigenfrequencies $\omega_n$, electric fields $\mathbf{E}_n$ and magnetic fields $\mathbf{H}_n$ must be properly normalized. We follow the normalization described in \cite{gorkunov2025substrate}:
\begin{equation}
1 = \| \mathbf{F}_n \|^2 = \int_V \left[ \varepsilon(\mathbf{r}) \mathbf{E}_n \cdot \mathbf{E}_n - \mathbf{H}_n \cdot \mathbf{H}_n \right] dV \\
+ \frac{i c}{\omega_n} \int_{\partial V} \left[ \mathbf{E}_n \times (\mathbf{r} \cdot \nabla) \mathbf{H}_n \mathbf{H}_n \times (\mathbf{r} \cdot \nabla) \mathbf{E}_n \right] \cdot d\mathbf{S},
\end{equation}
where $V$ is the volume of a unit cell with a surface boundary $\partial V$, and $\mathbf{F}_n(\mathbf{r}) = \{\mathbf{E}_n(\mathbf{r}), i \mathbf{H}_n(\mathbf{r})\}$ is the supervector defined for each mode. We assume nonmagnetic materials with permeability $\mu = 1$.

The hybridized states $\tilde{\mathbf{F}}_n$ are then expressed as a linear superposition of the unperturbed states $\mathbf{F}_1$ and $\mathbf{F}_2$ with $\Delta \varepsilon = 0$:
\begin{equation}
\tilde{\mathbf{F}} = \sum_{m=1}^{2} a_m \mathbf{F}_m,
\end{equation}
where we truncate the RSE matrix equation to the two relevant states and solve for the coefficients $a_n$:
\begin{equation}
\omega_n a_n = \tilde{\omega} \sum_{m=1}^2 (\delta_{nm} + V_{nm}) a_m.
\label{eq:RSE}
\end{equation}
Here, $\delta_{nm}$ is the Kronecker delta and elements of the perturbation matrix are given by:
\begin{equation}
V_{nm} = \int_V \Delta \varepsilon_{ij}(\mathbf{r}) E_{n,i}(\mathbf{r}) E_{m,j}(\mathbf{r}) dV = \frac{\Delta \varepsilon}{2} \int_V \left[ E_{n,x}(\mathbf{r}) E_{m,x}(\mathbf{r}) - E_{n,y}(\mathbf{r}) E_{m,y}(\mathbf{r}) \right] dV.
\end{equation}

The two initially degenerate modes with $\omega_1 = \omega_2 = \omega_0$ can be classified by their intrinsic symmetry. Applying the rotation operator $\hat{R}$ by $\pi/2$ around the $z$-axis gives:
\begin{equation}
\hat{R} \mathbf{F}_1(\mathbf{r}) = \mathbf{F}_2(\hat{R}^{-1} \mathbf{r}), \quad \hat{R} \mathbf{F}_2(\mathbf{r}) = -\mathbf{F}_1(\hat{R}^{-1} \mathbf{r}),
\end{equation}
which results in $V_{12} = -V_{21}$. Introducing elements of the perturbation matrix:
\begin{equation}
\begin{aligned}
v &= \int_V \left[ E_{1,x}(\mathbf{r})^2 - E_{1,y}(\mathbf{r})^2 \right] dV, \\
u &= \int_V \left[ E_{1,x}(\mathbf{r}) E_{2,x}(\mathbf{r}) - E_{1,y}(\mathbf{r}) E_{2,y}(\mathbf{r}) \right] dV,
\end{aligned}
\end{equation}
and solving Eq.~\eqref{eq:RSE}, we find a pair of hybrid eigenfrequencies:
\begin{equation} 
\omega_\pm = \omega_0 \left[ \frac{1 \pm \varkappa \Delta \varepsilon}{1 - \varkappa^2 (\Delta \varepsilon)^2} \right] \approx \omega_0 [1 \pm \varkappa \Delta \varepsilon],
\label{eq:RSE_gamma_point}
\end{equation}
where $\varkappa = \sqrt{v^2 + u^2}/2$ and empirically $|\varkappa| \lesssim 10^{-2}$.

\begin{figure}[hb]
    \centering
    \includegraphics[width=1\linewidth]{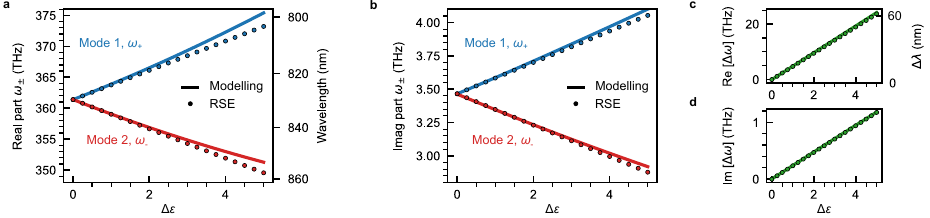}
    \caption{\textbf{Resonant State Expansion for Anisotropy-Induced Spectral Splitting of Eigenmodes.} \textbf{a} Real and \textbf{b} imaginary parts of the hybrid eigenfrequencies $\omega_\pm$ as functions of the in-plane anisotropy perturbation $\Delta \varepsilon$, obtained both from full-scale modelling using COMSOL Multiphysics and the analytical expression in Eq.~\eqref{eq:RSE_gamma_point} derived from RSE theory. The spectral splitting $\Delta\omega = \omega_+ - \omega_-$ for the real and imaginary parts is shown in \textbf{c} and \textbf{d}, respectively.}
    \label{fig:RSE}
\end{figure}

To validate that the in-plane anisotropy $\Delta \varepsilon$ indeed splits degenerate eigenmodes as predicted by Eq.~\eqref{eq:RSE_gamma_point}, we plot in Fig.~\ref{fig:RSE}a and Fig.~\ref{fig:RSE}b the real and imaginary parts of $\omega_\pm$ as functions of $\Delta \varepsilon$, using both analytical expressions and full-wave COMSOL Multiphysics simulations. Discrepancies between theory and simulations appear at relatively large $\Delta \varepsilon \geq 3$ ($\Delta n \geq 0.35$). However, as shown in Fig.~\ref{fig:RSE}c and Fig.~\ref{fig:RSE}d, the anisotropy-induced spectral splitting remains linear as $\omega_+ - \omega_- = 2 \omega_0 \varkappa \Delta \varepsilon$ up to $\Delta \varepsilon = 5$ for both real and imaginary parts.

Therefore, the anisotropy-induced spectral mode splitting
\begin{equation}
\Delta \lambda = 2\pi c(\omega_-^{-1} - \omega_+^{-1}) \approx 4\pi c\omega_0^{-1} \varkappa \Delta \varepsilon
\end{equation}
can be considered as a linear function of $\Delta \varepsilon$. Using this, we estimate the spectral splitting for intrinsically anisotropic ReS\textsubscript{2} with $\Delta \varepsilon \approx 1.7$ ($\Delta n \approx 0.2$).

\clearpage
\section{Resonant State Expansion for Anisotropy-Induced Topological Transformations}\label{sec:analytical_model_kspace}
In the following section we present a simplified analytical model explaining the mechanisms of complex anisotropy-induced topological transformations of the eigenmode spectra and polarization. The model captures the essential physics while maintaining mathematical tractability by adopting the general properties of a C\textsubscript{4
}-symmetric metasurface and retaining all key features of the numerical results from the main text figures.
\subsection{Transformation of Photonic Modes}
We assume simplified dispersion relations for the two modes of interest. Mode 1 exhibits parabolic frequency dependence on the wavevector, while Mode 2 has a flat dispersion:
\begin{align}
\omega_1(k) &= \omega_0(1 - \chi k^2/k_0^2) \label{eq:ToyM_initial_omega1}\\
\omega_2 &= \omega_0 
\label{eq:ToyM_initial_omega2}
\end{align}
where $k = \sqrt{k_x^2 + k_y^2}$ is the magnitude of the wavevector. 

The high rotational symmetry of the metastructure determines the symmetry properties of the field distributions. It also determines the presence of four mirror planes ($k_x = 0$, $k_y = 0$ along the square lattice sides, and $k_x = \pm k_y$ along the diagonals) in the reciprocal $k$-space. The waves propagating along these high-symmetry directions are either TE or TM polarized.

However, the symmetry does not prescribe the polarization assignment to spectral branches. Literature shows various examples where TE and TM polarized states can belong to the same spectral branches for states degenerate at normal incidence, as observed in hexagonal cylinder arrays~\cite{doiron2024dual} and square lattices with square holes~\cite{chen2019observing}.

\begin{figure}[ht]
    \centering
    \includegraphics[width=1\linewidth]{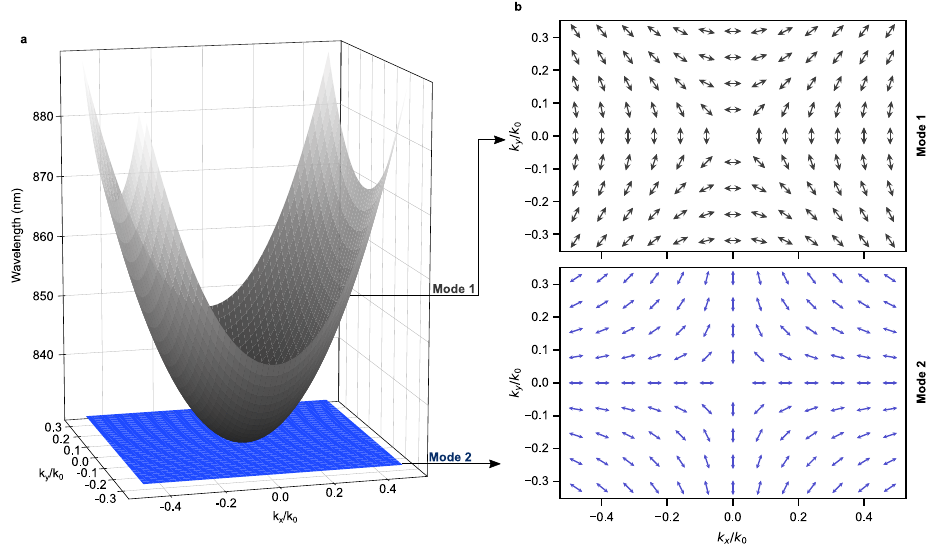}
    \caption{\textbf{Model Band Structure of Metasurface Made of Isotropic Material.} \textbf{a}.~Parabolic and flatband spectra of modes in $k$‑space given by Eqs.~\eqref{eq:ToyM_initial_omega1}-\eqref{eq:ToyM_initial_omega2}, plotted with model parameters $\omega_0/(2\pi) =361.1+2.7i$ THz and $\chi=0.2$. \textbf{b}.~Far‑field polarizations of Mode 1 and Mode 2, given by Eq.~\eqref{eq:ToyM_initial_pol_1} and Eq.~\eqref{eq:ToyM_initial pol_2}, respectively.}
    \label{fig:ToyModel1}
\end{figure}

According to COMSOL simulations (see Fig.~2(c) of the Main text), in our case, Mode 2 is TM-polarized along the principal axes ($k_x = 0$ or $k_y = 0$) and TE-polarized along diagonal directions ($k_x = \pm k_y$). Mode 1 exhibits the opposite behavior, maintaining orthogonality with Mode 2. The corresponding mutually orthogonal polarization unit vectors are:
\begin{align}
\mathbf{e}_1^k &= \frac{k_y}{k}\hat{\mathbf{x}} + \frac{k_x}{k}\hat{\mathbf{y}} \label{eq:ToyM_initial_pol_1}\\
\mathbf{e}_2^k &= \frac{k_x}{k}\hat{\mathbf{x}} - \frac{k_y}{k}\hat{\mathbf{y}} \label{eq:ToyM_initial pol_2}
\end{align}

In Fig.~\ref{fig:ToyModel1} we plot the band structure Eqs.~\eqref{eq:ToyM_initial_omega1}-\eqref{eq:ToyM_initial_omega2} and mode's farfield polarizations Eqs.~\eqref{eq:ToyM_initial_pol_1}-\eqref{eq:ToyM_initial pol_2} of our simplified model with $\chi=0.2$, which closely matches the parabolic dispersion of the Mode 1 in Fig. 2 of the main text. We also set 
$\omega_0/(2\pi)=361.1+2.7i$ THz, whose real part lies between two excitons obtained from TCMT fitting in the Section~\ref{section:TCMT fit} below, while the imaginary part takes a typical value provided by COMSOL.
By comparing Fig.~\ref{fig:ToyModel1} with Figs. 2(a-c) from the main text we see that our simple model reproduces the essential topological features observed in numerical simulations.

Material anisotropy induces coupling between the two modes from Eq.~\eqref{eq:ToyM_initial_omega1} and Eq.~\eqref{eq:ToyM_initial_omega2}. We model this effect by expressing the electric field distributions in the mixed modes as:
\begin{equation}
\tilde{\mathbf{E}}^k(\mathbf{r}) = \mathbf{E}_1^k(\mathbf{r})\cos\phi_k + \mathbf{E}_2^k(\mathbf{r})\sin\phi_k \label{eq:mixing}
\end{equation}
where $\mathbf{E}_{1,2}^k(\mathbf{r})$ are the field distributions for modes in an isotropic metasurface with transverse wavevector $\mathbf{k}$, and $\phi_k$ characterizes the degree of mixing.

We now generalize the system of equations within the RSE framework by allowing all coefficients to be $\mathbf{k}$-dependent. The decomposition coefficients of the mixed modes are given by $a_1 = \cos \phi_\mathbf{k}$ and $a_2 = \sin \phi_\mathbf{k}$. The resulting coupled equations become:
\begin{align}
\omega_1(\mathbf{k}) \cos \phi_\mathbf{k} &= \tilde{\omega}(\mathbf{k}) \left[ (1 + v_1(\mathbf{k})) \cos \phi_\mathbf{k} + u(\mathbf{k}) \sin \phi_\mathbf{k} \right], \label{eq:ToyM_RSE_eq1} \\
\omega_2 \sin \phi_\mathbf{k} &= \tilde{\omega}(\mathbf{k}) \left[ (1 + v_2(\mathbf{k})) \sin \phi_\mathbf{k} + u(\mathbf{k}) \cos \phi_\mathbf{k} \right], \label{eq:ToyM_RSE_eq2}
\end{align}
where the coupling parameters are defined as generalized overlap integrals along the $\mathbf{k}$-space directions:
\begin{align}
v_1(\mathbf{k}) &= \frac{\Delta \varepsilon}{2} \int_V \left[ \left(E_{1x}^\mathbf{k}(\mathbf{r})\right)^2 - \left(E_{1y}^\mathbf{k}(\mathbf{r})\right)^2 \right] \mathrm{d}V, \label{eq:SX7} \\
v_2(\mathbf{k}) &= \frac{\Delta \varepsilon}{2} \int_V \left[ \left(E_{2x}^\mathbf{k}(\mathbf{r})\right)^2 - \left(E_{2y}^\mathbf{k}(\mathbf{r})\right)^2 \right] \mathrm{d}V, \label{eq:SX8} \\
u(\mathbf{k}) &= \frac{\Delta \varepsilon}{2} \int_V \left[ E_{1x}^\mathbf{k}(\mathbf{r}) E_{2x}^\mathbf{k}(\mathbf{r}) - E_{1y}^\mathbf{k}(\mathbf{r}) E_{2y}^\mathbf{k}(\mathbf{r}) \right] \mathrm{d}V. \label{eq:SX9}
\end{align}
To avoid extensive simulations, we assume that for small transverse wavevectors $\mathbf{k}$ the field distributions are approximately constant and they remain similar to those of the degenerate modes at the normal incidence:
\begin{equation}
\mathbf{E}_{1,2}^\mathbf{k}(\mathbf{r}) \approx \mathbf{E}_{1,2}^{\mathbf{k} \to 0}(\mathbf{r}).
\end{equation}
Note that the latter cannot be unambiguously defined. For definiteness, we select orthogonal modes with far fields polarized along the $x$ and $y$ axes, $\mathbf{E}^{(x)}(\mathbf{r})$ and $\mathbf{E}^{(y)}(\mathbf{r})$, and express the modes as:
\begin{align}
\mathbf{E}_1^\mathbf{k}(\mathbf{r}) &\approx \frac{k_y}{k} \mathbf{E}^{(x)}(\mathbf{r}) + \frac{k_x}{k} \mathbf{E}^{(y)}(\mathbf{r}), \label{eq:ToyM_FarF_anis1} \\
\mathbf{E}_2^\mathbf{k}(\mathbf{r}) &\approx \frac{k_x}{k} \mathbf{E}^{(x)}(\mathbf{r}) - \frac{k_y}{k} \mathbf{E}^{(y)}(\mathbf{r}). \label{eq:ToyM_FarF_anis2}
\end{align}
which ensures that the mode far-fields are polarized as in Eqs.~\eqref{eq:ToyM_initial_pol_1} and Eq.~\eqref{eq:ToyM_initial pol_2}.
Substituting these into Eqs.~\eqref{eq:SX7}--\eqref{eq:SX9}, we obtain:
\begin{align}
v_1(\mathbf{k}) &\approx \frac{\Delta \varepsilon}{2} \left[ \int_V \left(\frac{k_y}{k} E_x^{(x)}(\textbf{r})\right)^2 \mathrm{d}V - \int_V \left(\frac{k_x}{k} E_y^{(y)}(\textbf{r})\right)^2 \mathrm{d}V \right] \notag\\
&= \frac{\Delta \varepsilon}{2} \ \frac{k_y^2 - k_x^2}{k^2} \int_V \left(E_x^{(x)}(\textbf{r})\right)^2 \mathrm{d}V = \frac{k_y^2 - k_x^2}{2k^2} \beta, \label{eq:ToyM_v1=beta} \\
v_2(\mathbf{k}) &\approx \frac{\Delta \varepsilon}{2} \ \int_V \left[\left(\frac{k_x}{k}E_x^{(x)}(\textbf{r})\right)^2-\left(\frac{k_y}{k}E_y^{(y)}(\textbf{r})\right)^2\right] \mathrm{d}V = -\frac{k_y^2 - k_x^2}{2k^2} \beta, \label{eq:ToyM_v2=beta} \\
u(\mathbf{k}) &= \frac{\Delta \varepsilon}{2} \ \frac{k_x k_y}{k^2} \left[ \int_V \left(E_x^{(x)}(\textbf{r})\right)^2 \mathrm{d}V + \int_V \left(E_y^{(y)}(\textbf{r})\right)^2 \mathrm{d}V \right] = \frac{k_x k_y}{k^2} \beta. \label{eq:ToyM_u=beta}
\end{align}
where the mode rotation symmetry at the normal incidence was used to insure the equality of the main integrals which determine the single coupling parameter:
\begin{equation}
\beta = \beta_0\Delta\varepsilon=\Delta \varepsilon \int_V \left( E_x^{(x)}(\mathbf{r}) \right)^2 \mathrm{d}V = \Delta \varepsilon \int_V \left( E_y^{(y)}(\mathbf{r}) \right)^2 \mathrm{d}V. \label{eq:ToyM_beta}
\end{equation}
Smaller contributions such as $\int_V [E_y^{(x)}]^2 \mathrm{d}V$ and $\int_V [E_x^{(y)}]^2 \mathrm{d}V$ are neglected.
Note that the actual field profiles of the modes shown in Fig.1d of the Main text indeed demonstrate the dominance of the field components along the corresponding farfield polarization vectors.


Solving Eqs.~\eqref{eq:ToyM_RSE_eq1} and \eqref{eq:ToyM_RSE_eq2}, we obtain the eigenfrequencies of the mixed modes:
\begin{equation}
\tilde{\omega}_\pm(\mathbf{k}) = \frac{1}{1 - \tilde{u}^2(\mathbf{k})} \left[ \frac{1}{2} \Big(\tilde{\omega}_1(\mathbf{k}) + \tilde{\omega}_2(\mathbf{k})\Big) \pm \sqrt{\frac{1}{4} \Big(\tilde{\omega}_1(\mathbf{k}) - \tilde{\omega}_2(\mathbf{k})\Big)^2 + \tilde{u}^2(\mathbf{k}) \tilde{\omega}_1(\mathbf{k}) \tilde{\omega}_2(\mathbf{k})} \right], \label{eq:ToyM_Dirac_full}
\end{equation}
with
\begin{align}\label{eq:omega and u tilde_RSE}
\tilde{\omega}_{1,2}(\mathbf{k}) = \frac{\omega_{1,2}(\mathbf{k})}{1 + v_{1,2}(\mathbf{k})}, \ \ \ \ \ \tilde{u}^2(\mathbf{k}) = \frac{u^2(\mathbf{k})}{(1 + v_1(\mathbf{k}))(1 + v_2(\mathbf{k}))}.
\end{align}
By using COMSOL Multiphysics we find that the modes self-action integral $\beta_0=0.015$ and therefore $\beta \ll 1$. Assuming $\chi k^2/k^2_0 \ll 1$, and substituting the model dispersion $\omega_1(\mathbf{k}) = \omega_0(1 - \chi k^2/k^2_0)$ and $\omega_2 = \omega_0$, we get:
\begin{equation}
\tilde{\omega}_\pm(\mathbf{k}) \approx \omega_0\left[1 - \frac{\chi k^2}{2k_0^2} \pm \sqrt{\left( \frac{\chi k^2}{2k_0^2} - \beta\frac{ k_x^2 - k_y^2}{2k^2} \right)^2 + \beta^2\frac{ k_x^2 k_y^2}{k^4}}\ \right]. \label{eq:ToyM_Dirac_simplified}
\end{equation}
This expression predicts both Dirac points and flatbands. Dirac points at $\omega_+=\omega_-$ arise when both real and positive terms under the square root simultaneously vanish. The second term vanishes along the directions $k_x = 0$ or $k_y = 0$. If $\beta > 0$ (i.e., $\Delta \varepsilon > 0$), the first term also vanishes when $k_y = 0$ and $k_x/k_0 = \pm \sqrt{\beta / \chi}$, which defines the condition for the Dirac point. Moreover, when the first term under the root is nonzero, along $k_x=0$ and $k_y=0$ directions one of the modes remains dispersionless with a constant eigenfrequency $\omega_0\left[1-\beta/2\right]$ (or $\omega_0\left[1+\beta/2\right]$), thus forming a flatband. 

The farfield polarization associated with both branches can also be directly derived. Expressing the mixing angle from \eqref{eq:ToyM_RSE_eq1} as:
\begin{equation}
\phi_\mathbf{k} = \atan \left[ \frac{\omega_1(\mathbf{k}) - \tilde{\omega}(\mathbf{k})(1 + v_1(\mathbf{k}))}{\tilde{\omega}(\mathbf{k}) u(\mathbf{k})} \right]. \label{eq:ToyM_phi_no_exc}
\end{equation}
and substituting the spectral branches $\tilde{\omega}_\pm(\mathbf{k})$ from Eq.~\eqref{eq:ToyM_Dirac_full} into Eq.~\eqref{eq:ToyM_phi_no_exc} yields the corresponding mixing angles $\phi_\pm^\mathbf{k}$, which directly determine the farfield polarization of the modes as described by Eq.~\eqref{eq:mixing}. In particular, taking into account the known polarizations of the initial modes given by Eqs.~\eqref{eq:ToyM_FarF_anis1} and \eqref{eq:ToyM_FarF_anis2}, we can express:
\begin{equation}
\tilde{\mathbf{e}}_+^\mathbf{k} = \mathbf{e}_1^\mathbf{k} \cos \phi_\mathbf{k} + \mathbf{e}_2^\mathbf{k} \sin \phi_\mathbf{k}, \label{eq:ToyM_FarF_unit_vectors_Anis}
\end{equation}
with $\mathbf{e}_1^\mathbf{k}$ and $\mathbf{e}_2^\mathbf{k}$ from Eq.~\eqref{eq:ToyM_initial_pol_1} and Eq~\eqref{eq:ToyM_initial pol_2}.
Since the mixed modes are linearly polarized in the far field and the mixing angle $\phi_\mathbf{k}$ is real, the resulting farfield polarization remains linear. The angle $\phi_\mathbf{k}$ thus represents a relative rotation of the original orthogonal farfield polarization directions, induced by the material anisotropy. 

\begin{figure}[ht]
    \centering
    \includegraphics[width=1\linewidth]{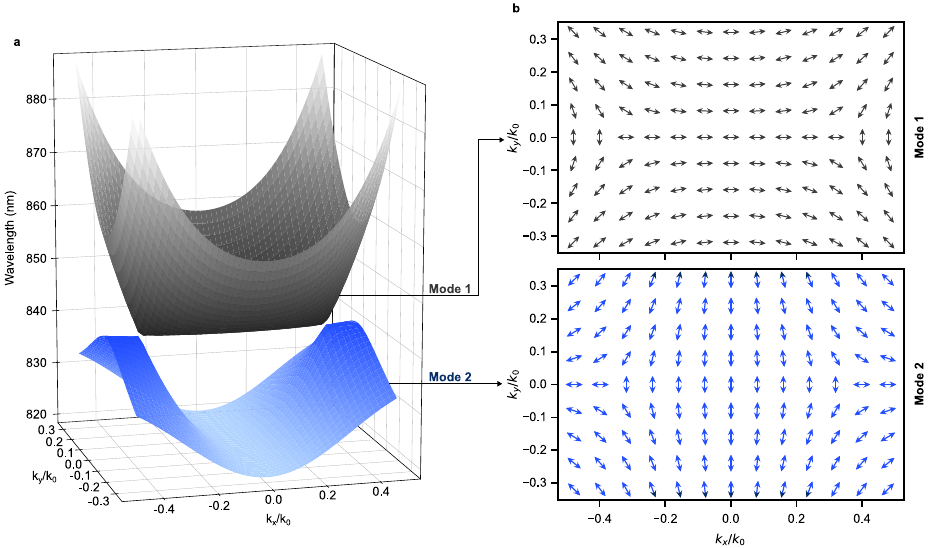}
    \caption{\textbf{Model Band Structure of Metasurface Made of Anisotropic Material.} 
    \textbf{a}.~Two spectral branches in $k$-space given by simplified form Eq.~\eqref{eq:ToyM_Dirac_full}, plotted with model parameters $\omega_{0}/(2\pi)=361.1+2.7i$ THz, $\chi=0.2$, $\beta_0=0.015$ and material anisotropy is $\Delta\varepsilon=1.7$.  \textbf{b}.~Farfield polarizations of Mode 1 (plus sign in Eq.~\eqref{eq:ToyM_Dirac_full}) and of Mode 2 (minus sign in Eq.~\eqref{eq:ToyM_Dirac_full}). The polarization unit vectors are evaluated using Eq.~\eqref{eq:ToyM_FarF_unit_vectors_Anis}. }\label{fig:ToyModel2}
\end{figure}

In Fig.~\ref{fig:ToyModel2} we plot both band structures given by the Eq.~\eqref{eq:ToyM_Dirac_full} and the modes' farfield polarizations according to Eq.\eqref{eq:ToyM_FarF_unit_vectors_Anis}. As previously, we take the model parameters $\omega_0/(2\pi) =361.1+2.7i$ THz and $\chi=0.2$. The material anisotropy is set to $\Delta\varepsilon=1.7$, while $\beta_0=0.015$. We can clearly see that this analytical model, based on only a few parameters, accurately reproduces all the key features given by the full-scale numerical simulations in Fig.~2(d–f) of the main text, and it captures all the effects caused by anisotropy in the metasurface material.

Finally, it is important to note that the derived expressions do not yield a well-defined limit at the $\Gamma$ point, as some terms become indeterminate in the form $0/0$ when $\mathbf{k} \to 0$. This ambiguity reflects the fundamental degeneracy of the eigenmodes at normal incidence. To resolve this, one can decide to approach the $\Gamma$ point along a specific direction, namely along $k_x = 0$. In this case, Mode 1 from Eq.~\eqref{eq:ToyM_FarF_anis1} becomes $x$-polarized, while Mode 2 from Eq.~\eqref{eq:ToyM_FarF_anis2} is $y$-polarized. Substituting into Eq.~\eqref{eq:ToyM_Dirac_full}, the spectral branches in the limit $\mathbf{k} \to 0$ reduce to:
\begin{equation}
\tilde{\omega}_\pm(\mathbf{k} = 0) \approx \omega_0 \left( 1 \pm \frac{\beta_0}{2}\Delta\varepsilon \right),
\end{equation}
which is equivalent to Eq~\eqref{eq:RSE_gamma_point} and Eq.~(2) of the main text.

\subsection{Model Including Excitonic Effects}

It is relatively straightforward to extend the preceding model to include, within a simple approximation, the effects of excitons and the associated features of light-matter strong coupling. A general theoretical framework for this has already been formulated in Ref.~\cite{S3}, where the RSE approach was shown to accurately predict the photonic eigenmodes of a system by leveraging the knowledge of another system of the same shape but composed of different materials. In particular \emph{nonlinear dispersive RSE} approach developed in Ref.~\cite{S3} establishes the connection between two systems characterized by negligible and strong permittivity dispersion. As we demonstrate below, this allows explicit accounting for the polaritonic effects within a remarkably simple analytical framework.

In our case, we begin with an initial system consisting of a metasurface composed of an isotropic, non-dispersive material with permittivity $\varepsilon_{ij} = \delta_{ij} \varepsilon$, whose mode eigenfrequencies are given by Eqs.~\eqref{eq:ToyM_initial_omega1}-\eqref{eq:ToyM_initial_omega2}, and whose farfield polarizations are described by Eqs.~\eqref{eq:ToyM_initial_pol_1}-\eqref{eq:ToyM_initial pol_2}. The system of interest, in contrast, is a metasurface of identical geometry but made of anisotropic excitonic material characterized by a permittivity tensor with the principal in-plane values having the Lorentzian frequency dispersion \eqref{eq:epsLorentz}. In the vicinity of their poles close to the positive frequency axis, they can be conveniently presented as:
\begin{equation}
\varepsilon_{xx}(\omega) = \varepsilon + \frac{\Delta \varepsilon}{2} + \frac{f_x}{\omega_x - \omega}, \quad
\varepsilon_{yy}(\omega) = \varepsilon - \frac{\Delta \varepsilon}{2} + \frac{f_y}{\omega_y - \omega},
\label{eq:SX19}
\end{equation}
where, to comply with observations in Ref.~\cite{S4}, we assume $\mathrm{Re}\,\omega_x < \mathrm{Re}\,\omega_y$, with similar oscillator strengths $f_{x}=f_{y}$. The anisotropic non-dispersive background is represented by $\Delta \varepsilon > 0$, and we treat both, this term and the excitonic contribution in Eq.~\eqref{eq:SX19}, as a perturbation $\Delta \varepsilon_{ij}$.

Nonlinear dispersive RSE then prescribes using the eigenvalue equation (Eq.~9 in Ref.~\cite{S3}), adapted to this context as:
\begin{equation}
[\omega_n(\mathbf{k}) - \tilde{\omega}] a_n = \tilde{\omega} \sum_m V_{nm}(\tilde{\omega}) a_m, \label{eq:ToyM_RSE_nonlinear_main}
\end{equation}
where $\omega_n(\mathbf{k})$ are the unperturbed eigenfrequencies and $V_{nm}(\tilde{\omega})$ are the perturbation matrix elements derived from overlap integrals with $\Delta \varepsilon_{ij}$.

Assuming again that only two modes are substantially mixed, we arrive at a simplified system analogous to Eqs.~\eqref{eq:ToyM_RSE_eq1}-\eqref{eq:ToyM_RSE_eq2}
\begin{align}
[\omega_1(\mathbf{k}) - \tilde{\omega}] \cos \phi_\mathbf{k} &= \tilde{\omega} \left[ v_1(\tilde{\omega}, \mathbf{k}) \cos \phi_\mathbf{k} + u(\tilde{\omega}, \mathbf{k}) \sin \phi_\mathbf{k} \right], \label{eq:ToyM_RSE_nonlinear1} \\
[\omega_2(\mathbf{k}) - \tilde{\omega}] \sin \phi_\mathbf{k} &= \tilde{\omega} \left[ v_2(\tilde{\omega}, \mathbf{k}) \sin \phi_\mathbf{k} + u(\tilde{\omega}, \mathbf{k}) \cos \phi_\mathbf{k} \right],
\label{eq:ToyM_RSE_nonlinear2}
\end{align}
where $\phi_\mathbf{k}$ is the mixing angle, and the frequency-dependent interaction parameters are given by:
\begingroup 
\makeatletter
  \setlength{\@mathmargin}{0pt}
  \makeatother
    \begin{align}
    v_1(\omega,\mathbf{k}) &= \int_V\!\left[
      \left[E^{\mathbf k}_{1x}(\mathbf r)\right]^{2}
       \left(\frac{\Delta\varepsilon}{2}+\frac{f_x}{\omega_x-\omega}\right)
     +\left[E^\textbf{k}_{1y}(\mathbf r)\right]^{2}
       \left(-\frac{\Delta\varepsilon}{2}+\frac{f_y}{\omega_y-\omega}\right)
     \right]\,
     \mathrm dV \label{eq:SX22}\\
    v_2(\omega,\mathbf{k}) &= \int_V\!\left[
      \left[E^{\mathbf k}_{2x}(\mathbf r)\right]^{2}
       \left(\frac{\Delta\varepsilon}{2}+\frac{f_x}{\omega_x-\omega}\right)
     +\left[E^\textbf{k}_{2y}(\mathbf r)\right]^{2}
       \left(-\frac{\Delta\varepsilon}{2}+\frac{f_y}{\omega_y-\omega}\right)
     \right]\,
     \mathrm dV \label{eq:SX23}\\
    u(\omega,\mathbf{k})   &= \int_V\!\left[
      E^{\mathbf k}_{1x}(\mathbf r)E^{\mathbf k}_{2x}(\mathbf r)
       \Bigl(\frac{\Delta\varepsilon}{2}+\frac{f_x}{\omega_x-\omega}\Bigr)
     +E^{\mathbf k}_{1y}(\mathbf r)E^{\mathbf k}_{2y}(\mathbf r)
       \Bigl(-\frac{\Delta\varepsilon}{2}+\frac{f_y}{\omega_y-\omega}\Bigr)
     \right]\,
     \mathrm dV \label{eq:SX24}
    \end{align}
\endgroup

Substituting the isotropic metasurface mode field distributions, approximately given by Eqs.~\eqref{eq:ToyM_FarF_anis1} and \eqref{eq:ToyM_FarF_anis2}, we obtain:
\begingroup 
\makeatletter
  \setlength{\@mathmargin}{0pt}
  \makeatother
\begin{flalign}
v_1(\omega,\mathbf k) &=
  \int_V\!\left\{
      \left[E^{(x)}_x(\mathbf r)\right]^{2}\frac{k_y^{2}}{k^{2}}
      \left(\frac{\Delta\varepsilon}{2}
            +\frac{f_x}{\omega_x-\omega}\right)
     +\left[E^{(y)}_y(\mathbf r)\right]^{2}\frac{k_x^{2}}{k^{2}}
      \left(-\frac{\Delta\varepsilon}{2}
            +\frac{f_y}{\omega_y-\omega}\right)
  \right\}\mathrm dV && \label{eq:SX25} \\[6pt]
v_2(\omega,\mathbf k) &=
  \int_V\!\left\{
      \left[E^{(x)}_x(\mathbf r)\right]^{2}\frac{k_x^{2}}{k^{2}}
      \left(\frac{\Delta\varepsilon}{2}
            +\frac{f_x}{\omega_x-\omega}\right)
     +\left[E^{(y)}_y(\mathbf r)\right]^{2}\frac{k_y^{2}}{k^{2}}
      \left(-\frac{\Delta\varepsilon}{2}
            +\frac{f_y}{\omega_y-\omega}\right)
  \right\}\mathrm dV && \label{eq:SX26} \\[6pt]
u(\omega,\mathbf k) &=
  \int_V\!\frac{k_xk_y}{k^{2}}\left\{\Big[
      E_x^{(x)}(\mathbf r)
  \Bigr]^{2}
      \left(\frac{\Delta\varepsilon}{2}
            +\frac{f_x}{\omega_x-\omega}\right)
     -\,\left[E_y^{(y)}(\mathbf r)\right]^{2}
      \left(-\frac{\Delta\varepsilon}{2}
            +\frac{f_y}{\omega_y-\omega}\right)
  \right\}\mathrm dV && \label{eq:SX27}
\end{flalign}
\endgroup
Introducing the parameter characterizing the nondispersive anisotropy, by analogy with \eqref{eq:ToyM_beta},
\begin{equation}
\beta = \Delta \varepsilon \int_V \left[ E_x^{(x)}(\mathbf{r}) \right]^2 \dd{V}
= \Delta \varepsilon \int_V \left[ E_y^{(y)}(\mathbf{r}) \right]^2 \dd{V}=\Delta\varepsilon\beta_0,
\label{eq:ToyM_beta_nonlinearRSE}
\end{equation}
as well as the effective exciton-polariton coupling strengths for the two orthogonal polarizations,
\begin{equation}
p_x = f_x \int_V \left[ E_x^{(x)}(\mathbf{r}) \right]^2 \dd{V}, \quad
p_y = f_y \int_V \left[ E_y^{(y)}(\mathbf{r}) \right]^2 \dd{V},
\label{eq:ToyM_g_x,y}
\end{equation}
allows us to express the coupling parameters in a more compact form as:
\begin{align}
v_1(\omega, \mathbf{k}) &= v_1(\mathbf{k})
+ \frac{k_y^2}{k^2} \frac{p_x}{\omega_x - \omega}
+ \frac{k_x^2}{k^2} \frac{p_y}{\omega_y - \omega}, 
\label{eq:ToyM_v1=g_x,y} \\
v_2(\omega, \mathbf{k}) &= v_2(\mathbf{k})
+ \frac{k_x^2}{k^2} \frac{p_x}{\omega_x - \omega}
+ \frac{k_y^2}{k^2} \frac{p_y}{\omega_y - \omega}, 
\label{eq:ToyM_v2=g_x,y} \\
u(\omega, \mathbf{k}) &= u(\mathbf{k})
+ \frac{k_x k_y}{k^2} \left[\frac{p_x}{\omega_x - \omega}
- \frac{p_y}{\omega_y - \omega}\right].
\label{eq:ToyM_u=g_x,y}
\end{align}
Here, \(v_1(\mathbf{k})\), \(v_2(\mathbf{k})\), and \(u(\mathbf{k})\) are defined as previously in Eqs.\eqref{eq:ToyM_v1=beta}-\eqref{eq:ToyM_u=beta} and describe the effects of nondispersive anisotropy, while the remaining terms are of clear excitonic origin.

Introducing also again $\tilde{\omega}_{1,2}(\mathbf{k}) = \omega_{1,2}(\mathbf{k}) \left( 1 + v_{1,2}(\mathbf{k}) \right)^{-1},$
one can write the corresponding form of the system of Eqs.~\eqref{eq:ToyM_RSE_nonlinear1}, \eqref{eq:ToyM_RSE_nonlinear2} as:
\begin{align}\label{eq:SX34}
&\left[ \left(1 + v_1(\mathbf{k}) \right) (\tilde{\omega}_1(\mathbf{k}) - \tilde{\omega})(\omega_x - \tilde{\omega})(\omega_y - \tilde{\omega})
- \tilde{\omega}(\omega_y - \tilde{\omega})p_x\frac{k_y^2}{k^2 }
- \tilde{\omega}(\omega_x - \tilde{\omega})p_y\frac{k_x^2}{k^2 }\right]\cos\phi_\textbf{k}\\
&\left. = \left[\tilde{\omega}u(\textbf{k})(\omega_x -\tilde{\omega})(\omega_y-\tilde{\omega})+\frac{k_xk_y}{k^2}\tilde{\omega}\Big((\omega_y-\tilde{\omega})p_x-(\omega_x-\tilde{\omega})p_y\right)\right]\sin\phi_\textbf{k},\notag \\
&\left[ \left(1 + v_2(\mathbf{k}) \right) (\tilde{\omega}_2(\mathbf{k}) - \tilde{\omega})(\omega_x - \tilde{\omega})(\omega_y - \tilde{\omega})
- \tilde{\omega}(\omega_y - \tilde{\omega})p_x\frac{k_x^2}{k^2 }
- \tilde{\omega}(\omega_x - \tilde{\omega})p_y\frac{k_y^2}{k^2 }\right]\sin\phi_\textbf{k}\\ 
&=\left[\tilde{\omega}u(\textbf{k})(\omega_x -\tilde{\omega})(\omega_y-\tilde{\omega})+\frac{k_xk_y}{k^2}\tilde{\omega}\Big((\omega_y-\tilde{\omega})p_x-(\omega_x-\tilde{\omega})p_y\Big)\right]\cos\phi_\textbf{k}\notag,
\end{align}

The corresponding eigenmode dispersion equation following from here is:
\begin{align}\label{eq:SX35}
&\left[ \left(1 + v_1(\mathbf{k}) \right) (\tilde{\omega}_1(\mathbf{k}) - \tilde{\omega})(\omega_x - \tilde{\omega})(\omega_y - \tilde{\omega})
-\tilde{\omega}(\omega_y-\tilde{\omega})p_x\frac{k_y^2}{k^2}-\tilde{\omega}(\omega_x-\tilde{\omega})p_y\frac{k_x^2}{k^2})\right]
 \notag \\
&\times \left[ \left(1 + v_2(\mathbf{k}) \right) (\tilde{\omega}_2(\mathbf{k}) - \tilde{\omega})(\omega_x - \tilde{\omega})(\omega_y - \tilde{\omega})
-\tilde{\omega}(\omega_y-\tilde{\omega})p_x\frac{k_x^2}{k^2}-\tilde{\omega}(\omega_x-\tilde{\omega})p_y\frac{k_y^2}{k^2})\right] \notag \\
&= \left[ \tilde{\omega} u(\mathbf{k})(\omega_x - \tilde{\omega})(\omega_y - \tilde{\omega})
+ \frac{k_x k_y}{k^2} \tilde{\omega} \Big( (\omega_y - \tilde{\omega}) p_x - (\omega_x - \tilde{\omega}) p_y \Big) \right]^2.
\end{align}

\begin{figure}[h]
    \centering
    \includegraphics[width=1\linewidth]{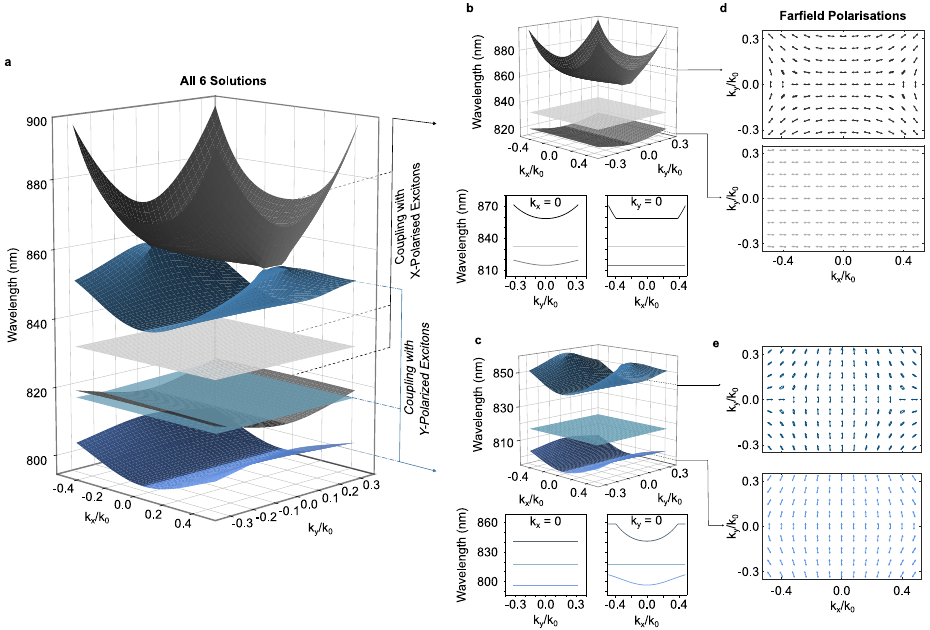}
    \caption{\textbf{Model Band Structure of Metasurface Made of Anisotropic Excitonic Material in the Strong Coupling Regime.} 
    \textbf{a}~Six spectral branches given as solutions of the eigenmode dispersion equation Eq.~\eqref{eq:4-order-polyn}, including two planes corresponding to the $x$-polarized ($\omega_x$) and $y$-polarized excitons ($\omega_y$), and four self-hybridized exciton-polaritons branches. Separate solutions corresponding to coupling with $x$-polarized (\textbf{b}) and $y$-polarized (\textbf{c}) excitons, together with cuts at $k_x=0$ and $k_y=0$. \textbf{d} and \textbf{e} show farfield polarizations of hybrid exciton-polaritons, obtained by evaluating the polarization unit vectors in Eq.~\eqref{eq:ToyM_FarF_unit_vectors_Anis} with the corresponding mixing angle for each mode from Eq.~\eqref{eq:ToyM_phi_exc}. The model parameters are set to $\omega_{0}=361.1+2.7i$ THz, $\chi=0.2$, $\beta_0=0.015$ and material anisotropy is $\Delta\varepsilon=1.7$. Parameters for the excitons  $\omega_x/(2\pi)=360.33+8.97i$ THz, $\omega_y/(2\pi)=366.95+5.77i$ THz, $f_x/(2\pi)=20.0$ THz, $f_y/(2\pi)=19.6$ THz were taken directly from the TCMT fit of the experimental results.} 
    \label{fig:ToyModel3}
\end{figure}

It is of sixth order in frequency \( \tilde{\omega} \), i.e., generally yields up to six spectral branches (surfaces in \(\mathbf{k}\)-space). However, it can be rewritten in the following form:
\begin{equation}
    (\omega_x-\tilde{\omega})(\omega_y-\tilde{\omega})\left[ b_4(\textbf{k})\tilde{\omega}^4+b_3(\textbf{k})\tilde{\omega}^3+b_2(\textbf{k})\tilde{\omega}^2+b_1(\textbf{k})\tilde{\omega}+b_0(\textbf{k})\right]=0,
    \label{eq:4-order-polyn}
\end{equation}
which always has solutions corresponding to what is commonly referred to as "uncoupled excitons" with $\tilde{\omega}=\omega_x$ and $\tilde{\omega}=\omega_y$, which seemingly do not participate in the coupling process. Notably, in the present formulation, these solutions emerge naturally from the analysis. The remaining fourth-order polynomial equation from Eq.\eqref{eq:4-order-polyn} is \textbf{k}-dependent and has the following coefficients: 
\begin{align}
    & b_4(\textbf{k})=(1+v_1(\textbf{k}))(1+v_2(\textbf{k}))-u^2(\textbf{k}),\\
    & b_3(\textbf{k})=u^2(\textbf{k})(\omega_x+\omega_y)-(1+v_1(\textbf{k}))(1+v_2(\textbf{k}))(\omega_x+\omega_y+{\tilde\omega_1}(\textbf{k})+{\tilde\omega_2}(\textbf{k})) \notag \\
    &-(1+v_1(\textbf{k}))\left(p_x\frac{k_x^2}{k^2}+p_y\frac{k_y^2}{k^2}\right)-(1+v_2(\textbf{k}))\left(p_x\frac{k_y^2}{k^2}+p_y\frac{k_x^2}{k^2}\right)+2u(\textbf{k})\frac{k_xk_y}{k^2}(p_x-p_y), \\
    & b_{2}(\textbf{k}) =p_{x}p_{y} + (1+v_{1}(\textbf{k}))(1+v_{2}(\textbf{k}))\big[ \omega_{x}\omega_{y} + \tilde\omega_{1}(\textbf{k})\tilde\omega_{2}(\textbf{k}) 
+ (\tilde\omega_{1}(\textbf{k})+\tilde\omega_{2}(\textbf{k}))(\omega_{x}+\omega_{y}) \big]\notag\\& - 2u(\textbf{k})\frac{k_{x}k_{y}}{k^{2}}
(p_{x}\omega_{y} - p_{y}\omega_{x}) + (1+v_{1}(\textbf{k}))\left[ p_{x}\frac{k_{x}^{2}}{k^{2}}(\tilde\omega_{1}(\textbf{k})+\omega_{y}) 
+ p_{y}\frac{k_{y}^{2}}{k^{2}}(\tilde\omega_{1}(\textbf{k})+\omega_{x}) \right] \notag \\&+ (1+v_{2}(\textbf{k}))\left[ p_{x}\frac{k_{y}^{2}}{k^{2}}(\tilde\omega_{2}(\textbf{k})+\omega_{y}) 
+ p_{y}\frac{k_{x}^{2}}{k^{2}}(\tilde\omega_{2}(\textbf{k})+\omega_{x}) \right]-u^{2}(\textbf{k})\omega_{x}\omega_{y}, \\
    & b_{1}(\textbf{k}) = - (1+v_{1}(\mathbf{k}))(1+v_{2}(\mathbf{k})) 
\Big[ (\tilde\omega_{1}(\mathbf{k})+\tilde\omega_{2}(\mathbf{k}))\omega_{x}\omega_{y} 
+ \tilde\omega_{1}(\mathbf{k})\tilde\omega_{2}(\mathbf{k})(\omega_{x}+\omega_{y}) \Big] \notag\\
&- (1+v_{1}(\mathbf{k}))\tilde\omega_{1}(\mathbf{k}) 
\left[ p_{x}\omega_{y}\frac{k_{x}^{2}}{k^{2}} + p_{y}\omega_{x}\frac{k_{y}^{2}}{k^{2}} \right] - (1+v_{2}(\mathbf{k}))\tilde\omega_{2}(\mathbf{k}) 
\left[ p_{x}\omega_{y}\frac{k_{y}^{2}}{k^{2}} + p_{y}\omega_{x}\frac{k_{x}^{2}}{k^{2}} \right], \\
    & b_0(\textbf{k})=(1+v_1(\textbf{k}))(1+v_2(\textbf{k}))\omega_x\omega_y\tilde{\omega}_1(\textbf{k})\tilde{\omega}_2(\textbf{k}).
\end{align}
We solve this equation numerically using the model parameters from the previous section. For the excitonic part, we use the data obtained from our TCMT fit of the experimental results (see Section \ref{section:TCMT fit}): $\omega_x/(2\pi)=360.33+8.97i$ THz, $\omega_y/(2\pi)=366.95+5.77i$ THz, $f_x/(2\pi)=20.0$ THz, $f_y/(2\pi)=19.6$ THz. Fig.~\ref{fig:ToyModel3}a displays the band structures of all six solutions of Eq.\eqref{eq:4-order-polyn}: four hybrid exciton–polaritons and two uncoupled excitons. For clarity, we split the figure into two parts, corresponding to the coupling with the $x$-polarized ($\omega_x$) and $y$-polarized ($\omega_y$) excitons (Exciton 1 and Exciton 2 in the main text). As shown in Fig.\ref{fig:ToyModel3}b and Fig.\ref{fig:ToyModel3}c the initial topological features of the photonic bands from Fig.2d of the main text and Fig.\ref{fig:ToyModel2}, such as flat and parabolic dispersions, are also present in the strong-coupling regime for both lower and upper polaritonic branches. This foundation is consistent with the experimental results in Fig.~4 of the main text. Moreover, even in the strong-coupling regime, two Dirac points at $k_x/k_0 = \pm 0.4$ can be found for two polaritonic solutions with the highest resonant wavelengths.
\clearpage
To find the farfield polarizations of the self-hybridyzed polaritons, the corresponding mixing angle can be evaluated for each mode, by analogy with Eq.\eqref{eq:ToyM_phi_no_exc}, as:
\begin{equation}
\phi_\mathbf{k} = \atan \left[ \frac{(1 + v_1(\mathbf{k}))(\tilde{\omega}_1(\mathbf{k}) - {\tilde\omega})(\omega_x - {\tilde\omega)}(\omega_y - {\tilde\omega})
- \tilde{\omega}(\omega_y - \tilde{\omega})p_x
\frac{k_y^2}{k^2} - {\tilde\omega}(\omega_x - \tilde{\omega}) p_y\frac{k_x^2}{k^2}  }
{\tilde{\omega} u(\mathbf{k})(\omega_x - \tilde{\omega})(\omega_y - \tilde{\omega})
+ \frac{k_x k_y}{k^2} \tilde{\omega} \left[ (\omega_y - \tilde{\omega}) p_x - (\omega_x - \tilde{\omega}) p_y \right]} \right].
\label{eq:ToyM_phi_exc}
\end{equation}
As in the previous section, in Fig.~\ref{fig:ToyModel3}d and Fig.~\ref{fig:ToyModel3}e we plot the  farfield distribution for all four polaritonic modes using the polarization unit vectors from Eq.~\eqref{eq:ToyM_FarF_unit_vectors_Anis}.
The figures show that, over almost the entire considered $k$-space region, the polarizations of all four polaritonic modes remain linear. The only exception occurs for the two branches with the highest wavelengths in the vicinity of the Dirac points, where a small ellipticity appears. The Dirac points themselves are accompanied by half-integer topological charges at $k_x/k_0=\pm0.4$, where the linear polarizations abruptly change to the orthogonal.

Finally, we separately consider a situation in the vicinity of the \(\Gamma\) point, which we again, for definiteness, approach along the line \( k_x = 0 \). Then, as in the previous section, \( v_1(\mathbf{k} \to 0) = \frac{\beta}{2} \), \( v_2(\mathbf{k} \to 0) = -\frac{\beta}{2} \), \( u(\mathbf{k} \to 0) = 0 \), and the dispersion Eq.~\eqref{eq:SX35} reduces to:
\begin{align}
&\left[ \left(1 + \frac{\beta}{2} \right) (\tilde{\omega}_1(\mathbf{k}) - \tilde{\omega})(\omega_x - \tilde{\omega})(\omega_y - \tilde{\omega})
- \tilde{\omega}(\omega_y - \tilde{\omega}) p_x \right] \notag \\
&\times \left[ \left(1 - \frac{\beta}{2} \right) (\tilde{\omega}_2(\mathbf{k}) - \tilde{\omega})(\omega_x - \tilde{\omega})(\omega_y - \tilde{\omega})
- \tilde{\omega}(\omega_x - \tilde{\omega}) p_y \right] = 0,
\label{eq:SX37}
\end{align}
which naturally splits into four separate equations:
\begin{align}
(\tilde{\omega}_- - \tilde{\omega})(\omega_x - \tilde{\omega}) &= \tilde{\omega} p_x, \label{eq:SX38} \\
(\tilde{\omega}_+ - \tilde{\omega})(\omega_y - \tilde{\omega}) &= \tilde{\omega} p_y, \label{eq:SX39} \\
\tilde{\omega} - \omega_x &= 0, \label{eq:SX40} \\
\tilde{\omega} - \omega_y &= 0. \label{eq:SX41}
\end{align}

The solutions of the first two equations, Eqs.~\eqref{eq:SX38} and \eqref{eq:SX39}, describe the strong coupling between photonic modes and excitons. These solutions are analogous to the expressions used for the polaritonic branches in the TCMT fit below (see Eq.~\eqref{eq:PolaritonBranches}). At the $\Gamma$-point, the branches $\tilde{\omega}_\pm$ are not degenerate, but are linearly polarized and interact selectively with the corresponding excitons.

Altogether, rather than relying on a phenomenological introduction of excitons within coupled-mode theory, the RSE approach allows deriving all spectral branches directly from the characteristics of the permittivity dispersion, without invoking empirical assumptions or fitting parameters.

\color{black}
\clearpage
\section{Stokes Parameters Analysis}
\subsection{Experimental Stokes Parameters Fit}

To unambiguously characterize the split photonic bands, we measure hyperspectral, $k$-space-resolved Stokes parameters $\mathbf{S}=(S_0,S_1,S_2)$. Here, $S_0$ denotes the total measured transmittance, while $S_1$ and $S_2$ quantify the polarization contrast in the horizontal ($0^\circ$)/vertical ($90^\circ$) and diagonal ($45^\circ$)/anti-diagonal ($-45^\circ$) bases, respectively,
\begin{equation}
    S_0 = T_H + T_V,
    \qquad
    S_1 = T_H - T_V,
    \qquad
    S_2 = T_{45^\circ} - T_{-45^\circ}.
\end{equation}
From these observables we obtain the degree of linear polarization
\begin{equation}
    P = \frac{\sqrt{S_1^2+S_2^2}}{S_0},
\end{equation}
and the polarization angle
\begin{equation}
    \phi=\frac{1}{2}\arctan\frac{S_2}{S_1}.
\end{equation}

\begin{figure}[ht]
    \centering
    \includegraphics[width=\linewidth]{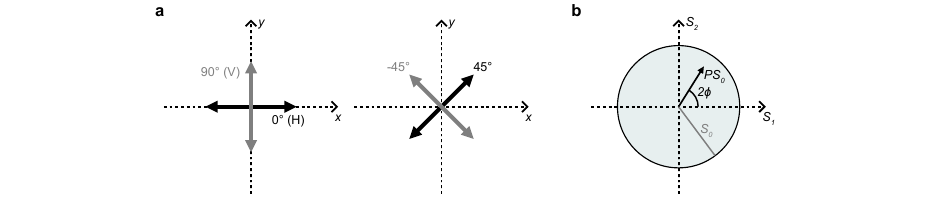}
    \caption{\textbf{Definition of the Measured  Stokes Parameters.}
    \textbf{a} Linear polarization bases used in the experiment. The horizontal/vertical basis yields $S_1 = T_H - T_V$, while the diagonal/anti-diagonal basis yields $S_2 = T_{45^\circ} - T_{-45^\circ}$.
    \textbf{b} Representation of linearly polarized states in the $(S_1,S_2)$ plane for $S_3=0$. Fully linearly polarized states lie on the outer circle of radius $S_0$, whereas partially polarized states occupy the interior with radial coordinate $\sqrt{S_1^2+S_2^2}=PS_0$. The azimuthal angle in Stokes space is $2\phi$, where $\phi$ is the real-space polarization angle.}
    \label{fig:stokes-definitions}
\end{figure}

As the two photonic basis modes are orthogonal, the measured polarization angle directly reflects their mixing angle Eq.~(\ref{eq:ToyM_FarF_unit_vectors_Anis}). The experimental polarization texture in $k$-space can therefore be obtained by evaluating $\phi$ on the dispersion surfaces of the split modes. To determine these mode surfaces, we first fit the RSE model from Eq.~(\ref{eq:ToyM_Dirac_full}) to the measured hyperspectral $S_0$ data.

For the experimental fit, the model is generalized to include anisotropic band curvatures, $\chi \rightarrow (\chi_x,\chi_y)$, global momentum offsets, $k_i \rightarrow k_i-k_{i,0}$ with $i=x,y$, and a global rotation $\theta_0$,
\begin{equation}
    (k_x,k_y)\rightarrow
    (k_x\cos\theta_0-k_y\sin\theta_0,\,
     k_x\sin\theta_0+k_y\cos\theta_0),
\end{equation}
in order to account for measurement misalignment. Prior to fitting, the data are smoothed by Gaussian filtering in the $(k_x,k_y)$ plane and by penalized least-squares smoothing along the frequency axis \cite{eilers2003perfect}. A dip-finding routine is then used to extract the two dispersion branches.

\begin{figure}[ht]
    \centering
    \includegraphics[width=\linewidth]{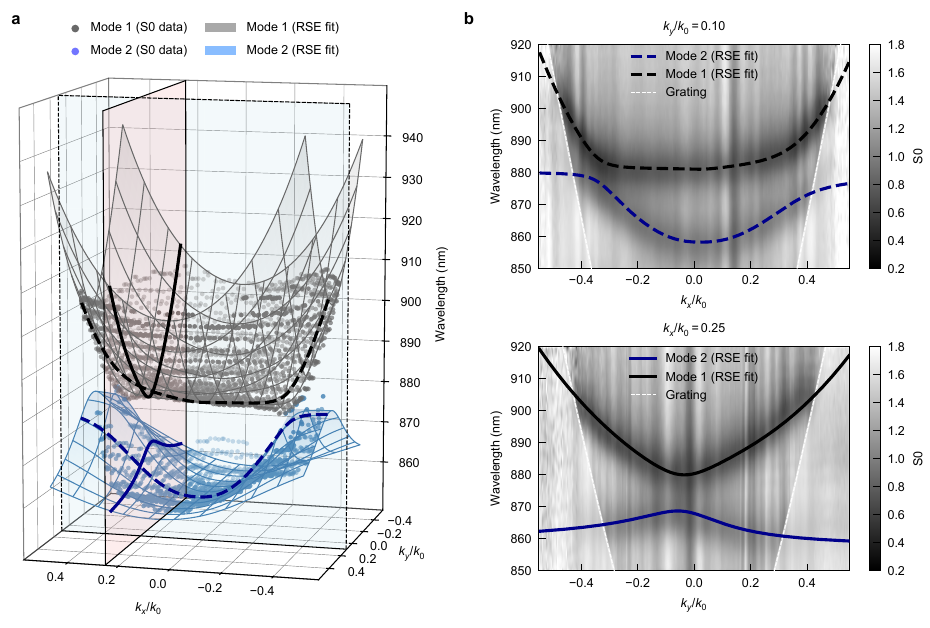}
    \caption{\textbf{Fit of the Generalized RSE Model to the Measured Hyperspectral $S_0$ Data.}
    \textbf{a} Extracted mode dispersion together with the fitted RSE surfaces in $(k_x,k_y,\lambda)$ space.
    \textbf{b} Representative cuts through the measured $S_0$ data and the corresponding fitted dispersions for selected $k$-space planes. Regions outside the light cone are masked in the visualization and excluded from the fit.}
    \label{fig:rse-fits}
\end{figure}

As discussed in Sec.~\ref{sec:flatness-analysis}, the measured dispersion is strongly affected by grating features outside the light cone. We therefore restrict the fit to the region inside the light cone. The resulting fit is in very good agreement with the data, as shown in Fig.~\ref{fig:rse-fits}, and the extracted parameters are listed in Table~\ref{tab:rse-fit-params}. In particular, the fitted values remain close to those assumed in the analytical model from Sec.~S3, namely $\chi=0.2$, $\beta = \beta_0\Delta\varepsilon = 0.0255$ with $\beta_0=0.015$ and $\Delta\varepsilon=1.7$. The curvatures $\chi$ along the principal k-space axes, extracted from experimental data and accounting for grating coupling, are detailed in \autoref{sec:flatness-analysis}.

\begin{table}[ht]
    \centering
    \begin{tabular}{l l S[table-format=2.4]}
        \toprule
        \textbf{Parameter} & \textbf{Value} & {\textbf{Rel. Unc. (\%)}} \\
        \midrule
        $\omega_0/(2\pi)$ (THz)           & 344.9858   & 0.0011 \\
        $\chi_x$                 & 0.2022    & 0.0744 \\
        $\chi_y$                 & 0.1095    & 0.1500 \\
        $\beta$                    & 0.0249    & 0.0850 \\
        $k_{x,0}/k_0$                  & 0.0098     & 1.2638 \\
        $k_{y,0}/k_0$                  & -0.0024    & 5.4222 \\
        $\theta_0$ (\si{\degree})  & 7.5133    & 0.5941 \\
        \bottomrule
    \end{tabular}
    \caption{Parameters of the generalized RSE fit to the experimental Stokes parameter data for the metasurface with scaling factor $S=1.0$ shown in Fig.~3 of the main text.}
    \label{tab:rse-fit-params}
\end{table}

To describe the measured polarization texture within the RSE model, we now express the hybrid eigenstates using Stokes parameters. We introduce the cartesian Jones vector through electric field components:
\begin{equation}
    \vec{J}=
    \begin{pmatrix}
        E_x \\ E_y
    \end{pmatrix},
\end{equation}
normalized such that:
\begin{equation}
    \|\vec{J}\|^2 = |E_x|^2+|E_y|^2 = S_0.
\end{equation}
The corresponding Stokes parameters are:
\begin{equation}
    S_1 = |E_x|^2 - |E_y|^2, \ \ \
    S_2 = 2\,\mathrm{Re}(E_xE_y^{\ast}), \ \ \
    S_3 = 2\,\mathrm{Im}(E_xE_y^{\ast}).
\end{equation}
In the present case the polarization is purely linear, so the field amplitudes are real and hence the last Stokes parameter $S_3=0$.

The Jones vector can be viewed as a pure state of a two-level system in the $(x,y)$ basis, with density matrix:
\begin{equation}
    \rho=\vec{J}\vec{J}^\dagger.
\end{equation}
For $S_3=0$, this density matrix can be expanded using the Pauli matrices formalism as:
\begin{equation}
    \rho=\frac{1}{2}\left( S_0 I + \vec{S}\cdot\vec{\sigma} \right)
    =
    \frac{1}{2}\left( S_0 I + S_1\sigma_3+S_2\sigma_1\right),
\end{equation}
where $I$ is the $2\times2$ identity matrix, $\vec{S}=(S_1,S_2)^T$, and:
\begin{equation}
    \sigma_3=
    \begin{pmatrix}
        1 & 0 \\
        0 & -1
    \end{pmatrix},
    \qquad
    \sigma_1=
    \begin{pmatrix}
        0 & 1 \\
        1 & 0
    \end{pmatrix}.
\end{equation}

The hybrid eigenstates $\vec{J}_\pm$ satisfy:
\begin{equation}
    \tilde{H}\vec{J}_\pm = \tilde{\omega}_\pm \vec{J}_\pm,
\end{equation}
with effective Hamiltonian:
\begin{equation}
    \tilde{H}=
    \begin{pmatrix}
        \tilde{\omega}_{1,\mathrm{eff}} & g_\mathrm{eff} \\
        g_\mathrm{eff} & \tilde{\omega}_{2,\mathrm{eff}}
    \end{pmatrix}.
\end{equation}
The renormalized eigenfrequencies $\tilde{\omega}_{1,\mathrm{eff}}$, $\tilde{\omega}_{2,\mathrm{eff}}$, and coupling strength $g_\mathrm{eff}$ are given by:
\begin{equation}
\tilde{\omega}_{1,\mathrm{eff}}=\frac{\tilde{\omega}_1}{1-\tilde{u}^2},
    \qquad
\tilde{\omega}_{2,\mathrm{eff}}=\frac{\tilde{\omega}_2}{1-\tilde{u}^2},
    \qquad
    g_\mathrm{eff}=
\frac{\tilde{u}}{1-\tilde{u}^2}\sqrt{\tilde{\omega}_1\tilde{\omega}_2},
\end{equation}
with $\tilde{\omega}_1$, $\tilde{\omega}_2$ and $\tilde{u}$ defined in accordance with Eq.~\eqref{eq:omega and u tilde_RSE}.
The eigenvalues of $\tilde{H}$ are:
\begin{equation}
    \tilde{\omega}_\pm=
    \frac{\tilde{\omega}_{1,\mathrm{eff}}+\tilde{\omega}_{2,\mathrm{eff}}}{2}
    \pm
    \sqrt{
        \left(\frac{\tilde{\omega}_{1,\mathrm{eff}}-\tilde{\omega}_{2,\mathrm{eff}}}{2}\right)^2 + g_\mathrm{eff}^2
    },
\end{equation}
such that the splitting becomes:
\begin{equation}
    \Delta\tilde{\Omega}
    =
    \tilde{\omega}_+ - \tilde{\omega}_-
    =
    2\sqrt{
        \left(\frac{\tilde{\omega}_{1,\mathrm{eff}}-\tilde{\omega}_{2,\mathrm{eff}}}{2}\right)^2 + g_\mathrm{eff}^2
    }.
\end{equation}

Since $\tilde{H}$ is Hermitian, it can likewise be decomposed as:
\begin{equation}
    \tilde{H}=h_0I+h_1\sigma_3+h_2\sigma_1,
\end{equation}
with
\begin{equation}
    h_0=\frac{\tilde{\omega}_{1,\mathrm{eff}}+\tilde{\omega}_{2,\mathrm{eff}}}{2},
    \qquad
    h_1=\frac{\tilde{\omega}_{1,\mathrm{eff}}-\tilde{\omega}_{2,\mathrm{eff}}}{2},
    \qquad
    h_2=g_\mathrm{eff}.
\end{equation}
We define $\vec{h}=(h_1,h_2)^T$ as an effective Stokes-space vector, such that:
\begin{equation}
    \tilde{H}=h_0I+\vec{h}\cdot\vec{\sigma}
    =
    h_0I+|\vec{h}|\,\hat{h}\cdot\vec{\sigma},
\end{equation}
where $\hat{h}=\vec{h}/|\vec{h}|$. Since the operator $\hat{h}\cdot\vec{\sigma}$ has eigenvalues $\pm1$, its eigenstates satisfy:
\begin{equation}
    (\hat{h}\cdot\vec{\sigma})\vec{J}_\pm=\pm \vec{J}_\pm.
\end{equation}
and the corresponding projectors are:
\begin{equation}
    P_\pm=\frac{1}{2}\left(I\pm \hat{h}\cdot\vec{\sigma}\right).
\end{equation}
For pure states normalized to $S_0$, the density matrix is proportional to the projector onto these states ($\rho_\pm=S_0P_\pm$). Looking at the upper branch:
\begin{equation}
    \rho_+ = S_0P_+
    =
    \frac{S_0}{2}\left(I+\hat{h}\cdot\vec{\sigma}\right).
\end{equation}
and comparing this expression with:
\begin{equation}
    \rho=\frac{1}{2}\left(S_0I+\vec{S}\cdot\vec{\sigma}\right),
\end{equation}
yields:
\begin{equation}
    \frac{\vec{S}}{S_0}=\hat{h}=\frac{\vec{h}}{|\vec{h}|},
\end{equation}
so, the initial Stokes vector $\vec{S}$ is parallel or anti-parallel to the effective Stokes vector $\vec{h}$. Thus, the Stokes vectors of the two eigenstates are:
\begin{equation}
    \vec{S}_\pm=\pm S_0\frac{\vec{h}}{|\vec{h}|}.
    \label{eq:rse-stokes-vecs}
\end{equation}
Equation~(\ref{eq:rse-stokes-vecs}) describes the polarization texture of the ideal Hamiltonian. To account for experimental misalignment and residual birefringence, we introduce a small correction \(\tilde{H}\rightarrow \tilde{H}+\delta \tilde{H}\), such that:
\begin{equation}
    \tilde{H}_{\mathrm{tot}}
    =
    h_{0,\mathrm{tot}}I+\vec{h}_{\mathrm{tot}}\cdot\vec{\sigma},
\end{equation}
with
\begin{equation}
    \vec{h}_{\mathrm{tot}}
    =
    \vec{h}+\delta\vec{h}
    =
    \frac{|\vec{h}|}{S_0}\vec{S}_\pm+\delta\vec{h}
    =
    \frac{\Delta\tilde{\Omega}}{2S_0}\vec{S}_\pm+\delta\vec{h}.
\end{equation}
Here, $\delta\vec{h}=(\delta h_1,\delta h_2)^T$ denotes a constant, zeroth-order correction. This approximation is sufficient because the residual deviations from the ideal RSE polarization texture are small over the fitted $k$-space range and mainly act as a rigid shift of the Stokes texture rather than a strong distortion. The new Stokes vectors then read as:
\begin{equation}
    \vec{S}_{\pm,\mathrm{tot}}
    =
    \pm S_0\frac{\vec{h}_{\mathrm{tot}}}{|\vec{h}_{\mathrm{tot}}|},
\end{equation}
and the corresponding polarization angle is:
\begin{equation}
    \phi_{\mathrm{tot}}
    =
    \frac{1}{2}\arctan\!\left(\frac{S_{2,\mathrm{tot}}}{S_{1,\mathrm{tot}}}\right).
\end{equation}

\begin{figure}[H]
    \centering
    \includegraphics[width=\linewidth]{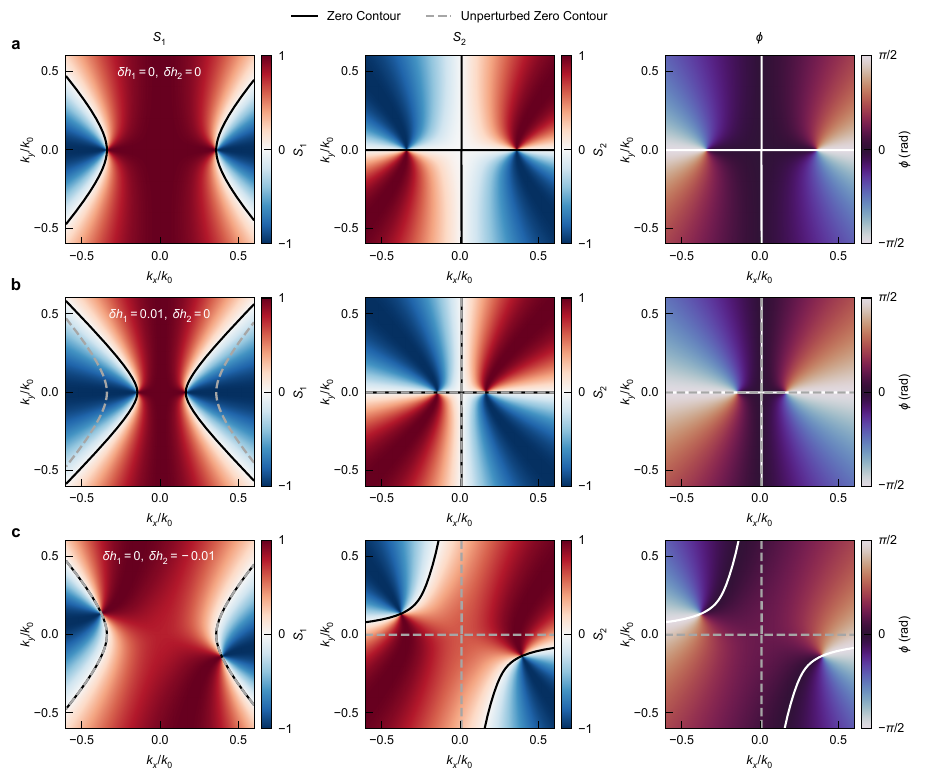}
    \caption{
        \textbf{Influence of Birefringent Terms ($\delta h_1$ and $\delta h_2$) on Stokes Parameters.}
        \textbf{a} Ideal RSE solutions for $S_1$, $S_2$, and the corresponding polarization angle $\phi$, calculated using the parameters from Table~\ref{tab:rse-fit-params}. For clarity, the global $k$-space rotation was set to $\theta_0=0^\circ$. Solid lines indicate the zero-crossings of the respective Stokes parameters.
        \textbf{b} Effect of a correction $\delta h_1$. A constant offset in $h_1$ shifts the polarization texture along the horizontal/vertical basis and moves the far-field singularities farther apart or closer together.
        \textbf{c} Effect of a correction $\delta h_2$. An offset in $h_2$ shifts the singularities along the nodal line of the ideal solution and lifts the zero-crossing of $S_2$ at the $\Gamma$ point, resulting in two diverging branches of the mixing angle.}
    \label{fig:stokes-vector-offsets}
\end{figure}

Figure~\ref{fig:stokes-vector-offsets} illustrates how $\delta\vec{h}$ modifies the Stokes texture and the associated mixing angle. For clarity, we compare the ideal solution ($\delta h_1=\delta h_2=0$), solutions with horizontal ($\delta h_1=0.01$ and $\delta h_2=0$) and vertical ($\delta h_1=0$ and $\delta h_2=-0.01$) offsets. As demonstrated in Fig.~\ref{fig:stokes-vector-offsets} $\delta h_1$ primarily changes the separation of the far-field singularities along the $k_x$ direction, $\delta h_2$ mainly shifts them in $k_y$. Together, these two terms account for the dominant experimental deviations from the ideal RSE solution, including small sample rotations and residual birefringence.

\begin{figure}[H]
    \centering
    \includegraphics[width=\linewidth]{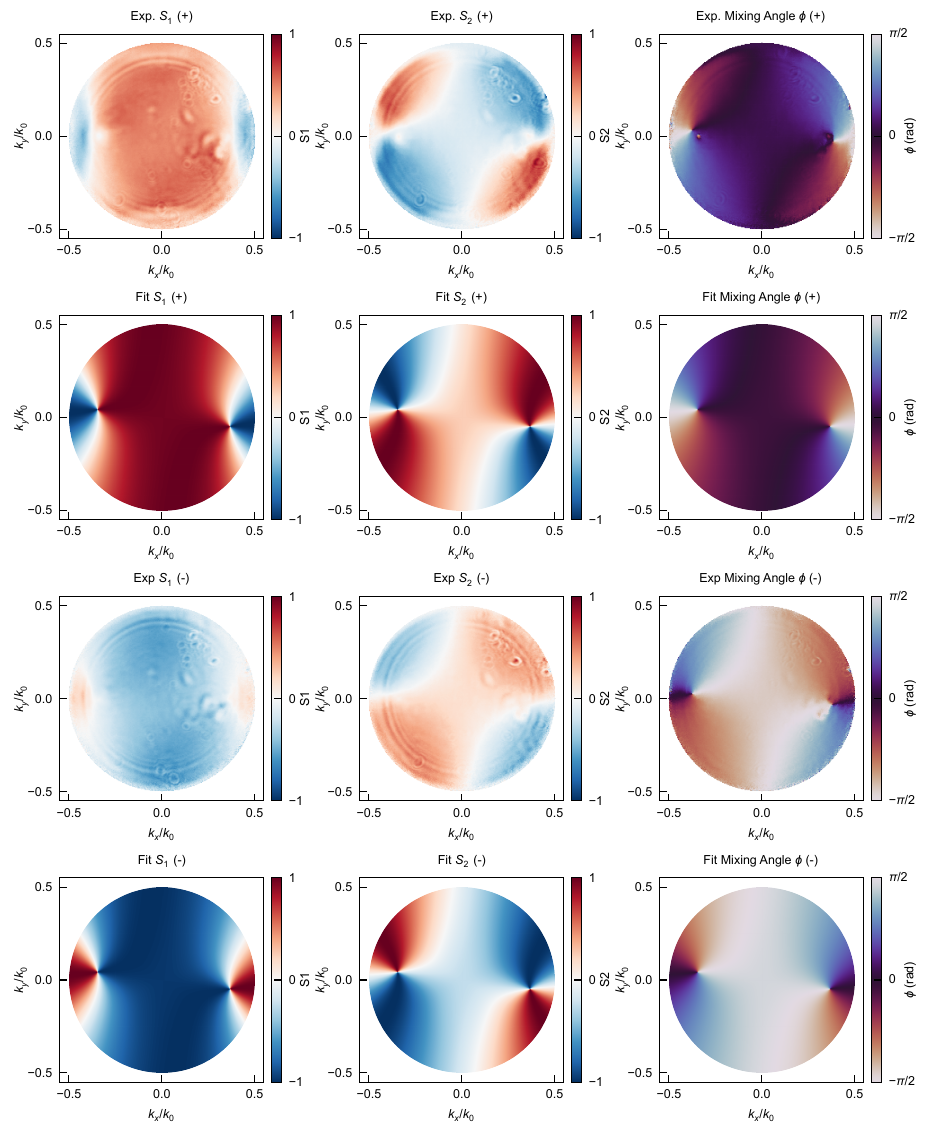}
    \caption{\textbf{Fit of the Effective Stokes-Vector Offset to the Experimental Stokes Parameters.}
    Experimental $S_1$, $S_2$, and derived polarization-angle maps evaluated on the mode surfaces obtained from the dispersion fit in Table~\ref{tab:rse-fit-params}, together with the corresponding model results. The dispersion parameters are kept fixed, the global rotation is set to $\theta_0=0^\circ$, and only the constant offset $\delta\vec{h}$ is fitted.}
    \label{fig:stokes-fits}
\end{figure}

Having established this model, we extract the experimental Stokes parameters by interpolating the measured hyperspectral data on the mode surfaces obtained from the dispersion fit in Table~\ref{tab:rse-fit-params}. In the subsequent Stokes fit, the dispersion parameters are kept fixed and the global rotation is set to $\theta_0=0^\circ$, such that all remaining deviations are absorbed into $\delta\vec{h}$. Fitting the model to the experimental $S_1$ and $S_2$ maps yields $\delta h_1 = \SI{-1.7354 \pm 0.1245e-4}{}$ and $\delta h_2 = \SI{-3.2271 \pm 0.0157e-3}{}$.
The resulting agreement between experiment and model is shown in Fig.~\ref{fig:stokes-fits}.

\clearpage
\color{black}
\subsection{Evaluation of Topological Charges}

We extract the topological charge $q_0$ of each polarization vortex as a contour integral around the singularity in k-space \cite{zhen2014topological}:
\begin{equation}
    q_0 = \frac{1}{2\pi} \oint_{\tilde{C}} d\mathbf{k} \cdot \nabla_\mathbf{k} \phi(\mathbf{k}) = \frac{1}{4\pi} \int_0^{2\pi} \frac{d}{d\theta}\Big[2\phi(\theta)\Big] \, d\theta
      = \frac{\phi(2\pi) - \phi(0)}{2\pi}.
\end{equation}
We evaluate this numerically on the interpolated experimental mixing angle, unwrapped counterclockwise along the contour. Since a linear polarization angle repeats every $\pi$ (not $2\pi$), we unwrap $2\phi$ rather than $\phi$. The factor of $2$ cancels against the prefactor, so $q_0$ reduces simply to the unwrapped phase difference between the end and start of the contour, divided by $2\pi$.

\begin{figure}[H]
    \centering
    \includegraphics[width=\linewidth]{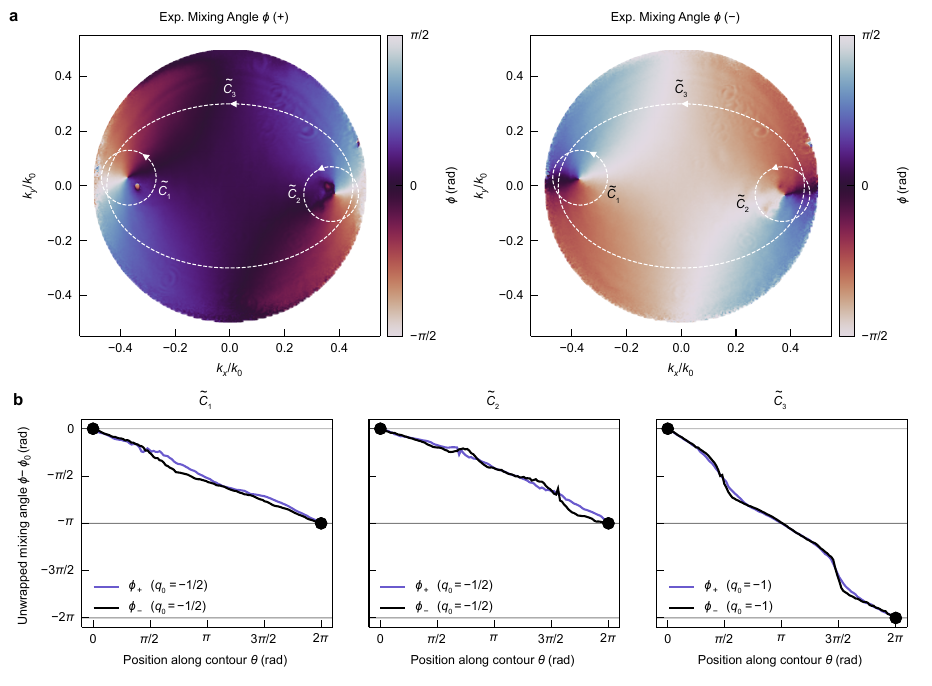}
    \caption{\color{black}\textbf{Extraction of Topological Charges from Experimental Polarization Data.}
    \textbf{a} Experimental mixing angle $\phi$ maps for both upper and lower mode surfaces and integration curves defined as closed loops around the individual singularities ($\tilde{C_1}$ and $\tilde{C_2}$) and around both singularities ($\tilde{C_3}$).
    \textbf{b} Unwrapped mixing angle along the integration curves. For the individual singularities ($\tilde{C_1}$, $\tilde{C_2}$) the phase decreases monotonically by $\pi$ over
one complete loop for both mode surfaces, corresponding to a topological charge of $-1/2$; for
the joint contour ($\tilde{C_3}$) the phase decreases by $2\pi$, giving a charge of $-1$, consistent
with conservation of total topological charge.
    }
    \label{fig:winding-number-calculation}
\end{figure}

\clearpage

\section{TCMT Fitting of Strong Coupling under Normal Incidence} \label{section:TCMT fit}
We utilize temporal coupled-mode theory (TCMT) \cite{, haus1984waves-96e} to model the strong coupling interactions between two independent photonic qBICs and two excitonic resonances supported by a ReS$_2$ metasurface. Each resonance set (Mode 1–Exciton 1 and Mode 2–Exciton 2) is independently coupled and described separately by the TCMT framework.

The metasurface resonant scattering process is described via the scattering matrix formalism:
\begin{equation}
\mathcal{S}(\omega) = C + K \left[i(\omega I - \Omega) + \Gamma\right]^{-1} K^\mathrm{T},
\end{equation}
where $C$ represents non-resonant background, given by:
\begin{equation}
C = e^{i\phi}\begin{pmatrix}
r_0 & i t_0 \\
i t_0 & r_0
\end{pmatrix},
\end{equation}
with global phase $\phi$, reflection $r_0$, and transmission $t_0$, and $|r_0|^2 + |t_0|^2 = 1$.
Each independent qBIC–exciton pair ($j=1,2$) characterized by their resonance frequencies: 
\begin{equation}
\Omega_j =
\begin{pmatrix}
\omega_{\mathrm{qBIC},j}+ i\gamma_{\mathrm{qBIC,int},j} & g_i & 0 \\
g_i & \omega_{\mathrm{EX},j}+ i\gamma_{\mathrm{EX,int},j} & 0 \\
0 & 0 & \omega_{\mathrm{EX},j}+ i\gamma_{\mathrm{EX,int},j}
\end{pmatrix},\quad 
\end{equation}
together with radiative and intrinsic damping rates:
\begin{equation}
\Gamma_j =
\begin{pmatrix}
\gamma_{\mathrm{qBIC,rad},j} & 0 & \sqrt{\gamma_{\mathrm{qBIC,rad},j}\gamma_{\mathrm{EX,rad},j}} \\
0 & 0 & 0 \\
\sqrt{\gamma_{\mathrm{qBIC,rad},j}\gamma_{\mathrm{EX,rad},j}} & 0 & \gamma_{\mathrm{EX,rad}j}
\end{pmatrix}.
\end{equation}

The port-mode coupling matrix $K$ for each resonance set is:
\begin{equation}
K_j =
\begin{pmatrix}
\sqrt{\gamma_{\mathrm{qBIC,rad},j}} & 0 & \sqrt{\gamma_{\mathrm{EX,rad},j}} \\
\sqrt{\gamma_{\mathrm{qBIC,rad},j}} & 0 & \sqrt{\gamma_{\mathrm{EX,rad},j}}
\end{pmatrix}.
\end{equation}

\begin{figure}[ht]
    \centering
    \includegraphics[width=0.5\linewidth]{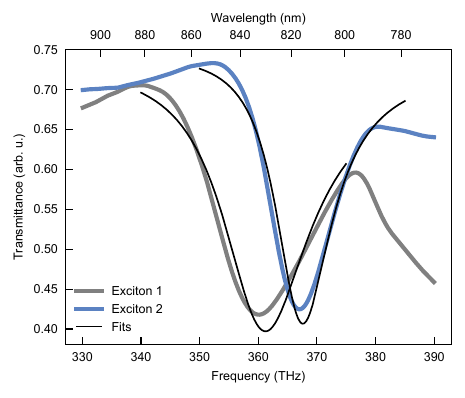}
    \caption{\textbf{Experimental Transmittance and TCMT Fits of Unstructured ReS$_2$ Flakes.}}
    \label{fig:exciton-fits}
\end{figure}

Experimental transmittance spectra $T(\omega)=|\mathcal{S}_{21}(\omega)|^2$ are simultaneously fit across multiple scaling factors using shared TCMT parameters\cite{nan2025angular, heimig2026chiral}. The exciton resonance frequencies and linewidths, extracted from independent single-resonance TCMT fits of unstructured ReS$_2$ flakes (Fig. \ref{fig:exciton-fits}), are fixed at:
\begin{itemize}
\item Exciton 1: $\omega_{\mathrm{EX},1}/(2\pi)=\SI{360.33}{\tera\hertz}$, $\gamma_{\mathrm{EX},1}/(2\pi)=\SI{8.97}{\tera\hertz}$,
\item Exciton 2: $\omega_{\mathrm{EX},2}/(2\pi)=\SI{366.95}{\tera\hertz}$, $\gamma_{\mathrm{EX},2}/(2\pi)=\SI{5.77}{\tera\hertz}$.
\end{itemize}

\begin{figure}[H]
    \centering
    \includegraphics[width=\linewidth]{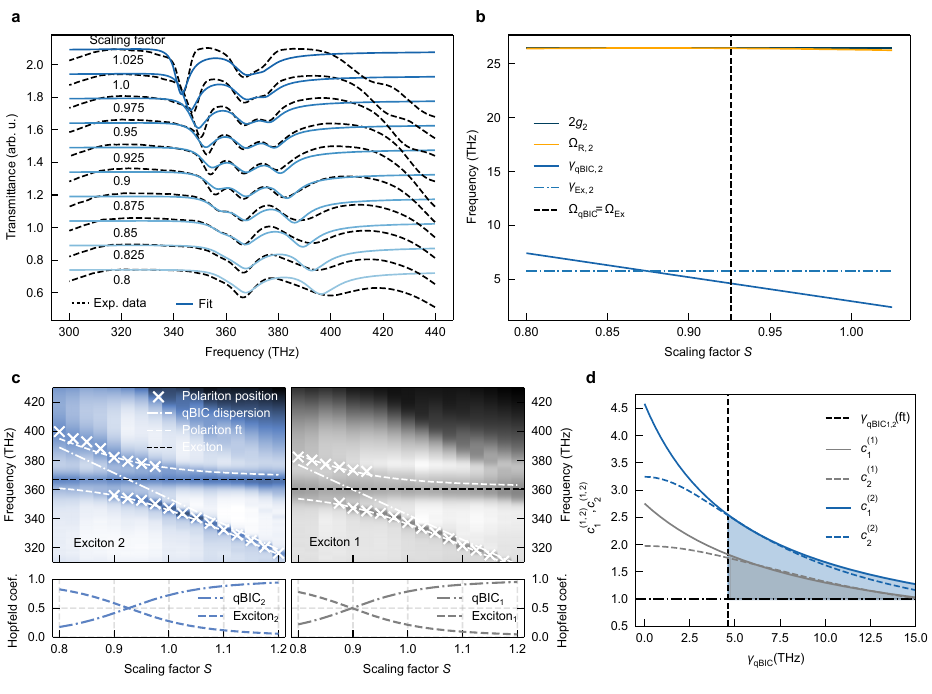}
    \caption{\textbf{TCMT Fit for Metasurfaces with Different Scaling Factors.} \textbf{a} Experimental transmittance spectra for different scaling factors in Mode 2/Exciton 2 configuration and TCMT fits. \textbf{b} Extracted fit parameters from \textbf{a}. The trend of the qBIC linewidth has been approximated with a linear function. The linewidth at the overlap position between qBIC and exciton is \SI{4.6}{\tera\hertz}. \textbf{c} Overlay of polariton positions extracted from experimental data and dispersion calculated with the parameters from \textbf{b}. The same qBIC linewidth has been assumed for both Exciton 1 and Exciton 2, showing good agreement. The lower panels show the Hopfield coefficients (Sec.~\ref{sec:hopfield}) representing the qBIC and excitonic fractions of the lower polariton branches. For the upper polariton branches, these fractions are reversed. \textbf{d} Strong coupling conditions with respect to the qBIC linewidths. The linewidth extracted at the spectral overlap with Exciton 2 serves as a conservative lower limit, confirming that our ReS$_2$ metasurface resides deeply in the strong-coupling regime with both excitons.}
    \label{fig:tcmt-fits}
\end{figure}

The uncoupled qBIC frequency and linewidth dependencies are approximated by linear functions of the metasurface scaling factor, with slope and offset parameters shared among all spectra during fitting (Fig.~\ref{fig:exciton-fits}a,b). We evaluate the linewidth of photonic Mode 2 at the overlap position with Exciton 2 to be \SI{4.61}{\tera\hertz}. The linewidth of the Mode 1 is assumed to be the same. Furthermore, the coupling strengths $g_i$ at optimal resonance overlap between the qBIC and the corresponding excitonic transitions are extracted to be \SI{12.5}{\tera\hertz} (Exciton 1) and \SI{13.2}{\tera\hertz} (Exciton 2). Subsequently, the Rabi splittings are calculated via:
\begin{equation}
\Omega_{\mathrm{R},j}=2\sqrt{g_j^2-\frac{(\gamma_{\mathrm{qBIC},j}-\gamma_{\mathrm{EX},j})^2}{4}},
\end{equation}
yielding $\hbar\Omega_{\mathrm{R},1} = \SI{102}{\milli\electronvolt}$ and $\hbar\Omega_{\mathrm{R},2} = \SI{109}{\milli\electronvolt}$, which we further use to estimate photonic and excitonic fractions within the Hopfield-coefficient formalism (Fig.~\ref{fig:tcmt-fits}c and Sec.~\ref{sec:hopfield}).

The $j$-th polariton dispersions ($j=1,2$) are given by:
\begin{equation}
\omega_{j\pm} = \frac{\omega_{\mathrm{qBIC},j} + \omega_{\mathrm{EX},j}}{2} + i\frac{\gamma_{\mathrm{qBIC},j} + \gamma_{\mathrm{EX},j}}{2} \pm \sqrt{g_j^2 - \frac{1}{4} \left( \gamma_{\mathrm{qBIC},j} - \gamma_{\mathrm{EX},j} + i(\omega_{\mathrm{qBIC},j} - \omega_{\mathrm{EX},j}) \right)^2}.
\label{eq:PolaritonBranches}
\end{equation}

Using the extracted parameters from Fig. \ref{fig:tcmt-fits}b we overlay the polariton dispersion with the measured polariton positions, which shows excellent agreement with the anti-crossing pattern of our experimental data (Fig.~\ref{fig:tcmt-fits}c). 

To further corroborate that the system operates in the strong-coupling regime, we evaluate the following established criteria for each $j$-th polariton ($j=1,2$) \cite{cao2020normal}:
\begin{equation}
c_1^{(j)} = \frac{\Omega_{R,j}}{\gamma_{\text{qBIC}_{j}} + \gamma_{\text{EX}_{j}}} > 1,
\end{equation}
which requires that the Rabi splitting exceed the sum of the linewidths.
The second, more stringent condition:
\begin{equation}
c_2^{(j)} = \frac{g_{j}}{\sqrt{\left(\gamma_{\text{EX}_{j}}^2 + \gamma_{\text{qBIC}_{j}}^2\right)/2}} > 1,
\end{equation}
compares the coupling strength to the root-mean-square of the individual losses.

Using the qBIC linewidth of \SI{4.61}{\tera\hertz} as a conservative lower limit, we evaluate these criteria to be $c_1^{(1)}=1.81$, $c_2^{(1)}= 1.75$ for the first polariton and $c_1^{(2)}=2.55$, $c_2^{(2)}= 2.53$ for the second one. Thus both the photonic Modes 1 and 2 couple strongly with their respective excitons (Fig. \ref{fig:tcmt-fits}d).

\clearpage
\color{black}
\section{Quantification of Curvature for Photonic and Polaritonic Bands}
\label{sec:flatness-analysis}
\subsection{Coupled qBIC-Grating Dispersion Fit}
To quantify the flatness of the experimental bands, we start from the simple photonic dispersions introduced in Eqs.~(\ref{eq:ToyM_initial_omega1}) and (\ref{eq:ToyM_initial_omega2}) within the RSE framework, i.e., parabolic and flatband dispersions. These models don't account for the hybridization of the qBICs with the first-order grating mode associated with the glass substrate, which leads to an abrupt change in the experimental dispersion slope at the edges of the measured $k$-range. 
To capture this hybridization, we use a simple Fano-Feshbach-type \cite{miroshnichenko2010fano} Hermitian Hamiltonian:
\begin{equation} \label{eq:H_Fano-Feshbach_simple}
    H_{FF} = \begin{pmatrix}
        \omega_\mathrm{M}(k) & q \\
        q & \omega_\mathrm{G}(k)
    \end{pmatrix}\;,
\end{equation}
where $\omega_\mathrm{M}(k)$ denotes the uncoupled qBIC dispersion, $\omega_\mathrm{G}(k)$ is the grating-mode dispersion, and $q$ is an empirical coupling strength. The corresponding lower-energy eigenvalue is given by:
\begin{equation}\label{eq:qBIC-grating_fit}
    \omega(k) = \frac{\omega_\mathrm{M}(k) + \omega_\mathrm{G}(k)}{2} - \sqrt{q^2 + \left( \frac{\omega_\mathrm{M}(k) - \omega_\mathrm{G}(k)}{2}\right)^2}\; ,
\end{equation}
which is used throughout the following analysis.

The qBIC dispersion is parameterized as $\omega_\mathrm{M}(k) = \omega_0\left(1 - \chi k^2/(k_0^{(0)})^2\right)$ with constant \mbox{$k^{(0)}_0=\omega_0/c$} (the superscript (0) is introduced to avoid confusion with the free-space wavenumber $k_0=\omega/c$) and $k=\sqrt{k_x^2+k_y^2}$. For flatband dispersion we set $\chi=0$, whereas for the parabolic band $\chi>0$ is treated as a fit parameter. The first-order grating mode is given by:
\begin{equation}\label{eq:omega_grating}
    \omega_\mathrm{G}(k) = \frac{c}{n_\mathrm{sub}}\left|G-k\right| \; ,
\end{equation}
with $G=2\pi/\Lambda$. $\Lambda = S\cdot \SI{465}{\nano\metre}$ is the metasurface periodicity, $n_\mathrm{sub}=1.45$ is the substrate refractive index and $S$ is the metasurface in-plane scaling factor.

\begin{figure}[ht]
    \centering
    \includegraphics[]{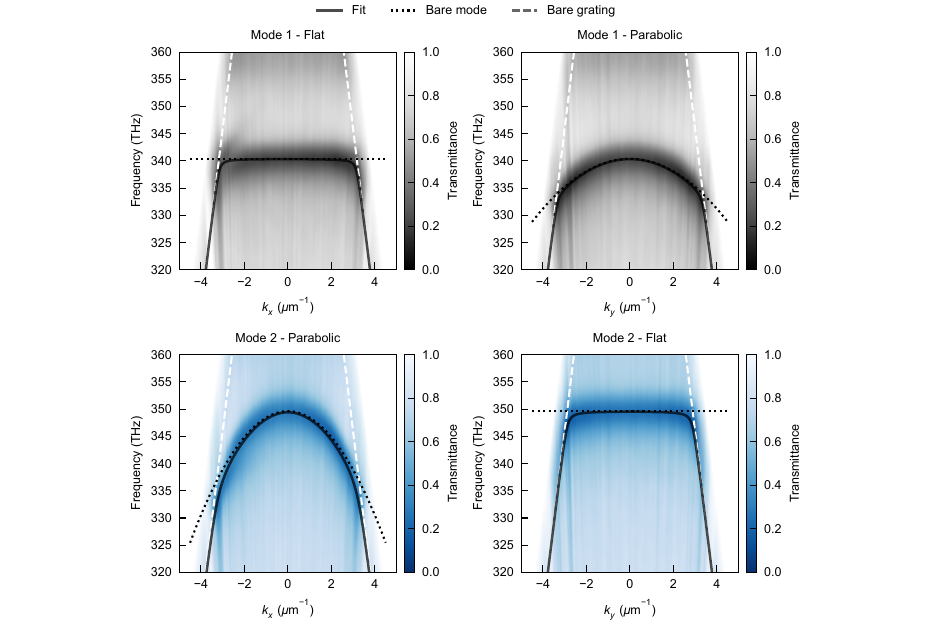}
    \caption{\textbf{Fano-Feshbach Fits to Experimental k-space-resolved Transmittance Spectra.} Backgrounds show transmittance for a metasurface with scaling factor $S=1.0$ from Figure~3 of the main text. Curves indicate the fitted grating-coupled mode (solid black), bare qBIC (black dotted), and grating branch (white dashed).}
    \label{fig:SI_fano_feshbach_fits}
\end{figure}

We fit this coupling model to dispersions extracted along the two principal measurement planes ($k_x=0$ and $k_y=0$). For cuts belonging to the same mode, we constrain the resonance frequency by sharing $\omega_0$ between the corresponding fits (Fig.~\ref{fig:SI_fano_feshbach_fits}). This procedure yields four independent fit parameters per crystal orientation, summarized in Tab.~\ref{tab:feshbach-fit-params-S1}.

\begin{table}[ht]
\centering
\begin{tabular}{lcc}
\toprule
Parameter & Mode 1 & Mode 2 \\
\midrule
$\omega_0/(2\pi)$ (THz) & $340.400 \pm 0.022$ & $349.634 \pm 0.036$ \\
$\chi$ & $0.086 \pm 0.002$ & $0.183 \pm 0.003$ \\
$q_{\mathrm{par}}$ (THz) & $2.267 \pm 0.134$ & $4.386 \pm 0.207$ \\
$q_{\mathrm{flat}}$ (THz) & $2.429 \pm 0.083$ & $2.993 \pm 0.123$ \\
\bottomrule
\end{tabular}
\caption{Fit parameters of the Fano--Feshbach model for the metasurface with scaling factor $S=1.0$. The parabolic branch is described by the curvature parameter $\chi$ and coupling strength $q_{\mathrm{par}}$, while the flat branch is characterized by $q_{\mathrm{flat}}$. Uncertainties denote one standard error returned by the fit.}
\label{tab:feshbach-fit-params-S1}
\end{table}

The curvature parameters $\chi$ align well with the RSE fit results (Tab. \ref{tab:rse-fit-params}). The slightly lower parabolic curvatures are mitigated by the additional curvature through grating coupling.

\subsection{Curvature Analysis}

From the fitted dispersions given by Eq.~\eqref{eq:qBIC-grating_fit} we obtain the group velocity and curvature by differentiation. The group velocity reads as:
\begin{equation}
    v_\mathrm{g} = \frac{d\omega}{dk} = \frac{\omega_\mathrm{M}^\prime + \omega_\mathrm{G}^\prime}{2} - \frac{\Delta\Delta^\prime}{4A}
\end{equation}
and the curvature is:
\begin{equation}
    \kappa = \frac{d^2\omega}{dk^2} = \frac{\omega_\mathrm{M}^{\prime\prime} + \omega_\mathrm{G}^{\prime\prime}}{2} - \left[ \frac{(\Delta^\prime)^2 + \Delta\Delta^{\prime\prime}}{4A} - \frac{\Delta^2 (\Delta^\prime)^2}{16A^3} \right],
\end{equation}
with:
\begin{align}
    \Delta(k) &= \omega_\mathrm{M}(k) - \omega_\mathrm{G}(k), \ \ \ \ \text{and} \ \ \ \
    A(k) = \sqrt{q^2 + \left[\Delta(k)/2 \right]^2},
\end{align}
where a single (double) prime denotes the first (second) derivative with respect to $k$.

Using the explicit derivatives with the assumption $k<G$ in \eqref{eq:omega_grating}:
\begin{align}
    \omega_\mathrm{M}^\prime &= -2\omega_0\chi k/(k_0^{(0)})^2,\; \ \ \omega_\mathrm{M}^{\prime\prime} = -2\omega_0 \chi/(k_0^{(0)})^2,\; \ \
    \omega_\mathrm{G}^\prime = -c/n_\mathrm{sub},\;  \ \ \omega_\mathrm{G}^{\prime\prime} = 0,
\end{align}
we obtain the expressions for the group velocity $v_g$ and curvature $\kappa$:
\begin{align}
    v_\mathrm{g} &= -\frac{\omega_0\chi k}{(k_0^{(0)})^2} -\frac{c}{2n_\mathrm{sub}}
-\frac{\Delta\,\bigl[c/n_\mathrm{sub}-2\omega_0\chi k/(k_0^{(0)})^2\bigr]}{4A} \; ,\\
    \kappa &= -\frac{\omega_0\chi}{(k_0^{(0)})^2} -\frac{\bigl[c/n_\mathrm{sub}-2\omega_0\chi k/(k_0^{(0)})^2\bigr]^2-2\omega_0\chi\Delta/(k_0^{(0)})^2}{4A} +\frac{\Delta^2\,\bigl[c/n_\mathrm{sub}-2\omega_0\chi k/(k_0^{(0)})^2\bigr]^2}{16A^3} \; ,
\end{align}
which depend only on the fitted parameters and the experimentally sampled $k$-values.

\begin{figure}[ht]
    \centering
    \includegraphics[width=\textwidth]{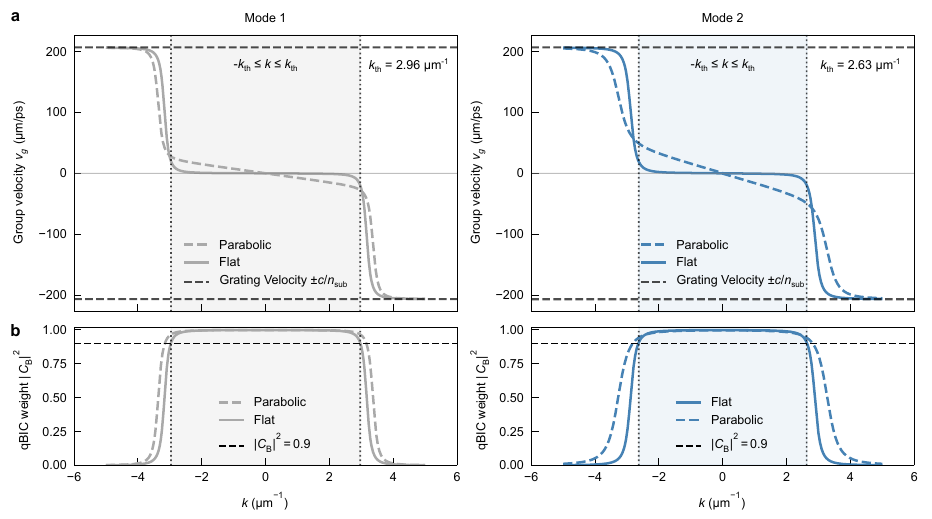}
    \caption{\textbf{Group-velocity and Flatband Extension Analysis.}
    \textbf{a} Group velocity for parabolic (dashed) and flat (solid) bands; black dashed lines indicate the grating mode velocity ($\pm c/n_{\mathrm{sub}}$).
    \textbf{b} qBIC weight $|C_{\mathrm{B}}|^2$ with the threshold given by $|C_{\mathrm{B}}|^2=0.9$ (black dashed line) corresponding to 90\% qBIC and 10\% grating mode contributions. Shaded regions and dotted vertical lines mark the $k$-ranges where the portion of flatband satisfies $|C_{\mathrm{B}}|^2 \ge 0.9$, and the inset reports the corresponding threshold $k$-values.
    }
    \label{fig:SI_group_velocity}
\end{figure}

By symmetry, $v_\mathrm{g}$ vanishes at $k=0$ for both flat and parabolic bands. Away from the $\Gamma$-point, the parabolic case exhibits the approximately linear increase expected in $|v_\mathrm{g}|$ for small $k$, while at larger in-plane momenta a pronounced increase towards the limiting slope set by the grating mode is observed. In the coupled-mode picture this crossover reflects the increasing admixture of the grating mode into the hybrid eigenstate, and $|v_\mathrm{g}|$ approaches the effective grating mode velocity $c/n_\mathrm{sub}$ in the grating-dominated regime (Fig. \ref{fig:SI_group_velocity}a).

To quantify the momentum range over which the mode remains predominantly qBIC-like, with negligible admixture of the grating mode, we evaluate the qBIC weight by analogy with the Hopfield-coefficient formalism (see Section~\ref{sec:hopfield}), as:
\begin{equation}
    \left| C_\mathrm{B}(k) \right|^2 = \frac{1}{2}\left(1-\frac{\Delta}{\sqrt{\Delta^2+4q^2}} \right).
\end{equation}
We define the qBIC-dominated region by the criterion $\left| C_\mathrm{B}(k) \right|^2 > 0.9$ (Fig.~\ref{fig:SI_group_velocity}b), yielding the corresponding thresholds
    $k_\mathrm{th,1} = \pm\SI{2.97}{\per\micro\metre}$ for Mode 1 and $k_\mathrm{th,2} = \pm\SI{2.64}{\per\micro\metre}$ for Mode 2.

Diffraction sets an upper bound on the accessible qBIC-like momentum range, determined by the grating mode. At the flatband frequency $\omega_0$, this limit is given by:
\begin{align}
    k_\mathrm{G} = G - \frac{\omega_0}{c} n_\mathrm{sub} \; .
\end{align}
We therefore define a flatband extent ratio as:
\begin{equation}
    \eta = \frac{k_\mathrm{th}}{k_\mathrm{G}},
\end{equation}
which yields k-space coverages of 
    $\eta_1 = \SI{93.7}{\percent}$ and $\eta_2 = \SI{91.5}{\percent}$.

\begin{figure}[H]
    \centering
    \includegraphics[width=1\linewidth]{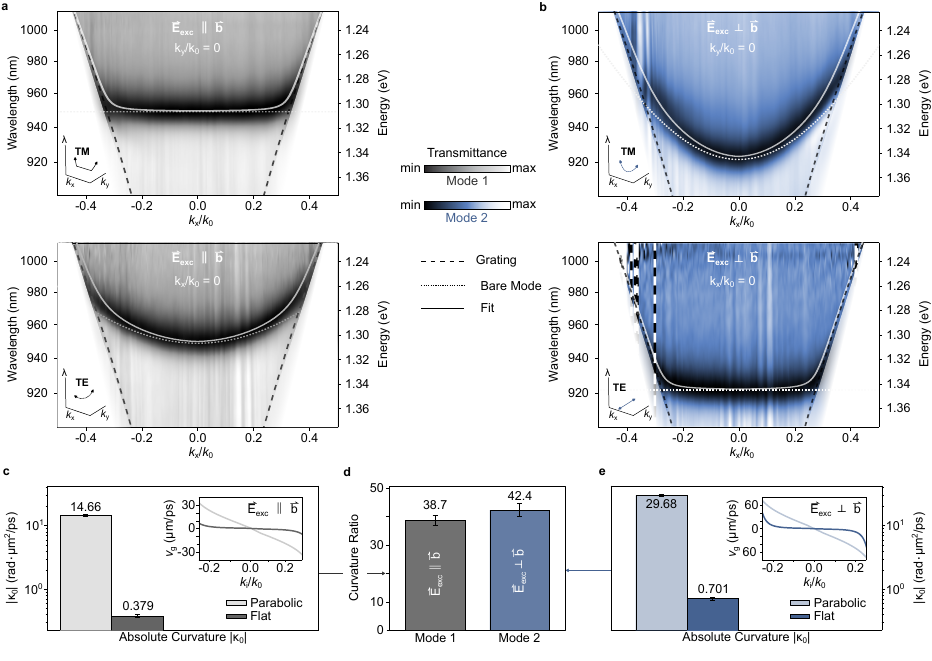}
    \caption{\color{black}\textbf{Experimental Photonic Flatbands of Far-detuned qBIC}.  
    Given that the hybrid modes remain predominantly photonic in character (photonic fraction exceeds 90\%, while excitonic fraction is less than 10\%, see Fig.~\ref{fig:HopfieldsMilano}), the observed dispersions can be identified as photonic flatbands. Compared to the main text with $S=1$, the increased scaling factor $S=1.15$ enlarges the lateral dimensions of the metasurface (including periodicity $\Lambda)$, shifting the grating cut-off spectrally closer to the qBICs. This enhances grating-induced curvature, reducing the momentum range of near-zero group velocity and increasing the residual curvature at the $\Gamma$-point. However, this effect and can be mitigated by increasing the slab thickness, which moves the qBIC farther from the diffraction threshold.
    Experimental angle-resolved Fourier-space transmittance for excitation polarization \textbf{a}
    $\vec{E}_\mathrm{exc}\parallel \vec{b}$ (Mode 1, grey) and \textbf{b}
    $\vec{E}_\mathrm{exc}\perp \vec{b}$ (Mode 2, blue), where $\vec{b}$ denotes the crystal axis. 
    For each excitation configuration, TM (upper panels, $k_x/k_0$) and TE (lower panels, $k_y/k_0$) dispersions are shown.  The metasurface parameters are $\Lambda=465\cdot S$~nm, $w_0=220\cdot S$~nm, $S=1.15$, $h=90$~nm, and asymmetry $\alpha=0.6$. The dashed lines indicate the grating mode introduced by the substrate. 
    The dotted lines represent the photonic mode in the absence of coupling to the grating mode. The solid lines correspond to a coupled-mode fit including hybridization between the intrinsic qBIC and the grating mode. 
    Depending on the propagation direction in $k$-space, the intrinsic qBIC exhibits either parabolic or strongly reduced curvature. 
    Absolute curvature $|\kappa_0|$ at the $\Gamma$-point for the parabolic and flat dispersions under \textbf{c} $\vec{E}_\mathrm{exc}\parallel\vec{b}$ (fitted values for parabolic $14.66 \pm 0.40$ and flat: $0.379 \pm 0.019$) and  \textbf{e}$~\vec{E}_\mathrm{exc}\perp\vec{b}$ excitation (fitted values for parabolic $29.68 \pm 0.92$ and flat: $0.701 \pm 0.037$), with the insets showing the corresponding calculated group velocity $v_g=\partial\omega/\partial k$. 
    \textbf{d}~Curvature ratio (Mode 1: $38.7 \pm 1.9$ and Mode 2: $42.4 \pm2.2$), directly quantifying the dispersion anisotropy. 
    Absolute curvature $|\kappa_0|$ and curvature ratio are presented as the mean fit values, $\pm$ the standard deviation of the fit.}
    \label{fig:FlatbandPhotExp}
\end{figure}

A flatness analysis based on averages of group velocity $v_\mathrm{g}$ or curvature $\kappa$ over a finite momentum interval is sensitive to the (necessarily somewhat arbitrary) choice of the $k$-space interval bounds over which the averaging is performed. We therefore use a local, boundary-independent metric and evaluate the curvature $\kappa_0$ at the $\Gamma$-point ($k=0$):
\begin{align}
    \kappa_0 = -\omega_0 \frac{\chi}{\left(k_0^{(0)}\right)^2} \left( 1 - \frac{\Delta_0}{\sqrt{\Delta_0^2 + 4q^2}} \right) - \frac{2q^2c^2}{n_\mathrm{sub}^2 \left( \Delta_0^2 + 4q^2 \right)^{3/2}}, \; \ \mathrm{with} \ \; \Delta_0 = \omega_0 - \frac{c}{n_\mathrm{sub}} G \; ,
\end{align}
which, for the flat bands ($\chi=0$), depends only on the coupling strength to the grating mode $q$. This highlights an important point: even for an ideally flat bare mode, hybridization with a dispersive grating channel introduces a finite curvature of the observed lower branch. 
Comparing the $\Gamma$-point curvatures $\kappa_0$, we can define the curvature ratio as $\kappa_{0,\mathrm{par}}/\kappa_{0,\mathrm{flat}}$,
that quantifies the directional curvature anisotropy between parabolic and flat branches. For a metasurface with $S=1.0$ we reach curvature ratio more than $100$ for both photonic modes (see Main Text Fig. 3).
An analogous study for a far-detuned qBIC metasurface with scaling factor $S=1.15$ is summarized in Fig.~\ref{fig:FlatbandPhotExp} with the corresponding fit parameters listed in Tab.~\ref{tab:feshbach-fit-params-S1p15}. Curvature analysis of the polaritonic metasurface with $S=0.95$ from Figure~5 of the main text, using a new fit that accounts for the excitonic contribution (see next Sec.~\ref{sec:S6_strong_coupling}), can be found in Fig.~\ref{fig:FlatbandExp} and Tab.~\ref{tab:feshbach-fit-params-S0p95}.

\begin{table}[H]
\centering
\begin{tabular}{lcc}
\toprule
Parameter & Mode 1 & Mode 2 \\
\midrule
$\omega_0/(2\pi)$ (THz) & $315.915 \pm 0.019$ & $325.205 \pm 0.037$ \\
$\chi$ & $0.1514 \pm 0.004$ & $0.3172 \pm 0.010$ \\
$q_{\mathrm{par}}$ (THz) & $4.998 \pm 0.168$ & $6.064 \pm 0.258$ \\
$q_{\mathrm{flat}}$ (THz) & $3.157 \pm 0.078$ & $3.488 \pm 0.095$ \\
\bottomrule
\end{tabular}
\caption{Fit parameters of the Fano--Feshbach model for far-detuned qBIC metasurface with $S=1.15$. The parabolic branch is described by the curvature parameter $\chi$ and coupling strength $q_{\mathrm{par}}$, while the flat branch is characterized by the coupling strength $q_{\mathrm{flat}}$. Uncertainties denote one standard error returned by the fit.}
\label{tab:feshbach-fit-params-S1p15}
\end{table}

\subsection{Curvature Analysis for Strongly Coupled Samples} \label{sec:S6_strong_coupling}

We now incorporate excitonic contribution into the Fano-Feshbach Hamiltonian ~\eqref{eq:H_Fano-Feshbach_simple} through the real part of the excitonic resonance frequencies $\omega_{\mathrm{EX}}$ and near-field coupling $g$. In comparison to the normal-incidence TCMT fits (Sec.~\ref{section:TCMT fit}), we add a $k$-dependence to the qBIC-exciton coupling strength $g(k)$ as:
\begin{equation}\label{eq:g(k)_fit}
    g(k)=g_0(1+\zeta k^2/k_0^2).
\end{equation}
The new Hamiltonian then reads:
\begin{equation} \label{eq:4x4_H}
    H_{\mathrm{full}}=\begin{pmatrix}
        \omega_\mathrm{M}(k) & q & g(k) & 0 \\
        q & \omega_\mathrm{G}(k) & 0 & 0 \\
        g(k) & 0 & \omega_{\mathrm{EX}} & 0 \\
        0 & 0 & 0 & \omega_{\mathrm{EX}}
    \end{pmatrix}.
\end{equation}
To account for so-called uncoupled excitons, $\omega_{\mathrm{EX}}$ appears twice in the Hamiltonian, allowing the bare excitonic resonance frequency to be fitted to the experimental data.

Let $\omega_n$ be the eigenvalues of $H_{\mathrm{full}}$. To extract the group velocities $v_{g,n}$ and curvatures $\kappa_n$ of the new eigenbranches we use the Hellmann-Feynman theorem \cite{hirao1992analytic} to calculate the derivatives with respect to $k$:
\begin{align}
    v_{\mathrm{g},n} = \frac{\partial\omega_n}{\partial k}=\bra{n}\frac{\partial H_{\mathrm{full}}}{\partial k}\ket{n}, \ \ \text{and}\ \ 
    \kappa_n = \frac{\partial^2\omega_n}{\partial k^2} = \bra{n}\frac{\partial^2 H_{\mathrm{full}}}{\partial k^2}\ket{n} +
    2\sum_{m\neq n}\frac{\left| \bra{m}\frac{\partial H_{\mathrm{full}}}{\partial k} \ket{n} \right|^2}{\omega_n - \omega_m}.\label{eq:hellmann-feynman}
\end{align}
We numerically diagonalize $H_{\mathrm{full}}$ to obtain the eigenvalues and eigenvectors, and fit the eigenvalues to the experimental data. We then use Eqs.~\ref{eq:hellmann-feynman} to analytically compute the curvature metric, directly constrained by the fit parameters.

For the flat polariton branches, we achieve the best agreement for $g(k)=g_0=$const., corroborating our conclusion regarding the formation of polaritonic flatbands. As for the parabolic polaritons, we fit both a constant $g$ ($\zeta=0$) and $g$ given by the parabolic dispersion in Eq.~\eqref{eq:g(k)_fit}. The fit results of the experimental data and the comparison with full-wave simulations are presented in Fig.~\ref{fig:SI-feshbach-gk-model}. To verify that the additional curvature of $g$ is not a feature of a material dispersion gradient close to the bandgap, we simulate with a simplified material model using constant background and single Lorentzian exciton peaks from Eq.~\eqref{eq:epsLorentz}, in the presence of a glass substrate with $n_\mathrm{sub}=1.45$

The complete fits and curvature analysis of the strongly coupled sample can be found in Main~Text~Fig.~5 and Fig.~\ref{fig:FlatbandExp}, with the respective fit parameters listed in Tab.~\ref{tab:feshbach-fit-params-S0p95}.

\begin{table}[H]
\centering
\begin{tabular}{lcc}
\toprule
Parameter & Mode 1 & Mode 2 \\
\midrule
$\omega_0/(2\pi)$ (THz) & $358.291 \pm 0.055$ & $367.921 \pm 0.033$ \\
$\omega_{\mathrm{EX}}/(2\pi)$ (THz) & $361.051 \pm 0.022$ & $367.509 \pm 0.015$ \\
$q$ (THz) & $3.884 \pm 0.207$ & $3.384 \pm 0.078$ \\
$g_0$ (THz) & $12.243 \pm 0.021$ & $12.431 \pm 0.013$ \\
$\zeta$ & $-0.515 \pm 0.077$ & $-2.208 \pm 0.051$ \\
$\chi$ & $0.119 \pm 0.004$ & $0.296 \pm 0.002$ \\
\bottomrule
\end{tabular}
\caption{Fit parameters of the Hamiltonian model from Eq.~\eqref{eq:4x4_H} for the strongly coupled sample with the scaling factor $S=0.95$. The fit was performed jointly for the flat and parabolic $k$-space cuts with shared couplings $q$ and $g_0$. The $\Gamma$-point coupling strength $g_0$ translates to a (vacuum) Rabi splitting $\hbar\Omega_{\mathrm{R},0}=2g_0\approx\SI{102}{\milli\electronvolt}$, which agrees well with the values extracted from normal-incidence TCMT fits (Sec.~\ref{section:TCMT fit}). Uncertainties denote one standard error returned by the fit .}
\label{tab:feshbach-fit-params-S0p95}
\end{table}

\begin{figure}[H]
    \centering
    \includegraphics[width=\textwidth]{SI_Figures/SI_Fig_curvature_parabolic.pdf}
    \caption{\textbf{Fitting Comparison of Parabolic Polaritons with Constant and k-dependent Coupling Strengths.}
    Coupled-mode fits for (\textbf{a}) experimental (S=0.95) and (\textbf{b}) simulated (S=0.99) transmission spectra with the constant $g_0$ (dashed black lines) and $k$-dependent $g(k)=g_0(1+\zeta k^2/k_0^2)$ (yellow solid lines) coupling strengths. A constant coupling strength model underestimates the curvature of the upper polariton branch, whereas as quadratic dependence describes its accurately, while retaining the same curvature for the lower polariton branch. Simulations were performed using a material model with constant background dispersion and a single Lorentzian exciton peak, as given by Eq.~\eqref{eq:epsLorentz}, in the presence of a glass substrate.
    }
    \label{fig:SI-feshbach-gk-model}
\end{figure}

\begin{figure}[H]
    \centering
    \includegraphics[width=1\linewidth]{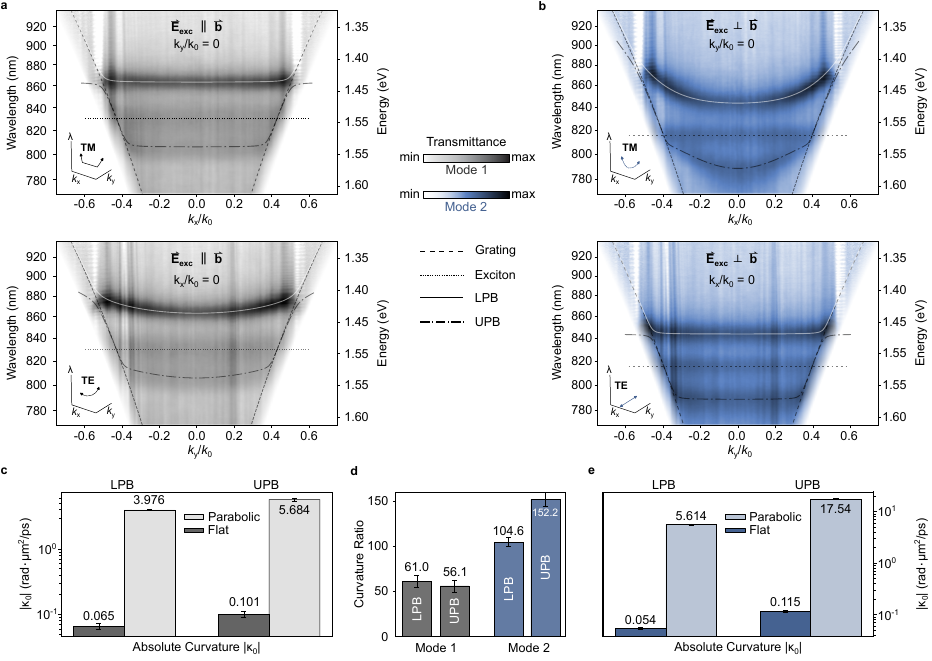}
    \caption{\color{black}\textbf{Experimental Polaritonic Flatbands.}  
    Given the near-ideal hybridization of photonic and excitonic contributions (photonic and excitonic contributions approaching equal weights, see Fig.~\ref{fig:HopfieldsMilano}), the observed dispersions can be identified as polaritonic.
    Experimental angle-resolved Fourier-space transmittance for \textbf{a}
    $\vec{E}_\mathrm{exc}\parallel \vec{b}$ (Mode 1, grey) and \textbf{b}
    $\vec{E}_\mathrm{exc}\perp \vec{b}$ (Mode 2, blue) excitation polarization, where $\vec{b}$ denotes the crystal axis. 
    For each excitation configuration, TM (upper panels, $k_x/k_0$) and TE (lower panels, $k_y/k_0$) dispersions are shown.  The metasurface parameters are $\Lambda=465\cdot S$ nm, $w_0=220\cdot S$ nm, $S=0.95$, $h=90$ nm, and asymmetry $\alpha=0.6$. 
    Dashed lines indicate the  grating mode introduced by the substrate. 
    Dotted lines represent the excitonic resonance. Solid and dot-dashed lines correspond to fit results of the lower (LPB) and upper (UPB) polariton branches, respectively. 
    Depending on the propagation direction in $k$-space, the polaritonic modes exhibit either parabolic or strongly reduced curvature. 
    \textbf{c}~Absolute curvature $|\kappa(0)|$ at the $\Gamma$-point for the parabolic and flat dispersions under $\vec{E}_\mathrm{exc}\parallel\vec{b}$ excitation. Fitted values for the LPB are $0.065 \pm 0.007$ and $3.976 \pm 0.103$ and for the UPB $0.101 \pm 0.001$ and $5.684 \pm 0.333$.
    \textbf{d}~Curvature ratio, directly quantifying the dispersion anisotropy (for $\vec{E}_\mathrm{exc}\parallel\vec{b}$ we find for the LPB $61.0 \pm 6.7$ and for the UPB $56.1 \pm 6.8$ and for $\vec{E}_\mathrm{exc}\perp\vec{b}$ we find for the LPB $104.6 \pm 5.0 $ and the UPB $152.2 \pm 7.3$). 
    \textbf{e} Absolute curvature $|\kappa(0)|$ at the $\Gamma$-point for the parabolic and flat dispersions under $\vec{E}_\mathrm{exc}\perp\vec{b}$ excitation.
    Fitted values for the LPB are $0.054 \pm 0.002$ and $5.614 \pm 0.087$ and for the UPB $0.115 \pm 0.005$ and $17.54 \pm 0.02$.
    Absolute curvature $|\kappa_0|$ and curvature ratio are presented as the mean fit values, $\pm$ the standard deviation of the fit.}
    \label{fig:FlatbandExp}
\end{figure}

\clearpage
\section{Experimental Hopfield Coefficients}
\label{sec:hopfield}
To quantify the ratio between excitonic and photonic modes in the resulting lower and upper polaritons, we express the polaritonic eigenstates $|LP\rangle$ and $|UP\rangle$ as linear combinations of the qBIC $|qBIC\rangle$ and exciton $|Ex\rangle$:
\begin{equation}
|LP\rangle = X_{\mathrm{LP}}|Ex\rangle + B_{\mathrm{LP}}|qBIC\rangle,
\qquad
|UP\rangle = X_{\mathrm{UP}}|Ex\rangle + B_{\mathrm{UP}}|qBIC\rangle,
\end{equation}
where $X_{\mathrm{LP}}$, $X_{\mathrm{UP}}$, $B_{\mathrm{LP}}$ and $B_{\mathrm{UP}}$ are the Hopfield coefficients with:
\begin{equation}
|X_{\mathrm{LP}}|^2 = |B_{\mathrm{UP}}|^2 = \frac{1}{2}
\left(
1 + \frac{\Delta}{\sqrt{\Delta^2 + \Omega_R^2}}
\right), \ \ \text{and} \ \
|X_{\mathrm{UP}}|^2 = |B_{\mathrm{LP}}|^2 = \frac{1}{2}
\left(
1 - \frac{\Delta}{\sqrt{\Delta^2 + \Omega_R^2}}
\right).
\end{equation}
Here $\Delta = \omega_{\mathrm{qBIC}} - \omega_{\mathrm{EX}}$ is the detuning and $\Omega_R$ is the Rabi splitting extracted from TCMT fits (Sec.~\ref{section:TCMT fit}).
\subsection{Hopfield Coefficients at the $\Gamma$-point}

\begin{figure}[H]
    \centering
    \includegraphics[width=1\linewidth]{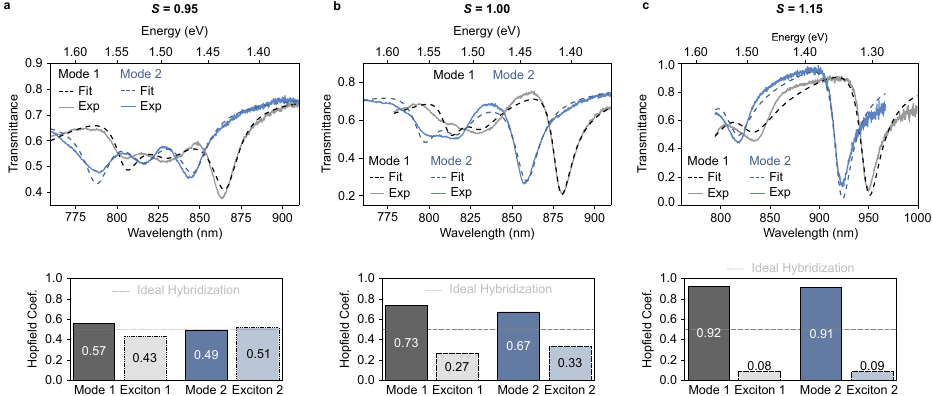}
    \caption{\color{black}\textbf{$\Gamma$-point transmittance fits and extracted Hopfield coefficients.}  
    \textbf{a}~Experimental transmittance spectra at the $\Gamma$-point for the nominally optimally coupled sample ($S=0.95$) shown in Fig.~5 of the main text, together with TCMT fits for both Mode~1 and Mode~2. The extracted Hopfield coefficients indicate near-balanced light--matter hybridization. The corresponding momentum-space dispersion and flatband formation are discussed in Fig.~\ref{fig:FlatbandExp}.  
    \textbf{b}~Same analysis for the more weakly coupled, predominantly photonic sample ($S=1.00$) shown in Fig.~3 of the main text. The extracted Hopfield coefficients reveal an increased photonic fraction. The associated momentum-space behavior is discussed in Fig.~3 of the main text.  
    \textbf{c}~Same analysis for the far-detuned qBIC sample ($S=1.15$), where the excitonic Hopfield coefficients drop below 0.1 for both modes. The corresponding dispersion and reduced flatness in momentum space are analyzed in Fig.~\ref{fig:FlatbandPhotExp}. These measurements were performed on a different sample than the geometric scaling sweep used for the $\Gamma$-point strong-coupling analysis in Fig.~\ref{fig:tcmt-fits}, which results in a slight shift of the scaling factor at which ideal hybridization occurs.
    }
    \label{fig:HopfieldsMilano}
\end{figure}

\clearpage
\subsection{Hopfield Coefficients in Momentum Space}
\begin{figure}[ht]
    \centering
    \includegraphics[width=1\linewidth]{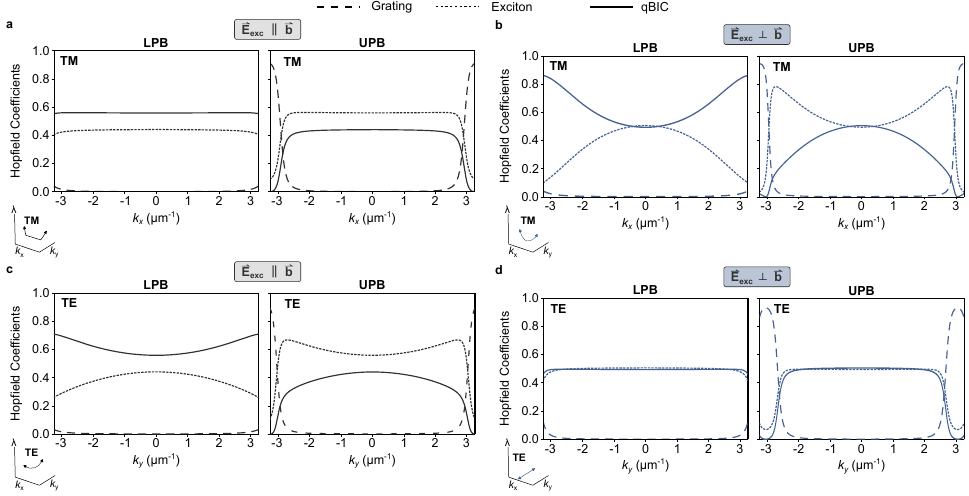}
    \caption{\textbf{Hopfield Coefficients in Momentum Space for Strongly Coupled Metasurface.}
    \textbf{a} TM-polarized modes excited with $\vec{E}_{\mathrm{exc}}\parallel\vec{b}$ showing the Hopfield coefficients of the upper (UPB) and lower (LPB) polariton branches as a function of momentum $k_x$. The dashed and solid lines represent the excitonic and photonic qBIC fractions, respectively. 
    \textbf{b} TM mode for $\vec{E}_{\mathrm{exc}}\perp\vec{b}$.
    \textbf{c} TE mode excited with $\vec{E}_{\mathrm{exc}}\parallel\vec{b}$ plotted as a function of $k_y$.
    \textbf{d} TE modes for $\vec{E}_{\mathrm{exc}}\perp\vec{b}$. Overall, the momentum-dependent Hopfield coefficients directly mirror the anisotropic polaritonic dispersions, with parabolic and flat dispersions leading to correspondingly momentum-dependent and momentum-independent polaritonic mixing fractions. The metasurface corresponds to the sample with $S=0.95$ shown in Fig.~5 of the main text.
    }
    \label{fig:Hopfield_Momentum}
\end{figure}

\clearpage
\section{Experimental k-Space Setup}

\begin{figure}[ht]
    \centering
    \includegraphics[width=0.5\linewidth]{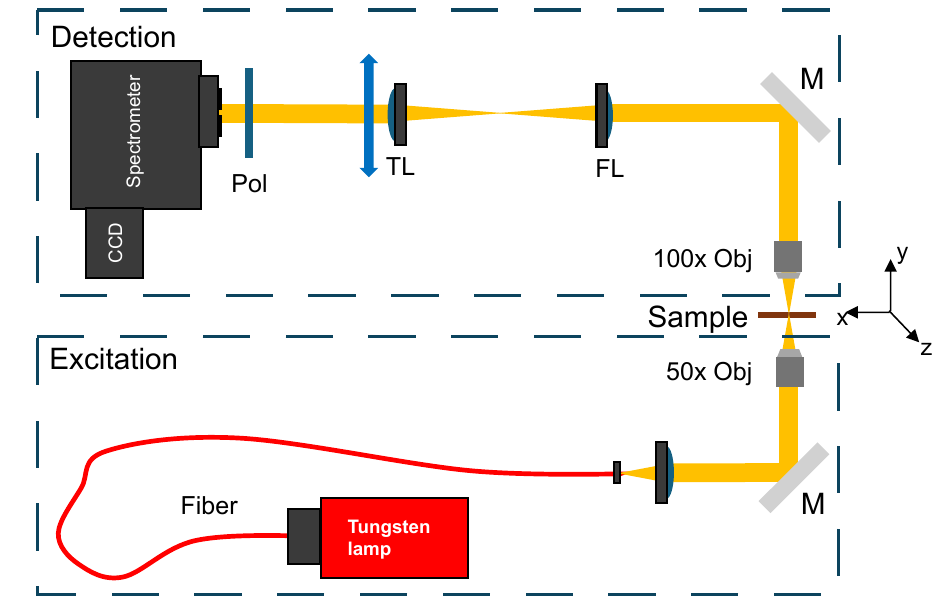}
    \caption{\textbf{Sketch of the Fourier-space Hyperspectral Microscope.} Scheme of the hyperspectral polarization-resolved k-space imaging setup in transmission, implemented with a push-broom approach. M: mirror; CL: collimation lens; Obj: objective lens; FL: Fourier lens; TL: tube lens; Pol: polarizer}
    \label{fig:setup}
\end{figure}

\clearpage

\color{black}
\section{Topological Photonic Landscape as a Function of the in-plane Anisotropy Orientation}
\begin{figure}[H]
    \centering
    \includegraphics[width=\linewidth]{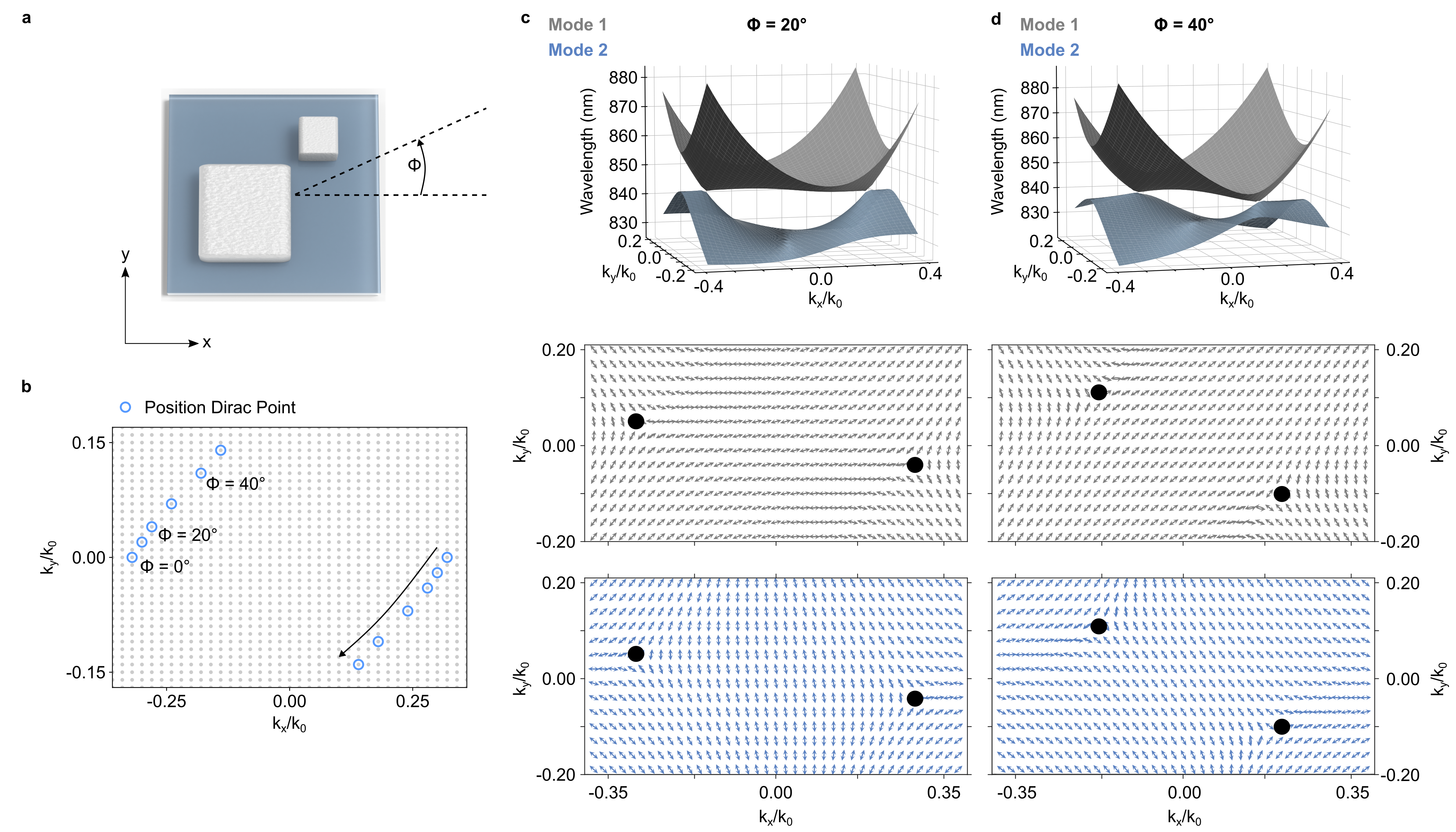}
    \caption{\color{black}\textbf{Topological Photonic Landscape as a Function of the In-plane Anisotropy
    Orientation $\Phi$}. \textbf{a} Schematic definition of the anisotropy orientation angle $\Phi$.
    \textbf{b} Position of the Dirac points in momentum space as $\Phi$ increases from $0^\circ$ to
    $45^\circ$. \textbf{c}, \textbf{d} Band structures (top) and far-field polarization maps for the two qBICs (middle, bottom) for \textbf{c} $\Phi = 20^\circ$ \textbf{d} and $\Phi = 40^\circ$. Black dots on the far-field polarization maps mark the Dirac points. The simulation was performed using the same metasurface as in Fig. 2 of the main text, with period $\Lambda$ = 535 nm, base side length $w_0$ = 250 nm, $\alpha$ = 0.6, height $h$ = 70 nm, refractive index $n_0=\sqrt{18}$, and $\Delta n=0.2 $. 
    }
    \label{fig:top_orientation}
\end{figure}
\clearpage

\printbibliography[title={Supplementary References}]
\end{refsection}

\end{document}